\documentclass[slac_one]{revtex4}
\usepackage{graphicx}
\usepackage{fancyhdr}

\begin{document}

\def\d{\partial}
\def\um{\,{\rm \mu m}}
\def\mm{\,   {\rm mm}}
\def\cm{\,   {\rm cm}}
\def \m{\,   {\rm  m}}
\def\ps{\,   {\rm ps}}
\def\ns{\,   {\rm ns}}
\def\us{\,\mu{\rm  s}}
\def\ms{\,   {\rm ms}}
\def\nA{\,   {\rm nA}}
\def\uA{\,\mu{\rm  A}}
\def\mA{\,   {\rm mA}}
\def\A {\,   {\rm  A}}
\def\mV{\,   {\rm mV}}
\def\V {\,   {\rm  V}}
\def\fF{\,   {\rm fF}}
\def\pF{\,   {\rm pF}}
\def\GeV{\, {\rm GeV}}
\def\MHz{\, {\rm MHz}}
\def\uW{\,\mu{\rm  W}}
\def\e {\,  {\rm e^-}}

\renewcommand{\labelenumi}{\arabic{enumi}}
\renewcommand{\labelitemi}{-}
\def\ppp {\hfill\break}


\title{Pixel Vertex Detectors} 

%

\author{N. Wermes}
\affiliation{Bonn University, Bonn, Germany}

\begin{abstract}
Pixel vertex detectors are THE instrument of choice for the
tracking of charged particles close to the interaction point at
the LHC. Hybrid pixel detectors, in which sensor and read-out IC
are separate entities, constitute the present state of the art in
detector technology. Three of the LHC detectors use vertex
detectors based on this technology. A development period of almost
10 years has resulted in pixel detector modules which can stand
the extreme rate and timing requirements as well as the very harsh
radiation environment at the LHC for its full life time and
without severe compromises in performance. This lecture reviews
the physics and technology of pixel detectors for tracking and
vertexing at the LHC.
\end{abstract}

\maketitle

\thispagestyle{fancy}

\section{Introduction}
\subsection{From gas-filled wire chambers to pixel vertex
detectors}
Advances in particle tracking have always lead to
breakthroughs in experimental methods and hence to a new quality
of experiments. A prominent example of this statement is the
Nobel-awarded invention of the multi-wire proportional chamber by
G. Charpak in 1968 \cite{charpak1,charpak2}, which for the first
time allowed the electronic detection of particle tracks to an
accuracy in the order of a mm and below. Later ($\sim$1975)
spatial resolutions in the order of 100$\um$ were obtained with
drift chambers. These detectors typically had a readout density of
0.05 channels/cm$^2$. So called vertex drift chambers (see e.g.
\cite{jaros}) improved the resolution of such chambers down to
about 50$\um$ with a readout density of 0.1 channels/cm$^2$. These
devices enabled the detection of decay vertices, and hence
measurements of the life times of long lived ($\sim$ps) particles.

Silicon micro strip detectors were developed in the early eighties
\cite{hyams}. With this new type of detectors spatial resolutions
in the order of 10$\um$ became accessible for the first time. The
identification of secondary vertices and and hence of particle
life times became precision measurements. These devices had a
channel density of the order of 100 channels/cm$^2$.

Pixel detectors~\cite{pixelbook}, finally, belong to this category
of instruments leading the way to a new frontier in measurement
techniques and hence in physics. At the LHC, close to the
interaction point, no other detector instrument is capable to cope
as well with the high density and rate of particle tracks and
stand the fierce radiation environment. There is no large
improvement in spatial resolution when comparing to strip
detectors, however pixel detectors return true 3D space points, a
necessity for pattern recognition and tracking in the LHC
environment near the interaction point. The channel density
increases by more than an order of magnitude compared to strip
detectors to about 5000 channels/cm$^2$. Fast readout of such a
large (in terms of channels) and complex system required new
technologies and methods, which have been developed during the
past decade and are subject to this lecture.

\subsection{Tracking detectors at the LHC}
At the LHC with a center of mass energy of 14 TeV on average 22
interactions with $\sim$1200 tracks occur during one bunch
crossing every 25 ns. This is a track rate of 50 GHz, about 10$^6$
times the track rate that LEP encountered. The total particle
fluence is 10$^{15}$ n$_{\rm eq}$/cm$^2$ over a projected 10 year
life time at the LHC, where the damage created by the charged and
neutral particles, mostly pions and protons near the IP, has been
normalized to the equivalent of the damage of 1 MeV neutrons
(n$_{\rm eq}$). This fluence of particles causes lattice damage by
collisions with the lattice atoms (non-ionizing energy loss), but
also by ionization of atoms which corresponds to a total dose of
600 kGy in 250$\um$ silicon bulk material assuming minimum
ionizing particles (mips).

\ppp Pixel detectors are most important for the following
measurement tasks, in the order of importance

\begin{figure}[!ht]
\begin{center}
  \includegraphics[width=.60\textwidth]{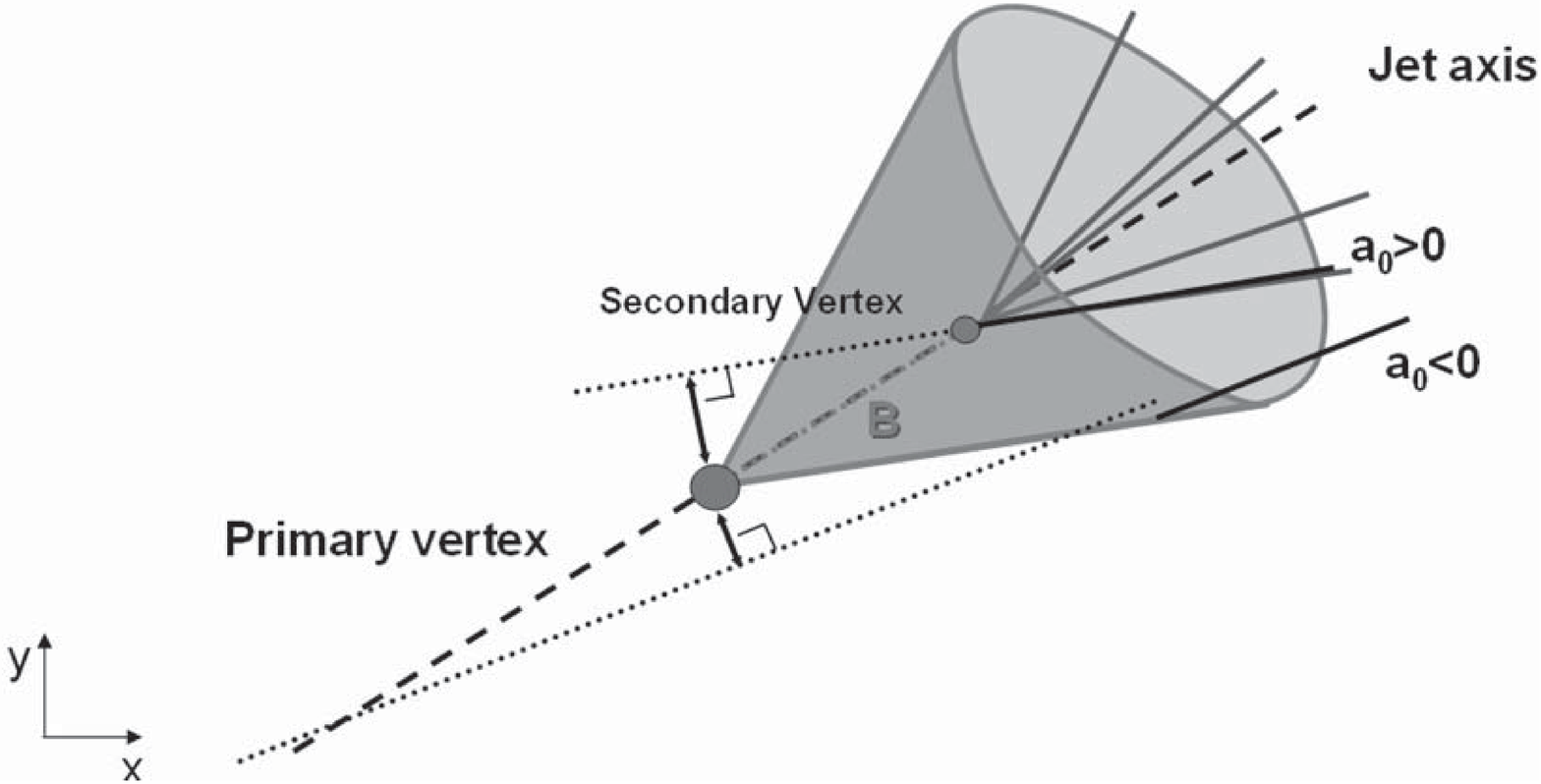}
\end{center}
\caption{\label{impactparam}}{Sketch of a secondary vertex which
is displaced from the primary vertex. The impact parameter is
defined as the perpendicular distance of closest approach of a
track to the primary vertex and has a positive sign if it lies in
the same hemisphere as the track, otherwise the sign is negative.}
\end{figure}

\begin{enumerate}

\item Pattern recognition and tracking

Precision tracking points are provided in three dimensions which
can be used as precise seeds for the reconstruction of tracks.
With respect to this task a pixel layer is probably as valuable as
3 to 4 layers of microstrip detectors which need x,y as well as
u,v strip orientations to resolve the ambiguities existing in an
environment with many tracks per uni area.

\item Vertexing (primary and secondary vertex, see
Fig.\ref{impactparam})

The ATLAS inner detector for example projects the following
performance values
\begin{itemize}
\item impact parameter resolution: $\sim$10$\mu$m (r$\phi$),
$\sim$70$\mu$m (z)
\item secondary vertex resolution: $\sim$50$\mu$m (r$\phi$),
$\sim$70$\mu$m (z)
\item primary (main) vertex resolution: $\sim$11$\mu$m (r$\phi$),
$\sim$45$\mu$m (z)
\item (life) time resoution: $\sim$70 ps
\end{itemize}
The impact parameter (c.f. Fig. \ref{impactparam}) resolution in
ATLAS is given by
$$
\hspace{-9cm} \sigma_b \approx 10 \oplus \frac{98}{p_T \sqrt{sin
\theta}} \, \, \mu m
$$

\item Momentum measurement

For ATLAS the expected momentum resolution obtained with the
complete Inner Detector is
$$
\hspace{-8cm}\frac{\sigma_{p_{T}}}{p_{T}} = 0.03\% \, p_{T} {\rm
(GeV)} \, + \, 1.2\%
$$
\end{enumerate}

\begin{figure}[!ht]
\begin{center}
  \includegraphics[width=.8\textwidth]{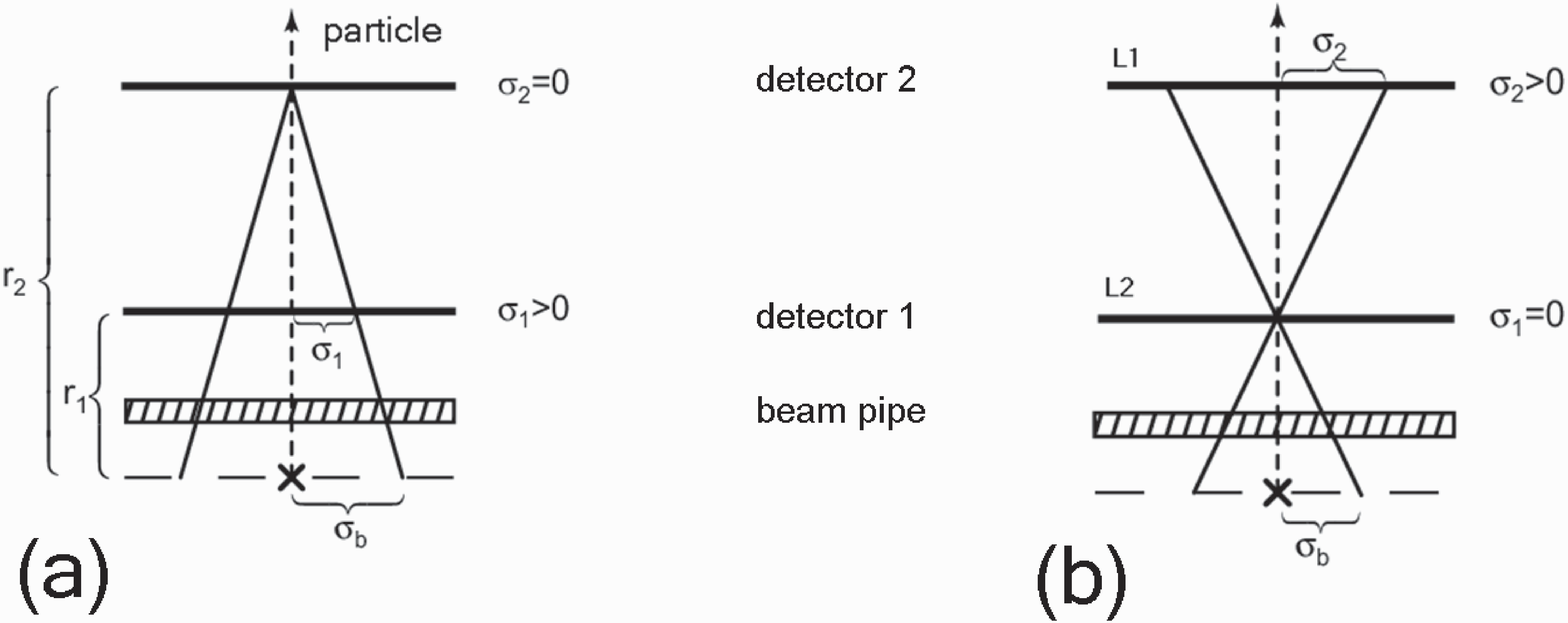}
\end{center}
\caption{\label{idealdet}}{Idealized two layer detector. $\times$
marks the interaction point. $\sigma_b$ is the extrapolated
interaction point of the track at the IP (impact parameter). (a)
configuration assuming an ideal layer 2 ($\sigma_2$=0), (b)
configuration assuming an ideal layer 1 ($\sigma_1$=0).}
\end{figure}

In order to assess which are the important parameters for a micro
vertex detector, a simple 2-layer detector example already
provides the most important insights. Consider in
Fig.~\ref{idealdet} configuration in which first the outer layer,
detector-2 is assumed to have perfect resolution ($\sigma_2$=0,
Fig. \ref{idealdet}(a)). The impact parameter error of the track
$\sigma_b$ is thus determined by $\sigma_1$, the resolution of the
first layer and the aspect ratio of the two layers
$$
\frac{\sigma_b}{\sigma_1} = \frac{r_2}{r_2 - r_1}
$$

Reversing now the situation and assuming that the first layer has
perfect resolution ($\sigma_1$=0, Fig. \ref{idealdet}(b)) yields
$$
\frac{\sigma_b}{\sigma_2} = \frac{r_1}{r_2 - r_1}
$$

Adding the two contributions to $\sigma_b$ in quadrature and
adding also a term for multiple scattering yields

$$
\sigma_b^2=\left( \frac{r_1}{r_2 - r_1} \sigma_2 \right)^2 +
\left( \frac{r_2}{r_2 - r_1} \sigma_1 \right)^2 + \sigma_{MS}^2
$$
from which it is obvious what to pay attention to for vertexing:
the innermost layer of a vertex detector must be as close to the
IP as possible, the resolution ($\sigma_1$) of this innermost
layer is most important, and the material measured in radiation
lengths $x/X_0$ must be as little as possible, because
$\sigma_{MS} \sim 1/p \sqrt{x/X_0}$.

\begin{figure}[!h]
\begin{center}
  \includegraphics[width=.45\textwidth]{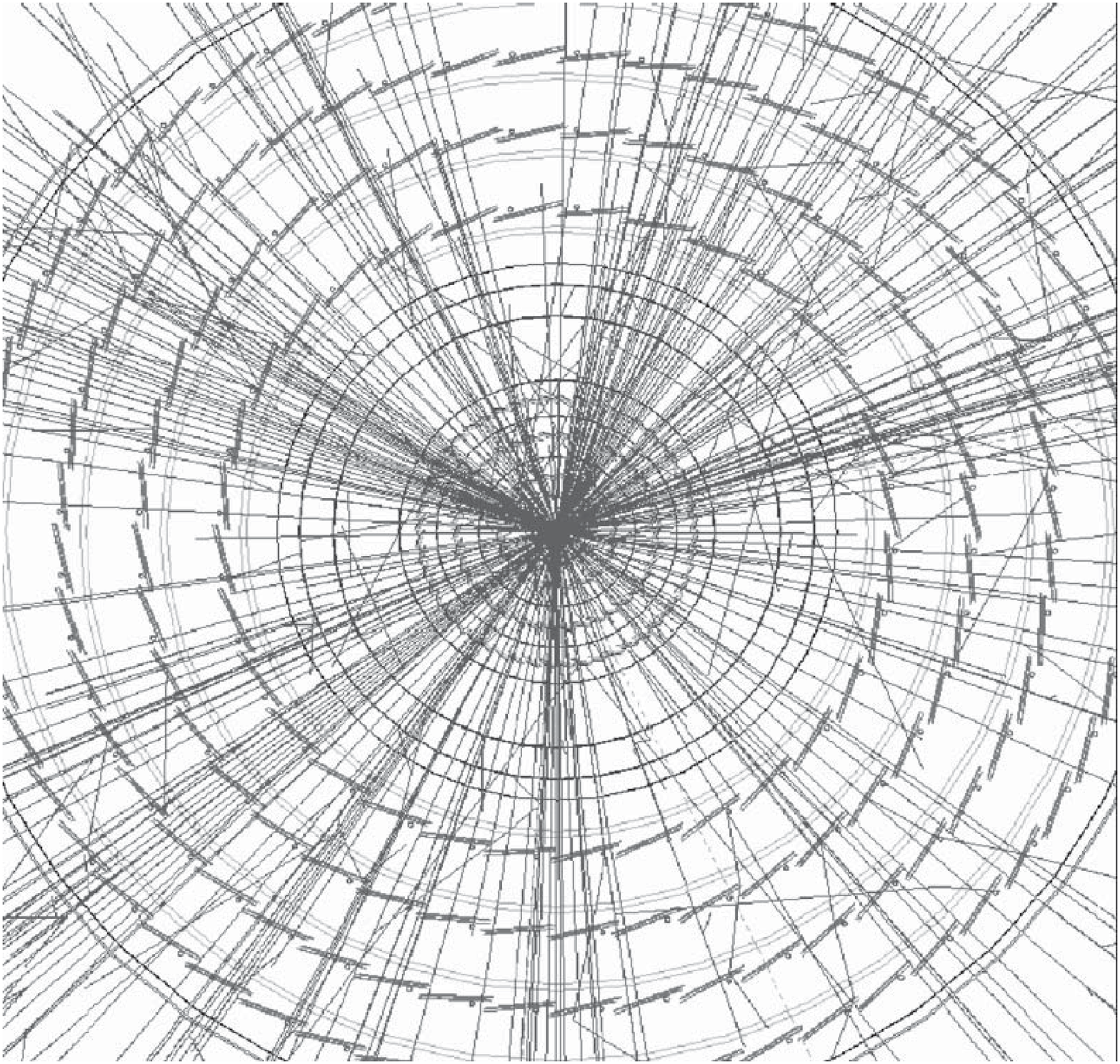}
\end{center}
\caption{\label{LHC-event}}{Simulated event of the reaction pp
$\rightarrow$ ttH, H $\rightarrow$ bb, tt $\rightarrow$ W(l$\nu$)b
W(qq)b} at the LHC. Also shown are the positions of the ATLAS
silicon tracking detectors: pixels and SCT.
\end{figure}

Figure \ref{LHC-event} shows a simulated event of the reaction pp
$\rightarrow$ ttH, which demonstrates again the importance of
3D-hits from pixel detectors for track reconstruction.

The inner tracking systems of ATLAS, CMS and ALICE all employ
semiconductor detectors, as the event rate and complexity does not
allow large volume gas-filled detectors which were still the best
suited tracking detectors for experiments at LEP and SLC. Some
essential parameters of the tracking systems are give in
Table~\ref{trackers}.

\begin{table}[!h]
\begin{center}
\caption{Central tracking systems of the LHC experiments}
\begin{tabular}{|l|l|c|c|c|c|}
\hline \textbf{} & & points & $\sigma$(r$\phi$) ($\um$) &
$\sigma$(Rz) ($\um$) & area (m$^2$) \\
\hline ATLAS & pixels   & 3  & 10 & 60 & $\sim$ 1.8 \\
\hline       & SCT      & 4  & 17 & 580 & $\sim$ 60 \\
\hline       & TRT      & 36 & 170& -- & $\sim$300 equiv.\\
\hline
\hline CMS   & pixels   & 3  & 10 & 20 & $\sim$ 1 \\
\hline       & strips   & 10  & 23 - 52 & 23 - 52 & $\sim$ 200 \\
\hline
\hline ALICE & pixels   & 2  & 12 & 100 & $\sim$ 0.2 \\
\hline       & Si-drift & 2  & 38 & 28 & $\sim$ 1.3 \\
\hline       & strips   & 2  & 20 & 830 & $\sim$ 4.9\\
\hline
\end{tabular}
\label{trackers}
\end{center}
\end{table}

The ATLAS inner detector has three systems: a 1.8 m$^2$ pixel
detector, a semiconductor tracker (SCT) consisting of Si micro
strip detectors of about 60 m$^2$ and a transition radiation
tracker (TRT) made of individual straw tube, providing on average
36 points on a track. The pixel detector comes as 3 barrel layers
and 3 disks on either side. It has a total of about 8 $\times$
10$^7$ pixels with dimensions of $50\times400$$\um^2$.

\begin{figure}[!h]
\begin{center}
  \vspace {0.5cm}
  \includegraphics[width=.90\textwidth]{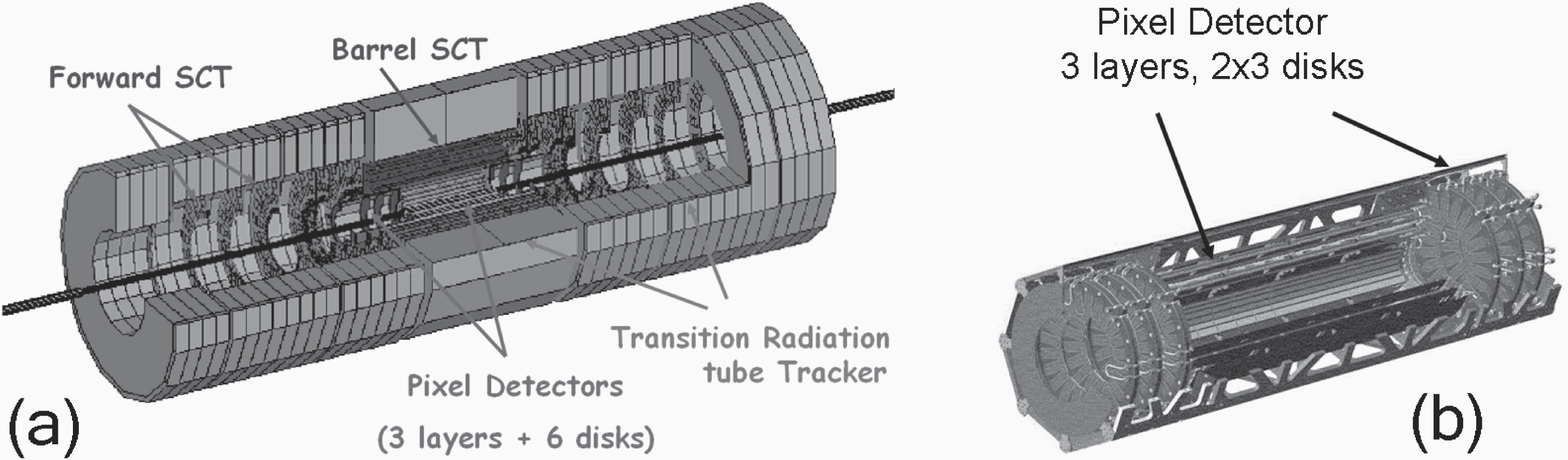}
\end{center}
\caption{\label{ATLAS-ID}}{The ATLAS Inner Detector (a) has a
pixel detector, a semiconductor tracker (SCT) and a Transition
Radiation Tracker (TRT). The pixel detector (b) consists of 3
barrel layers and 2$\times$3 disks.}
\end{figure}

\begin{figure}[!h]
\begin{center}
  \includegraphics[width=.70\textwidth]{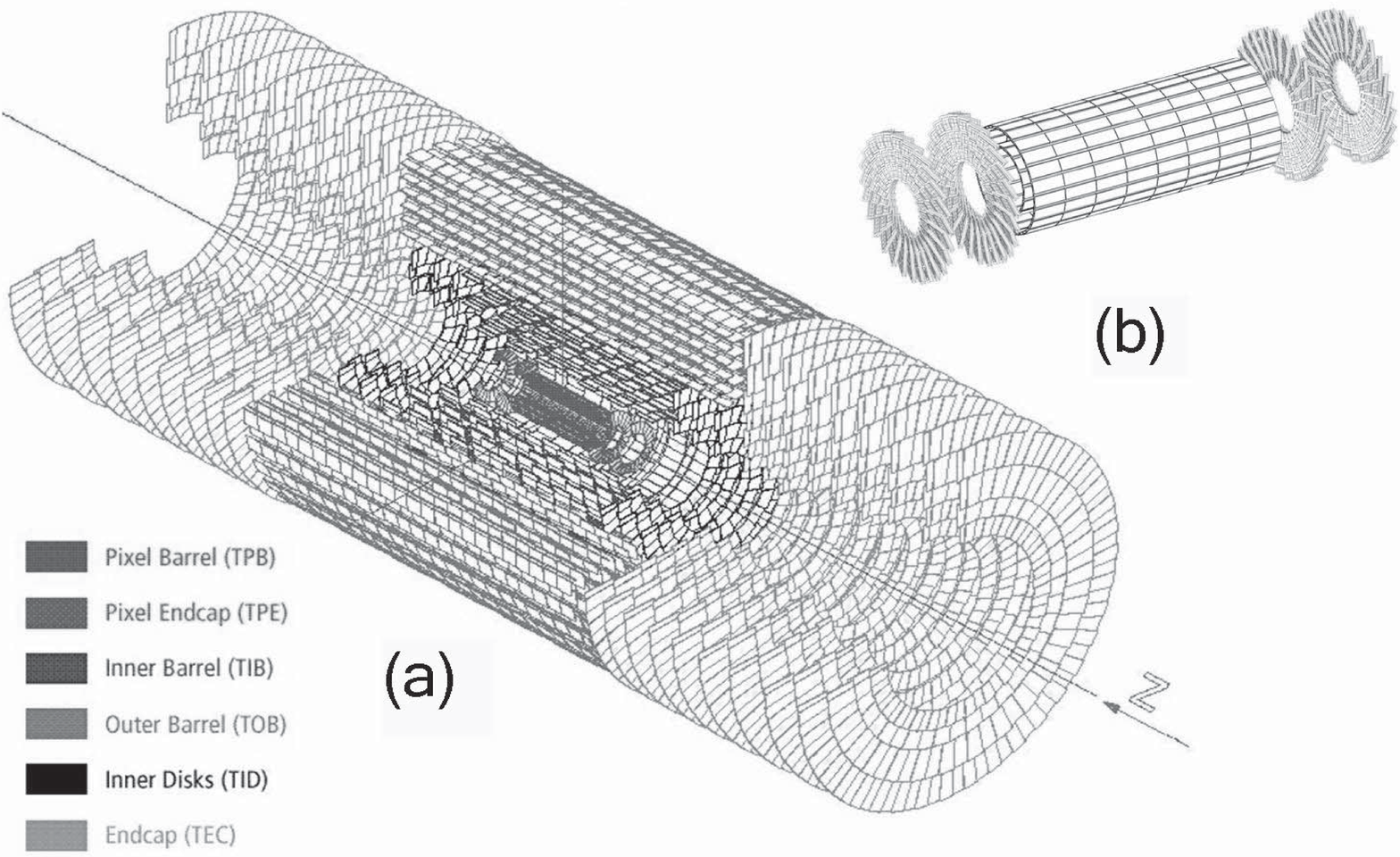}
\end{center}
\caption{\label{CMS-ID}}{The CMS Inner Detector (a) has a pixel
detector and a large Si-microstrip tracker. The pixel detector (b)
consists of 3 barrel layers and 2$\times$2 disks.}
\end{figure}

\begin{figure}[!h]
\begin{center}
  \includegraphics[width=.30\textwidth]{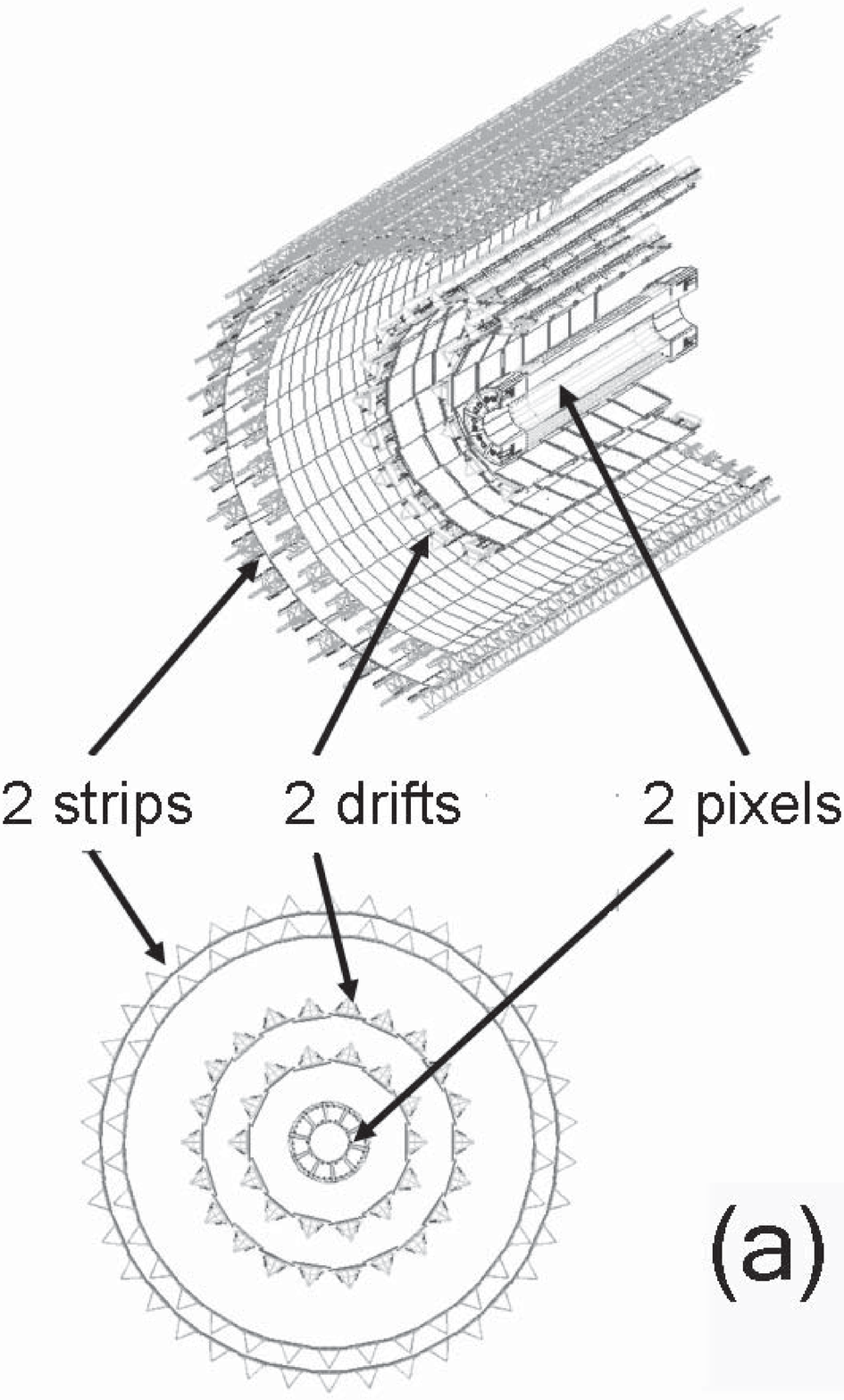}
  \hspace {2cm}
  \includegraphics[width=.35\textwidth]{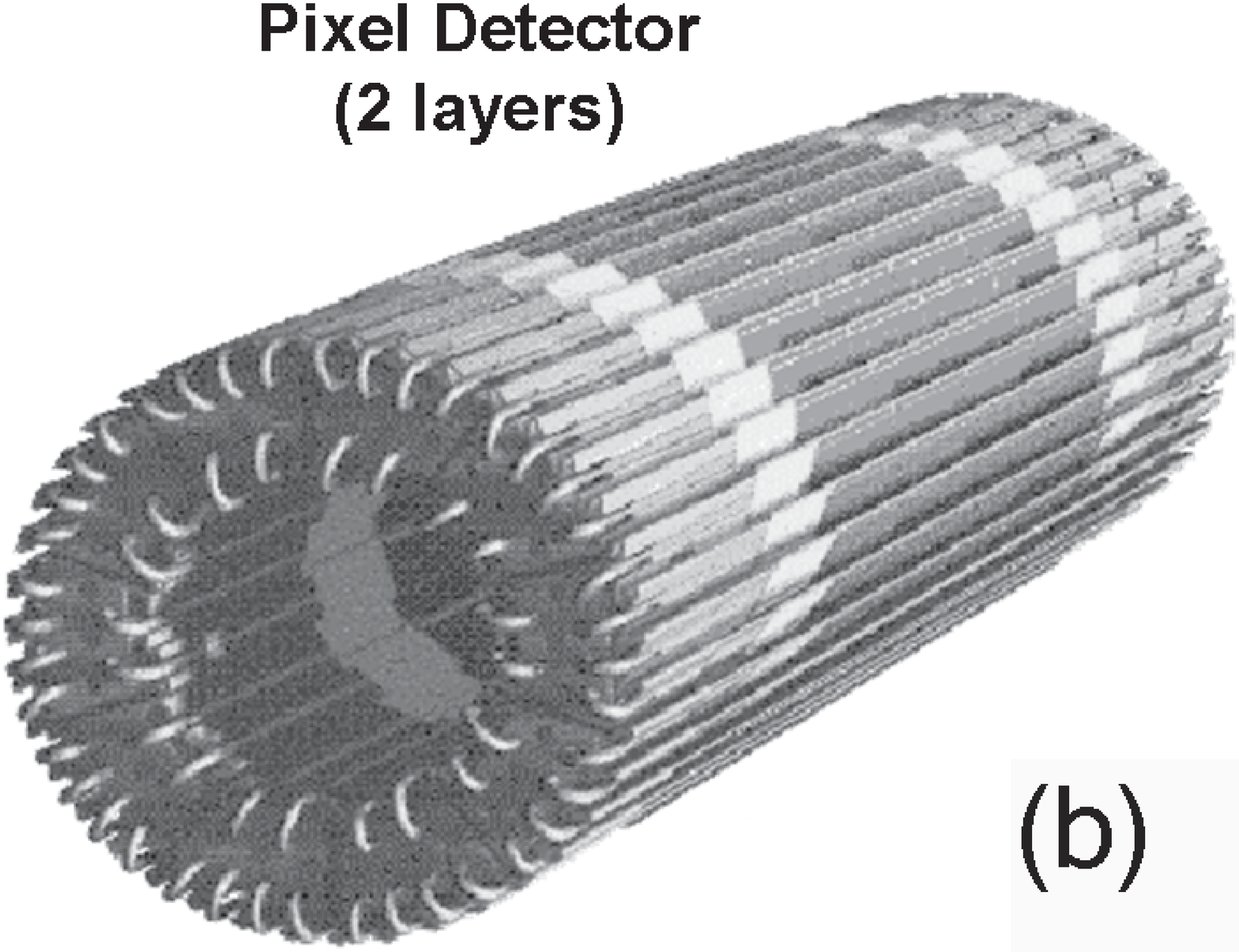}
\end{center}
\caption{\label{ALICE-ID}}{The ALICE Inner Detector has a large
TPC (not shown) plus a silicon tracking system consisting of (a) a
pixel detector (PD), a siicon drift detector (SDD) and a Si-Strip
detector (SSD). The pixel detector (b) consists of 2 barrel layers
and no disks.}
\end{figure}

CMS has a large all silicon tracker, composed of a pixel detector
with 3 barrel layers and 2$\times$2 disks ($\sim$1m$^2$) and a
$\sim$200 $m^2$ silicon microstrip detector with 10 measurement
points. The pixel detector has 3.3 $\times$ 10$^7$ cells with
dimensions of 100$\times$150 $\um^2$. Drawings are shown in
Fig.\ref{CMS-ID}. ALICE, finally, has two layers each of different
technologies~\cite{kotov2006}, pixels, Si drift detectors, and Si
strip detectors as shown in Fig.~\ref{ALICE-ID}.


\section{Hybrid Pixel Detectors for the LHC}
\subsection{The pixel detector principle}
\begin{figure}[!h]
\begin{center}
  \includegraphics[width=.30\textwidth]{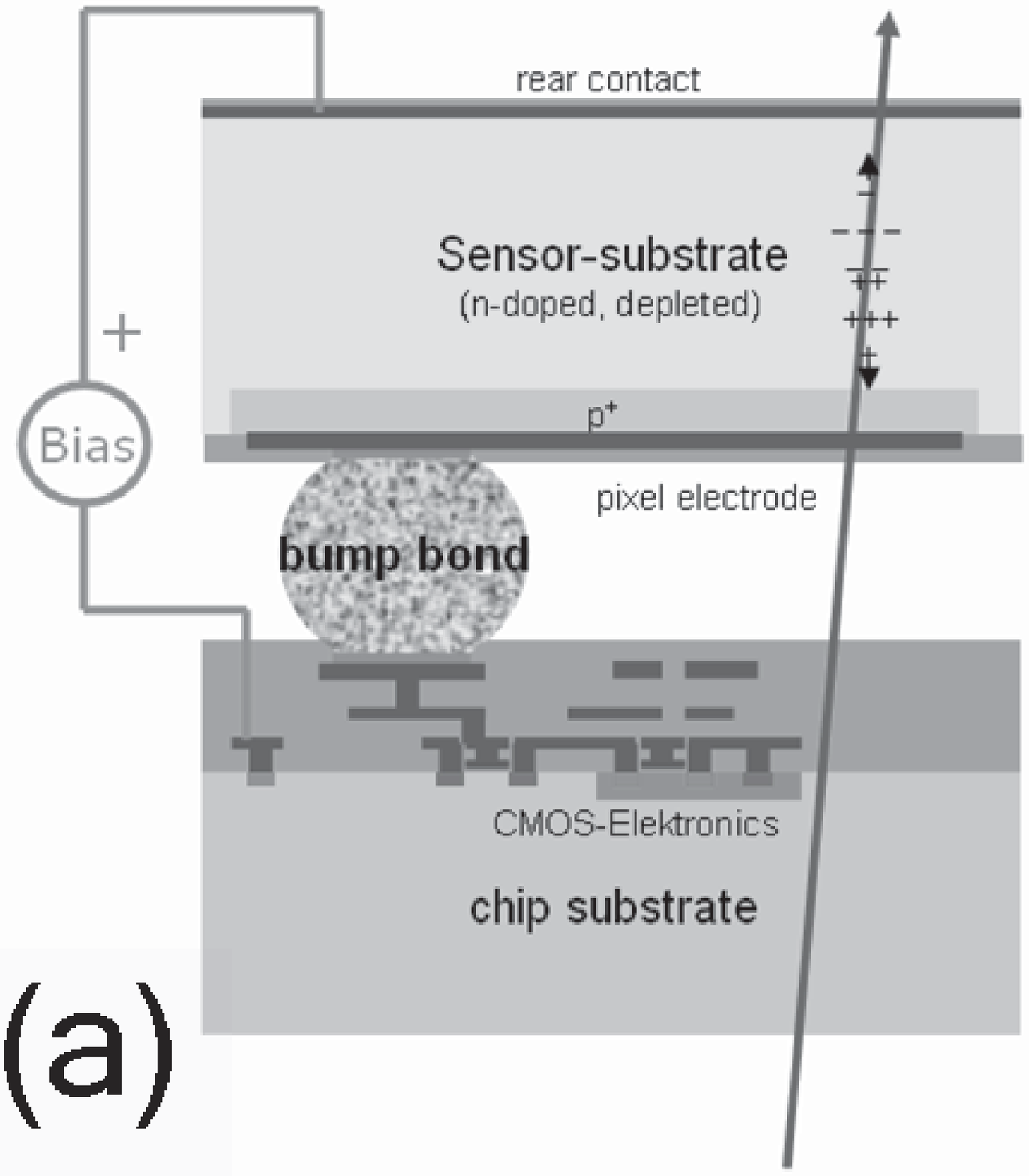}
  \hspace {2cm}
  \includegraphics[width=.55\textwidth]{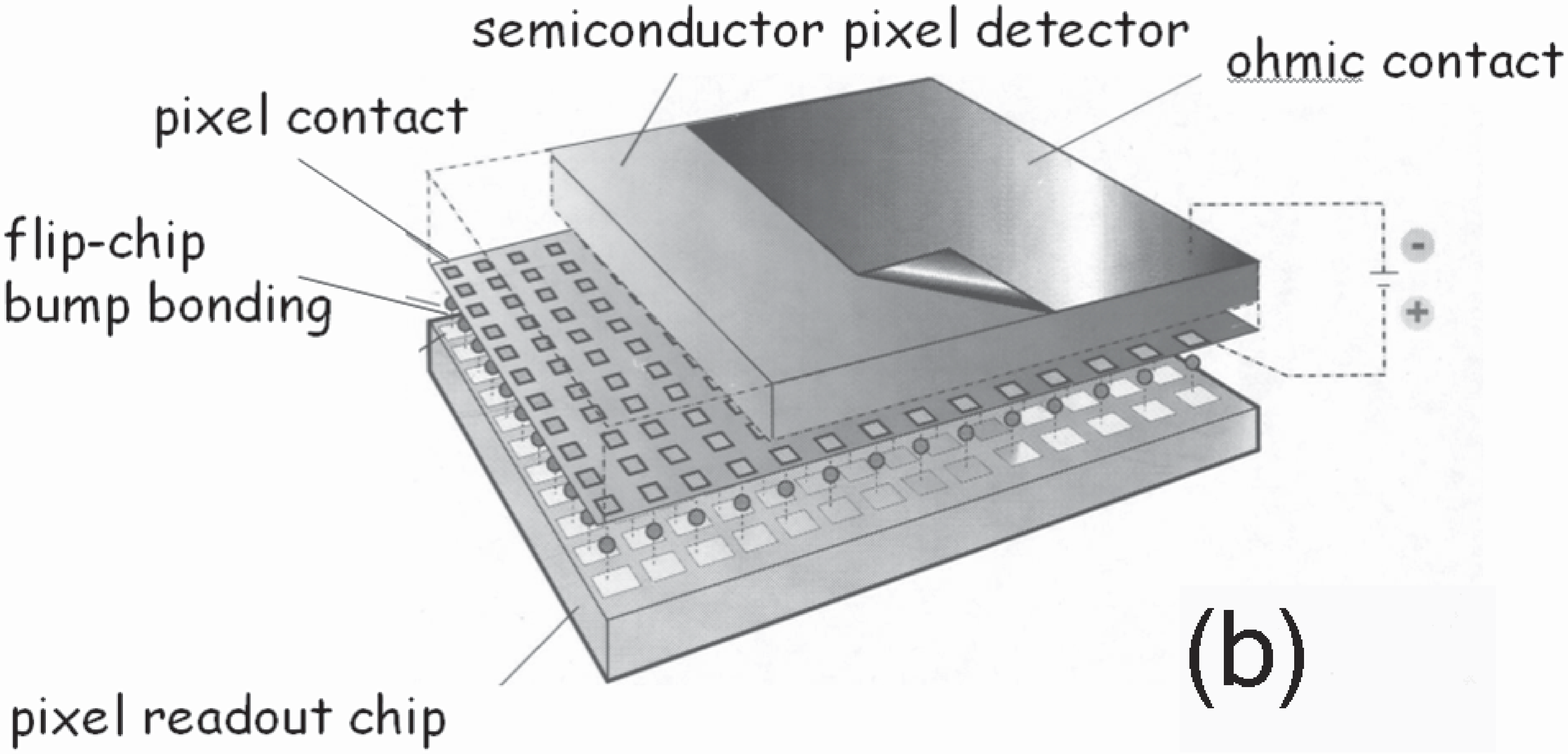}
\end{center}
\caption{\label{hybrid-pixels}}{The hybrid-pixel concept. (a) one
pixel cell showing the charge collecting sensor on top and the
signal amplifying circuitry on the bottom, both connected by a
bump connection.}
\end{figure}

Figures~\ref{hybrid-pixels}(a) and (b) show the principle of the
hybrid pixel technology. A pixellated silicon pn-diode as sensor
and a readout chip, 1-1 corresponding in every pixel cell, are
connected via tiny conductive bumps using the bumping and
flip-chip technology. A traversing particle creates charges
(electrons and holes) in the sensor by ionization. Separated by
the electric field applied to the sensor, the movements of
negative and positive charges induce a signal on the pixel
electrode above (and may be also on its neighbors). The signal is
amplified, discriminated and digitized by the electronics
circuitry in the pixel cell of the chip and is then transmitted to
the periphery.

\begin{figure}[!h]
\begin{center}
  \includegraphics[width=.45\textwidth]{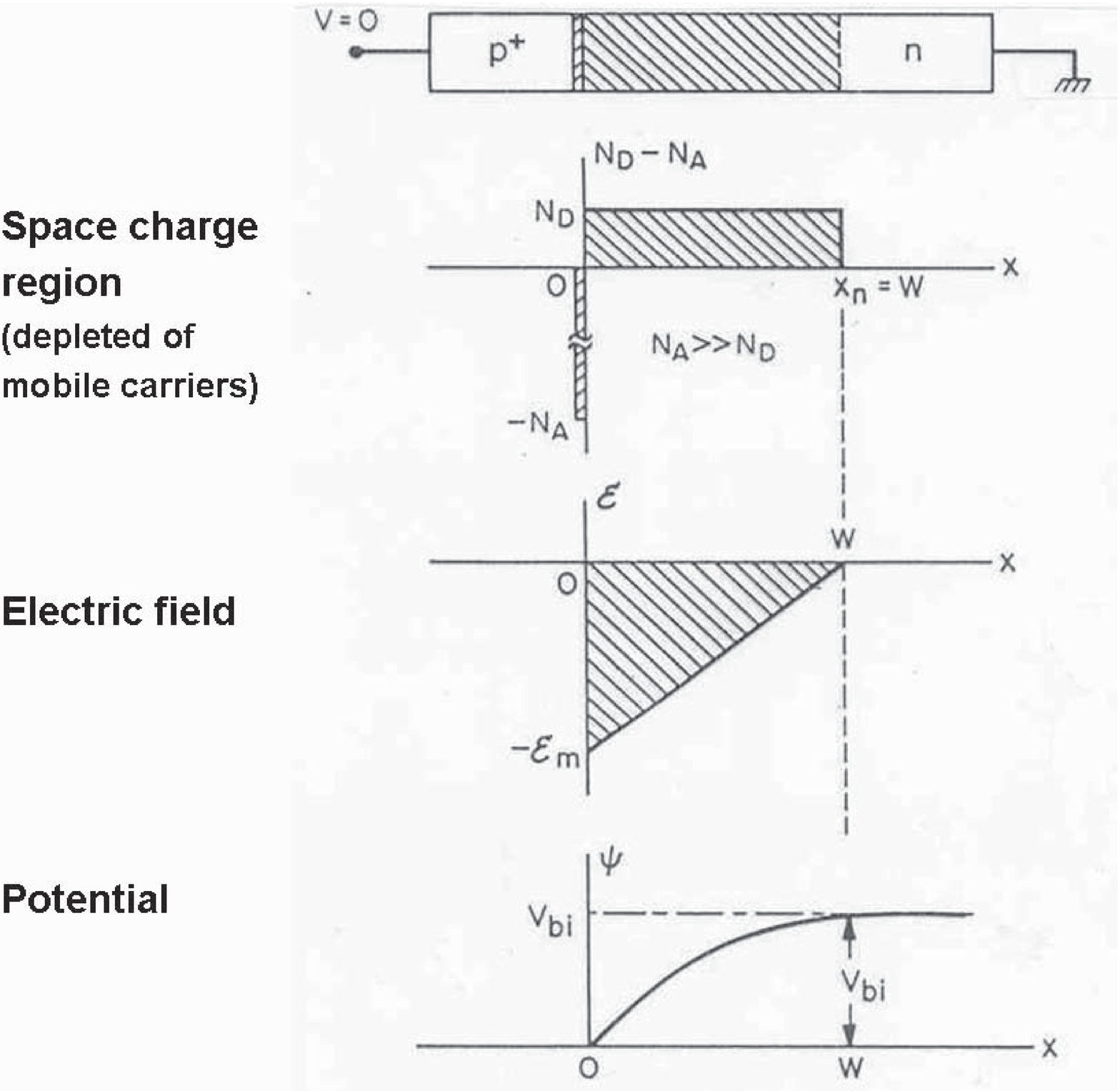}
\end{center}
\caption{\label{pn-diode}}{Characteristics of a reverse biased
abrupt pn-junction: space charge, electric field and potential
distributions.}
\end{figure}

The detection principle is - with one peculiarity, explained below
- the same as that of a pn-junction in reverse bias
(Fig.~\ref{pn-diode}). The characteristic distribution of the
space charge is directly derived from the neutrality condition
\begin{equation}
N_A x_p = N_D x_n
\end{equation}
in which $N_A$,$N_D$ are the number of acceptors and donors,
respectively, and $x_n$, $x_p$ denote the coordinates of the space
charge boundaries. Assuming an abrupt transition between n and p
regions this results in a rectangular shape of the respective
space charge regions with a constant charge density. The electric
field and the built-in potential result from two integrations of
the first Maxwell equation
\begin{eqnarray}
\frac{dE}{dx} = \frac{1}{\epsilon} \rho(x) \nonumber
\end{eqnarray}

\begin{eqnarray}
    \Rightarrow E(x) &=& { \left \{  \begin{array}{rl}
          \frac{-e N_{A}}{\epsilon} \left( x + x_{p} \right) & ;-x_{p} < x < 0\\[.2cm]
          \frac{+e N_{D}}{\epsilon} \left( x - x_{n} \right) & ;0 < x < x_{n}
                   \end{array} \right. }\\ \hfill \nonumber \\
    \Rightarrow V_{{bi}} &=& \frac{e}{2 \epsilon} \left(
N_{A} x_{p}^2 + N_{D} x_{n}^2 \right)
\end{eqnarray}

with a linearly rising and falling electric field with its maximum
at $x=0$.

In the case of a particle detector the doping of one side of the
junction, say the p-side, is very high ($\sim$10$^{19}$
cm$^{-3}$), while the n-side is lightly doped ($\sim$10$^{12}$
cm$^{-3}$). The depletion region thus basically grows from the
junction into the more lightly doped bulk. With an additional
reverse bias voltage V$_{ext}$ = V$_{dep}$ + V, where V$_{dep}$ is
the depletion voltage, the electric field becomes
\begin{eqnarray}\label{eq-field}
E(x) = - \frac{V + V_{dep}}{d} + \frac{2 V_{dep} \, x}{d^2}
\end{eqnarray}

\begin{figure}[h]
\begin{center}
  \includegraphics[width=.40\textwidth]{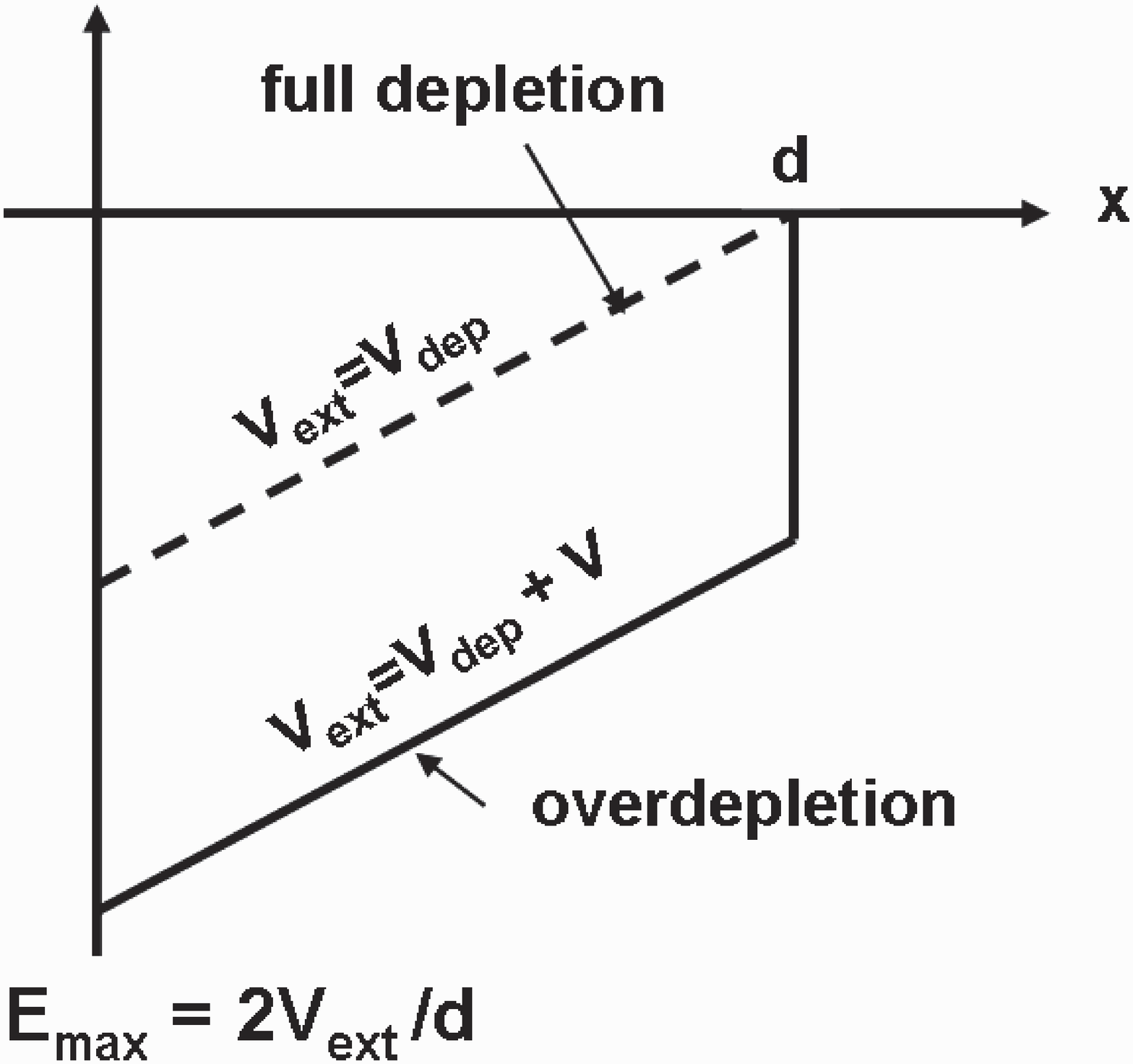}
  \hspace {0.5cm}
  \includegraphics[width=.45\textwidth]{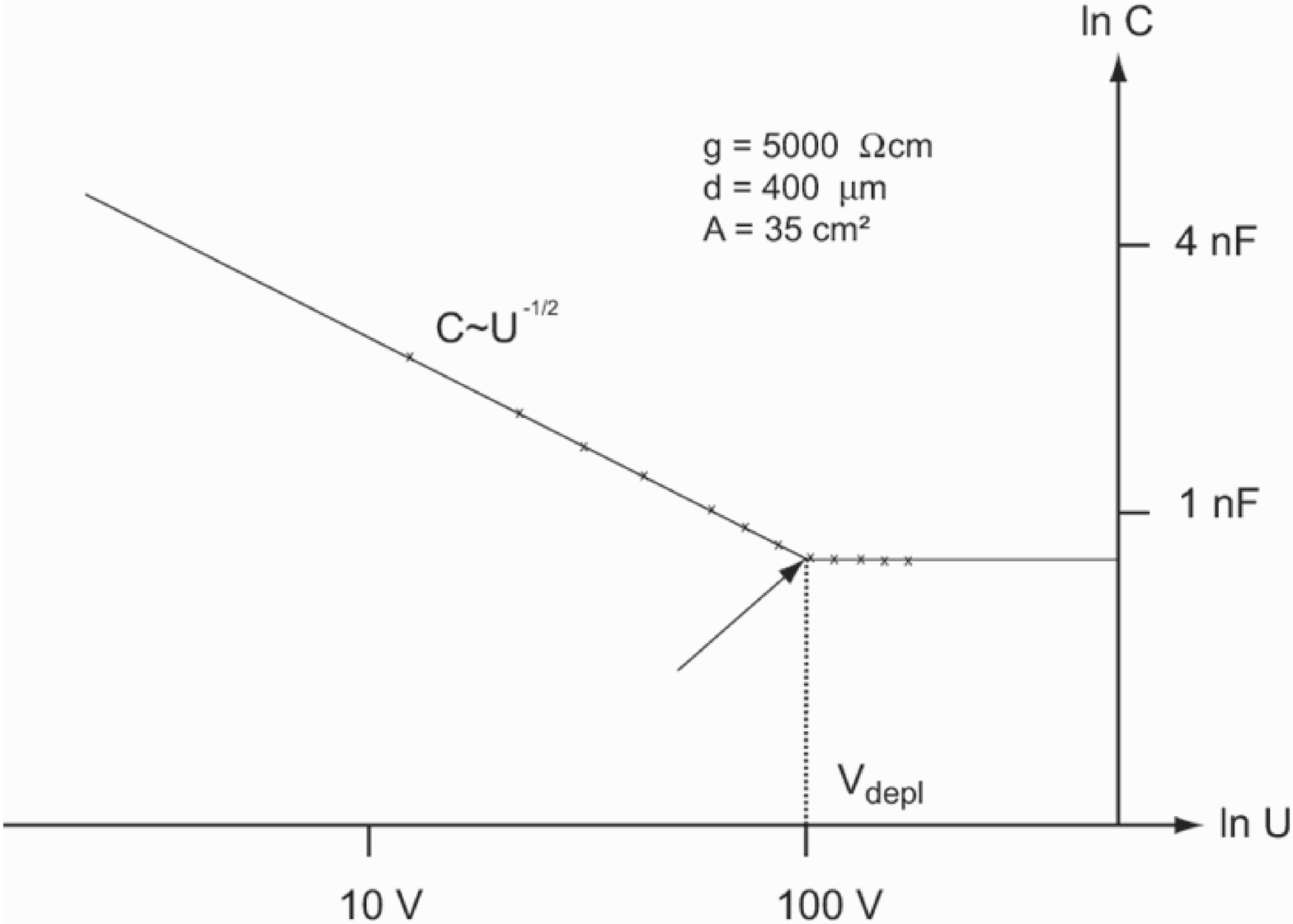}
\end{center}
\caption{\label{Efield}}{(left) Linear shape of the electric field
of a silicon detector in full depletion (dashed line) and in
over-depletion (full line). (right) Determination of the full
depletion voltage by measuring the capacitance.}
\end{figure}

The depletion depth d is proportional to the square root of the
applied external voltage
\begin{eqnarray}
d = x_n = \sqrt{\frac{2\epsilon}{e}\frac{1}{N_D} (V_{bi}-V_{ext} )
}
\end{eqnarray}
and the electric field falls linearly from the front to the back.
Normally sensors are operated with in over-depletion resulting in
a non-zero field at the back side which is beneficial for a fast
charge collection (see Fig.~\ref{Efield}(a)). The capacitance of
the device can be treated as a parallel plate capacitor
\begin{eqnarray}
\frac{\mathrm C}{\mathrm A}  =  \frac{\epsilon \epsilon_0}{d}
\propto \frac{1}{\sqrt{V_{ext}}}
\end{eqnarray}
and grows inversely propostional to $\sqrt{V_{ext}}$. This allows
to determine the full depletion voltage from a capacitance
measurement as shown in Fig.~\ref{Efield}(b).

\subsection{The signal} Although in silicon the band gap is 1.1 eV, 3.61 eV
are needed on average to create one e/h pair as energy is also
absorbed into phonons. A minimum ionizing particle hence creates
by energy loss $\sim$80 e/h pairs per $\um$ path length and
$\sim$20000 e/h in a 250$\um$ thick fully depleted sensor,
corresponding to a total charge of about 3 fC. Note that pixel
sensor are also suited to detect radiation: a 10 keV X-ray photon,
when absorbed, generates about 3000 e/h pairs or 0.5 fC. The
generated charges are separated by the electric field inside the
sensor and drift with drift velocity $v_{drift}(x) = \mu E(x)$,
where x is the depth coordinate, towards their respective
electrodes. The charge cloud widens with time by diffusion while
it drifts to the electrode. The gaussian width of the diffusion
cloud is given by $\sigma_{drift}(t) = \sqrt{2Dt}$, where
D~$\approx$~36~cm$^2$s$^{-1}$ is the diffusion coefficient of
silicon. This leads in typical sensor thicknesses of 250$\um$ and
with typical electric fields and drift velocities ($\sim$5mm/ns)
to drift times of about 10 ns and cloud widths in the order of
8-10$\um$. For pixel detectors the electrode on one side (usually
the one that collects electrons, n$^+$, see section~\ref{sensors})
is segmented into pixels with a typical size of $100 \times 150$
$\um^2$ (CMS) or $50 \times 400$ $\um^2$ (ATLAS and ALICE). The
charge cloud therefore spreads at most over 2 pixels, 4 pixels if
a corner of 4 pixels is hit. Note that for photon detection sensor
materials with a high Z, such as Cd(Zn)Te or HgI$_2$ are
desirable, due to the dependence of the photo effect
$\sim$Z$^{4-5}$, where the exponent changes with the photon
energy.

\begin{figure}[h]
\begin{center}
  \includegraphics[width=.40\textwidth]{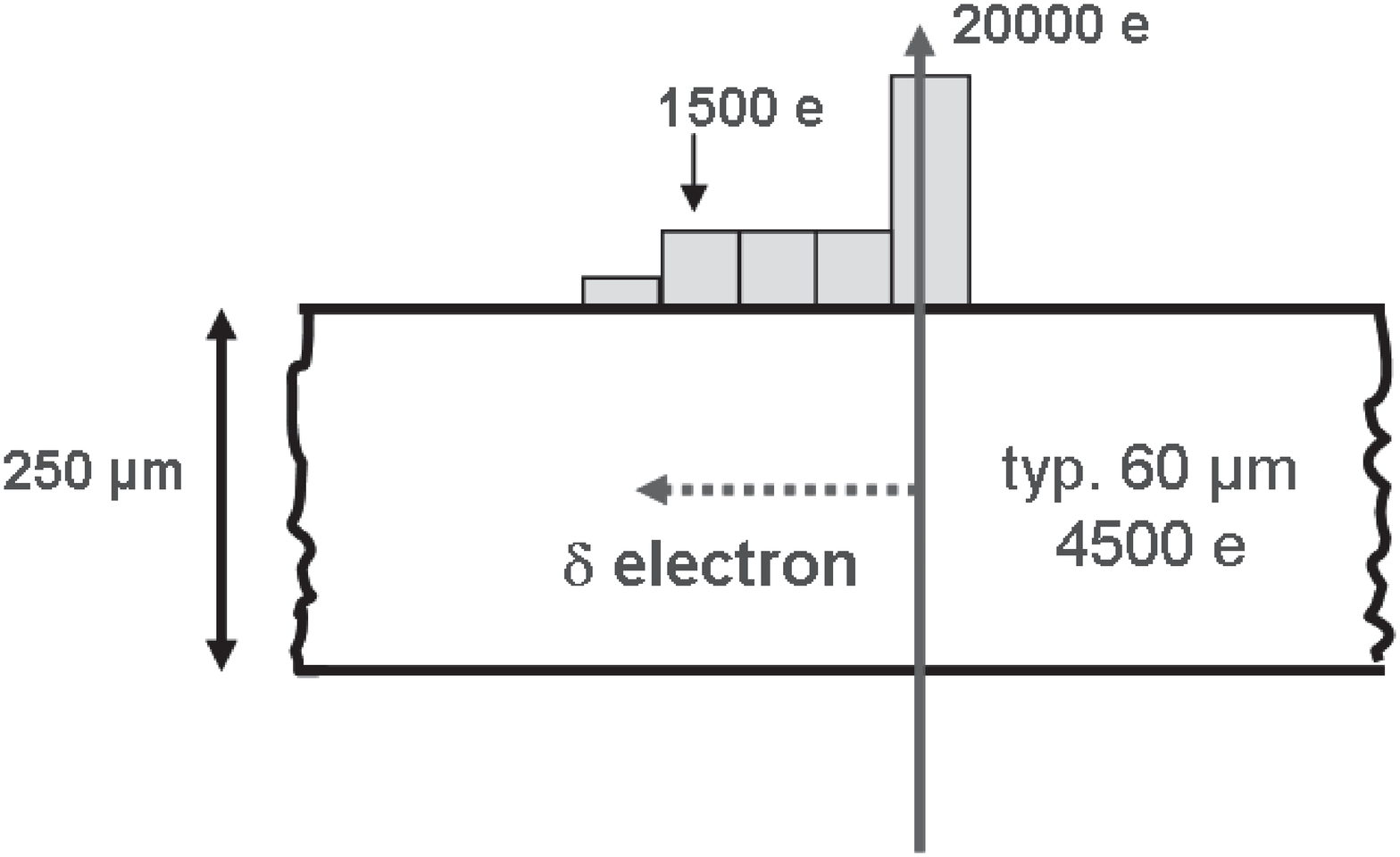}
\end{center}
\caption{\label{delta}}{Effect of a hit mis-measurement that
$\delta$ electrons can cause in a pixel detector.}
\end{figure}

A permanent nuisance in the ionization process for
experimentalists is the appearance of $\delta$ electrons, high
energy knock-on electrons. $\delta$ electrons follow a 1-1
relation between their kinetic energy and their emission angle
\begin{eqnarray}
\Theta_e(T) = \mathrm{arctan} \left[\frac{1}{\gamma} \left(
\frac{T_{max}}{T} - 1 \right) ^{\frac{1}{2}} \right] \simeq
\mathrm{arctan} \sqrt{\frac{2m}{T}}
\end{eqnarray}
where T denotes the $\delta$ electron's kinetic energy with
kinematically maximally possible value T$_{max}$ =
2mc$^2$$\beta^2\gamma^2$. $\gamma$=$\frac{E}{m}$ is the Lorentz
variable and m the electron mass. This relation is a very steeply
falling function: high energetic $\delta$ electrons (compared to
the energy of the projectile particle) are emitted in the forward
direction and slow, highly ionizing ($dE/dx \sim 1/\beta^2$)
$\delta$ electrons are emitted a right angles. The angular
distribution is
\begin{eqnarray}
\frac{dN}{d\Theta} = \frac{1}{2} D z^2 \frac{Z}{A} \rho x
\frac{sin\Theta}{cos^3\Theta}
\end{eqnarray}
where D=0.3071 MeV g$^{-1}$cm$^2$ is a constant, z the charge of
the projectile and Z, A, $\rho$, x are charge and mass number,
density and thickness of the traversed matter. This function peaks
very steeply by orders of magnitude at angles close to 90$^\circ$.
Therefore, as a rule of thumb, $\delta$-electrons are almost
always emitted under right angles and are highly ionizing. The
effect of a $\delta$ electron emission in a silicon pixel detector
is sketched in Fig.~\ref{delta}. A 100 keV $\delta$ electron
occurs in 300 $\um$ thick silicon with 6$\%$ probability and has a
typical range of 60 $\um$. Hence it created about 4500 e/h pairs
along its path, perpendicular to the parent particle. This causes
several neighbor pixels to fire. The reconstructed ´'hit`' will
thus be systematically shifted to the side of the $\delta$
electron emission.

\begin{figure}[h]
\begin{center}
  \includegraphics[width=.30\textwidth]{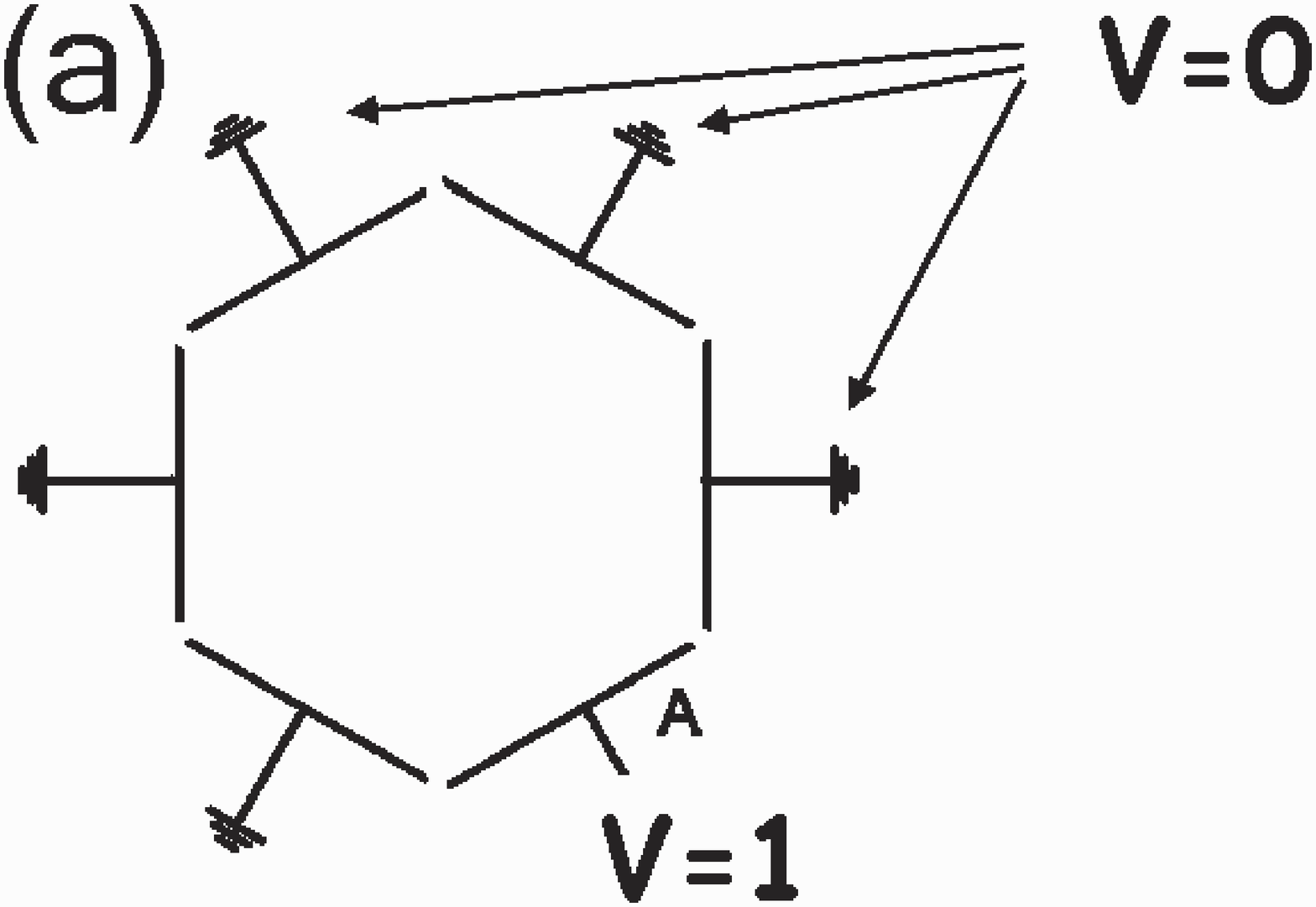}
  \includegraphics[width=.25\textwidth]{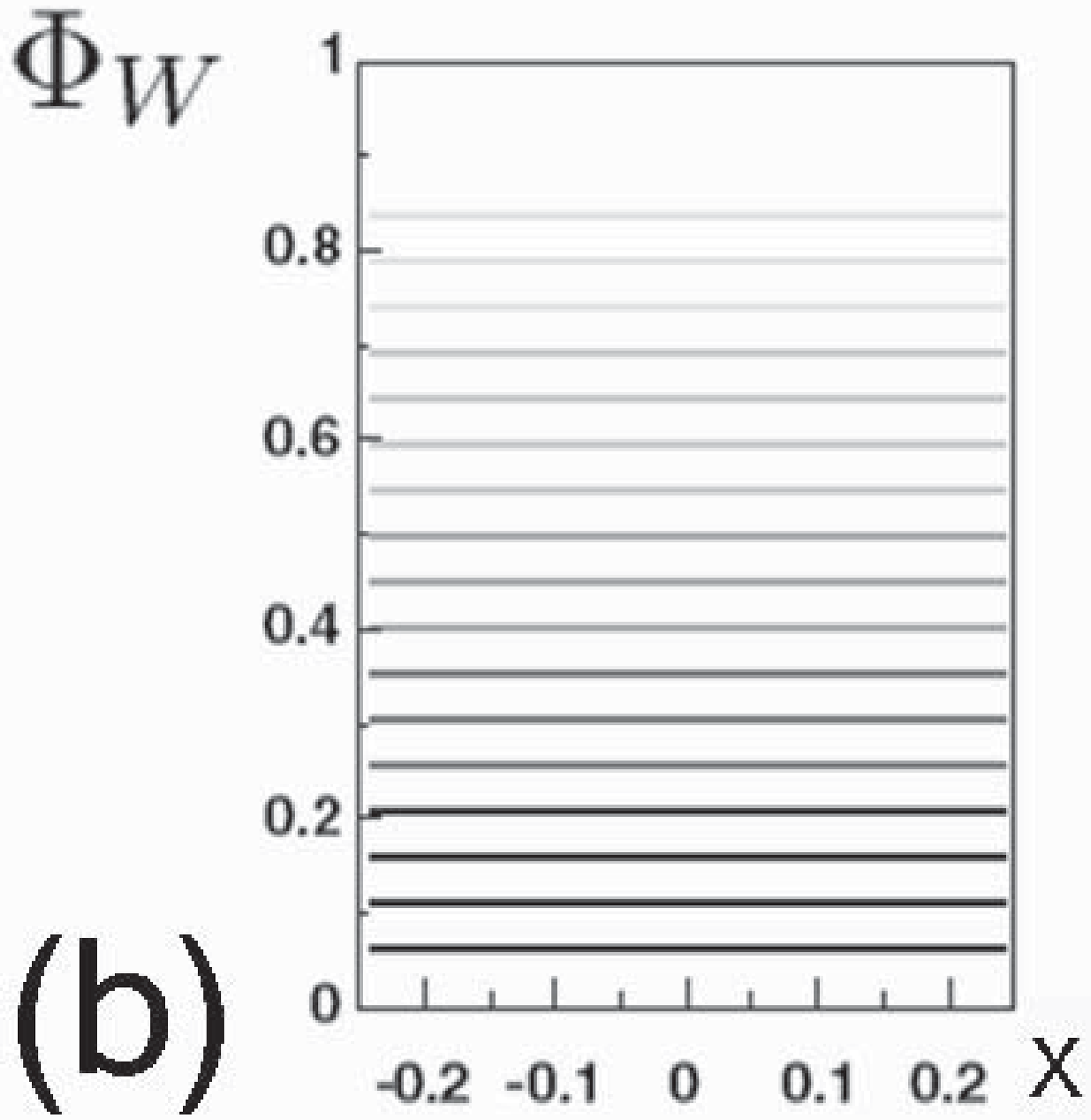}
  \includegraphics[width=.35\textwidth]{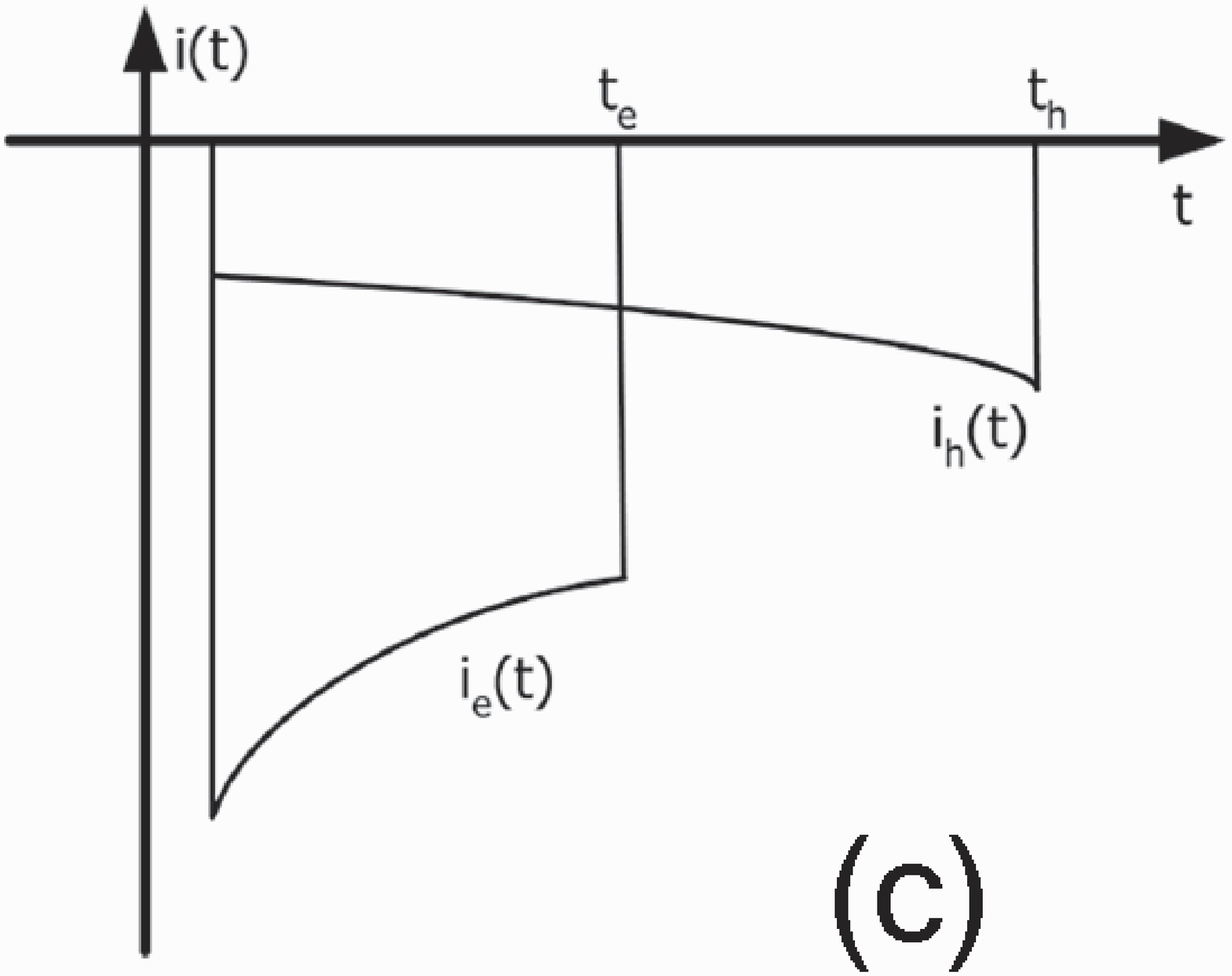}
\end{center}
\caption{\label{Ramo-1}}(a) How to calculate the weighting
potential/field in a multi electrode configuration. (b) Weighting
potential in a simple 2-electrode configuration, e.g. a parallel
plate detector or an area diode. (c) Current signals as a function
of time in a 2-electrode configuration.
\end{figure}

As mentioned before the signal on the electrodes is induced by the
movement of the separated charges in the electric field inside the
sensor. This is generally described by the Ramo theorem
\cite{Ramo} as
\begin{eqnarray}
i_{e/h}&=&-\frac{dQ}{dt}=q\vec{E}_W \cdot \vec{v}\\
dQ &=& q\vec\nabla\Phi_W d\vec r \nonumber
\end{eqnarray}
where $i_{e/h}$ and $Q$ are the current resp. charge induced on
the electrode by the movement of the electrons and holes, q is the
moving charge, $\vec{v}$ the velocity of the movement and
$\vec{E}_W$ and $\Phi_w$ are the so called weighting field resp.
weighting potential. They have nothing to do with the electric
field and potential inside the sensor, but rather determine how
the charge movement couples to a specific electrode. The weighting
field can be calculated by setting the potential of the electrode
under consideration to 1 (or 1V) and all other electrodes to 0
(0V) as indicated in Fig.~\ref{Ramo-1}(a).

In a 2-electrode configuration (e.g. an area diode) the
calculation is rather simple: set the top electrode of the sensor
to V=1 and the bottom electrode to V=0. The formulae for weighting
field and potential thus are
\begin{eqnarray}
\vec{E}_W = \frac{1}{d} \, \hat{e} \Rightarrow \Phi_W(x) =
\frac{x}{d}
\end{eqnarray}
as shown in Fig.~\ref{Ramo-1}(b) and the current induced on the
readout electrode reads
\begin{eqnarray}\label{eq-current}
 i(t) = - \frac{q}{d} \cdot v(t) = - \frac{q}{d} \, \mu
E(x(t))
\end{eqnarray}
with the electric field given by eq.~\ref{eq-field} with time
dependent x(t). Then solving eq.~\ref{eq-current} we receive the
characteristic exponential shape of the current with time, which
reflects the drift velocity change with time (here for the
electron signal)
\begin{eqnarray}
v_e(t) = \frac{dx}{dt} = - \mu_e \left( \frac{V + V_{dep}}{d} +
\frac{2 V_{dep} \, x(t=0)}{d^2} \right) \cdot e^{-\frac{2\mu_e
V_{dep}}{d^2}\, t}
\end{eqnarray}

\begin{figure}[h]
\begin{center}
  \includegraphics[width=.40\textwidth]{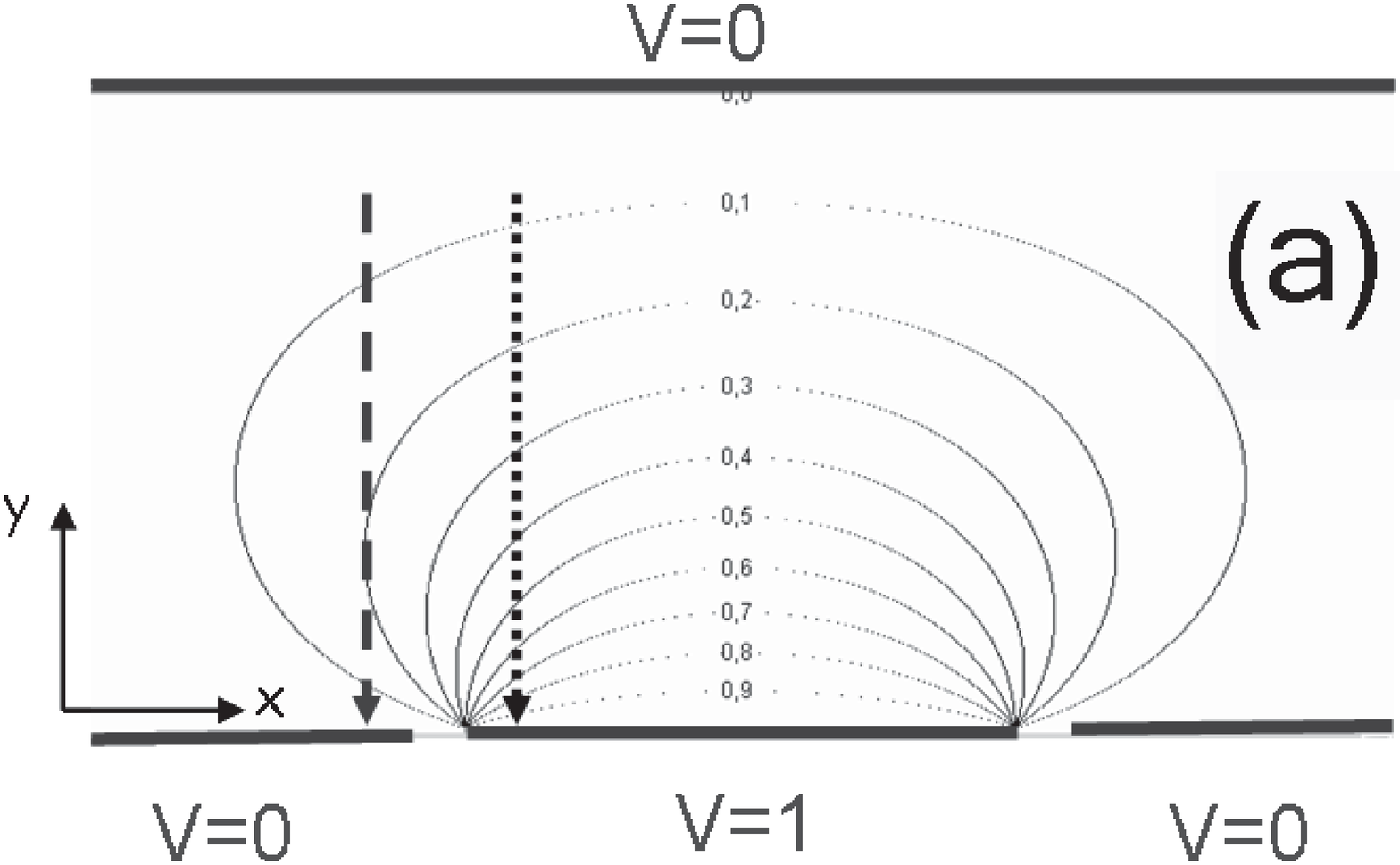}
  \hskip 2cm
  \includegraphics[width=.40\textwidth]{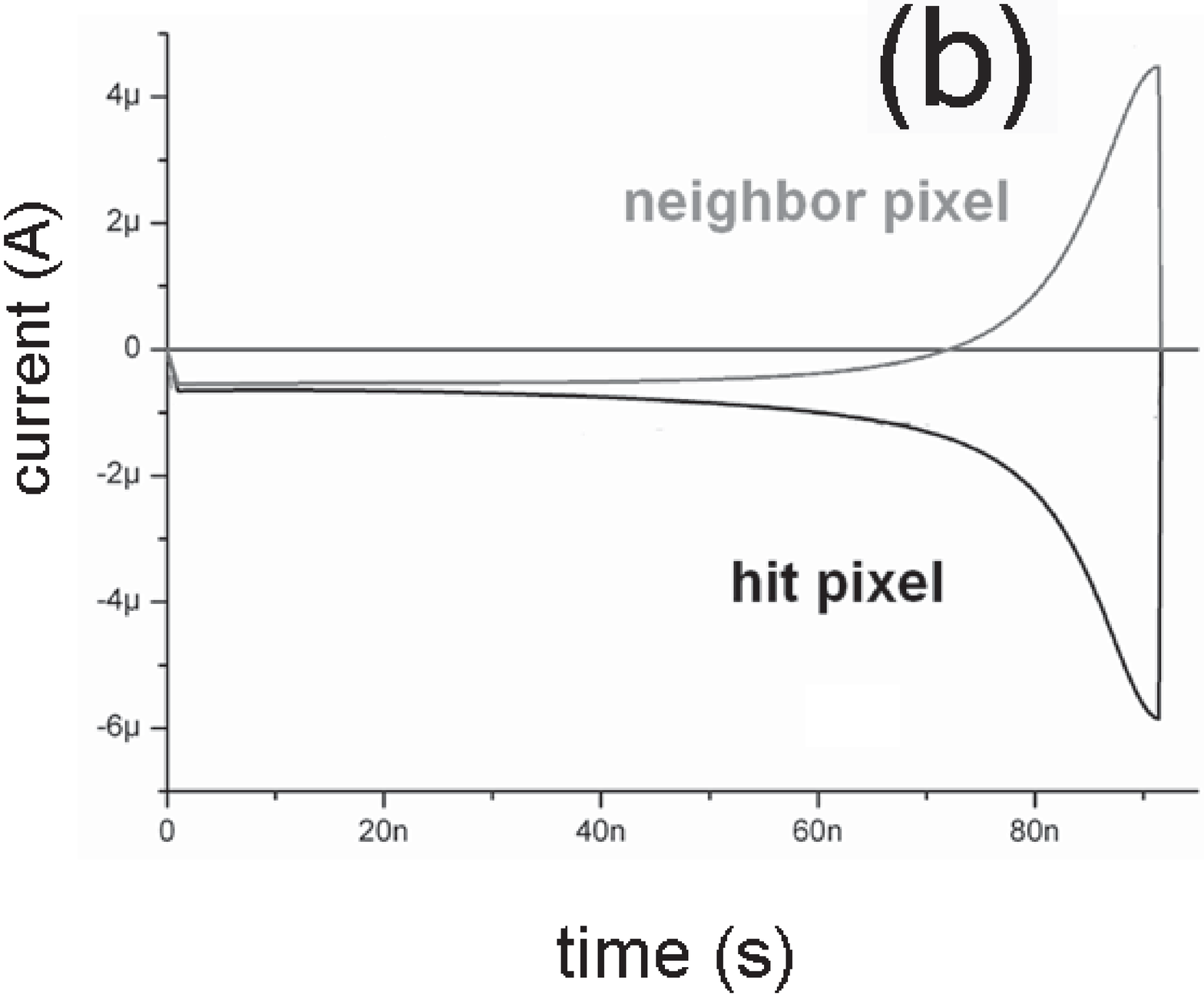}
\end{center}
\caption{\label{Ramo-2}}(a) Weighting potential in a strip/pixel
geometry (1-dim) together with two different paths of charge
movement, one arriving at the readout electrode, the other
arriving at the neighbor electrode, (b) current signals for the
two different charge drift paths showing the ´'small pixel
effect`.
\end{figure}

In a configuration with segmented electrodes like in a pixel
detector some caution is necessary. Figure~\ref{Ramo-2}(a) shows
the ansatz to calculate the weighting potential, and the result in
one dimension. The not easy calculation to solve the Poisson
equation using the Schwarz-Christoffel conformal
transformation~\cite{conformal} yields
\begin{equation}
\Phi(x,y)=\frac{1}{\pi}\arctan\frac{\sin(\pi y)\cdot
\sinh(\pi\frac{a}{2})}{\cosh(\pi x)-\cos(\pi y)\cosh(\pi
\frac{a}{2})} \quad ,
\end{equation}
where $x$ and $y$ are the coordinates and $a$ is the pixel width.
The form of the weighting potential is shown in
Fig.~\ref{Ramo-2}(a). Also shown are two parallel drift
trajectories, one ending on the electrode considered (dotted
arrow) and the other on the neighbor electrode (dashed arrow). The
corresponding current signals are displayed in
Fig.~\ref{Ramo-2}(b). For both pixels initially the current
signals are identical, independent of the electrode which
eventually collects the charge. While the trajectory of the hit
pixel, however, shows a rapidly increasing (negative) current
signal when approaching the electrode, the signal of the neighbor
trajectory crosses the zero line and then goes positive resulting
in a vanishing net integral of the pulse (i.e. the total charge).
Most of the signal is induced on the pixel electrodes only shortly
before the drifting charge cloud arrives at the pixel (small pixel
effect).

Finally, pixel detectors at LHC are operated in magnetic fields
(2T for ATLAS, 4T for CMS). In a B-field the drift motion of the
charge clouds is on a circle. This motion is stopped on average
after one mean free path length of the drifting electrons,
resulting in a complicated path which can be described by an
effective motion towards the readout electrode under an angle
$\alpha_L$, the Lorentz angle, as
\begin{eqnarray}
\mathrm{tan}\, \alpha_L = \mu_{{\mathrm{Hall}}} B_\bot
\end{eqnarray}
where $\mu_{{\mathrm{Hall}}}$ is the Hall mobility, which differs
by a factor 1.15 (0.72) from the normal mobility for electrons
(holes). The Lorentz angle has been measured~\cite{Lorentz-ATLAS,
Lorentz-CMS} in ATLAS (2T field) to be -15$^\circ$ before
irradiation with 150 V depletion voltage, and to -5$^\circ$ after
irradiation with 600 V depletion voltage, respectively. This
difference is mostly due to a decrease of the mobility with
increasing electric field inside the sensor. Note that the
effective incidence angle to the electrode is given by the sum of
the Lorentz angle and the tilt angle, by which the pixel modules
are tilted out of the perpendicular plane. This tilt angle in
ATLAS is +20$^\circ$.
\begin{figure}[h]
\begin{center}
  \includegraphics[width=.55\textwidth]{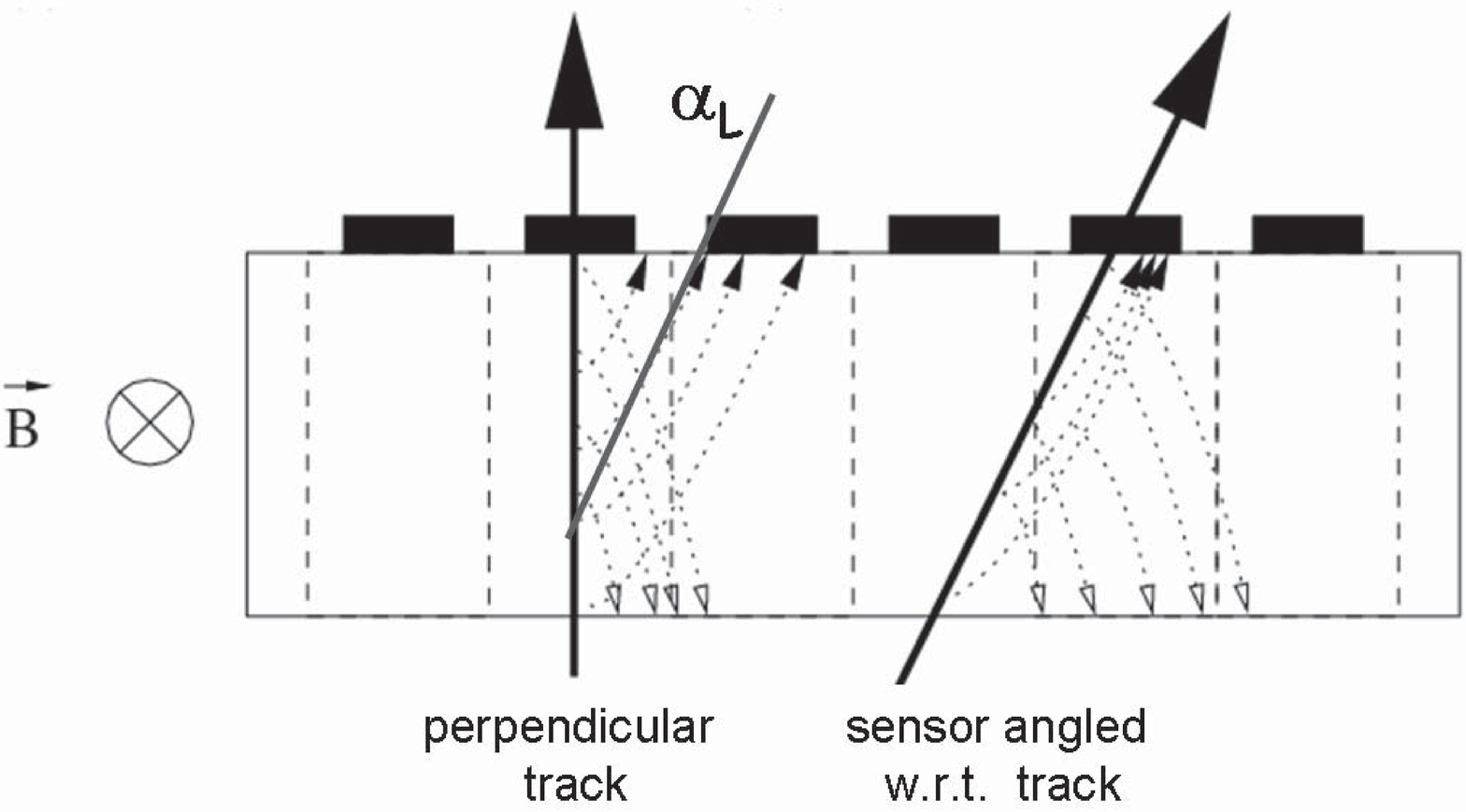}
\end{center}
\caption{\label{Lorentz}} Drift of charges in a pixel detector
under the effect of e perpendicular magnetic field. The drift
occurs under an angle $\alpha_L$, the Lorentz angle.
\end{figure}

\subsection{The noise}
For noise considerations in detector and electronics readout
configurations it is important to consider the following physical
noise sources
\begin{itemize}
\item number fluctuations of quanta, which lead to 1. shot noise
and 2. 1/f noise \item velocity fluctuations of quanta which lead
to 3. thermal noise.
\end{itemize}
Shot noise is generated when charge quanta are emitted over a
barrier, as is the case in a diode. Thermal noise occurs from
statistical thermal motion of charge carriers, here inside the
transistor channel of the amplifying transistor. 1/f noise in a
FET has its physical origin in the trapping and releasing of
charge carriers in traps in the Si-SiO$_2$ interface of the
transistor channel.

\begin{figure}[h]
\begin{center}
  \includegraphics[width=.45\textwidth]{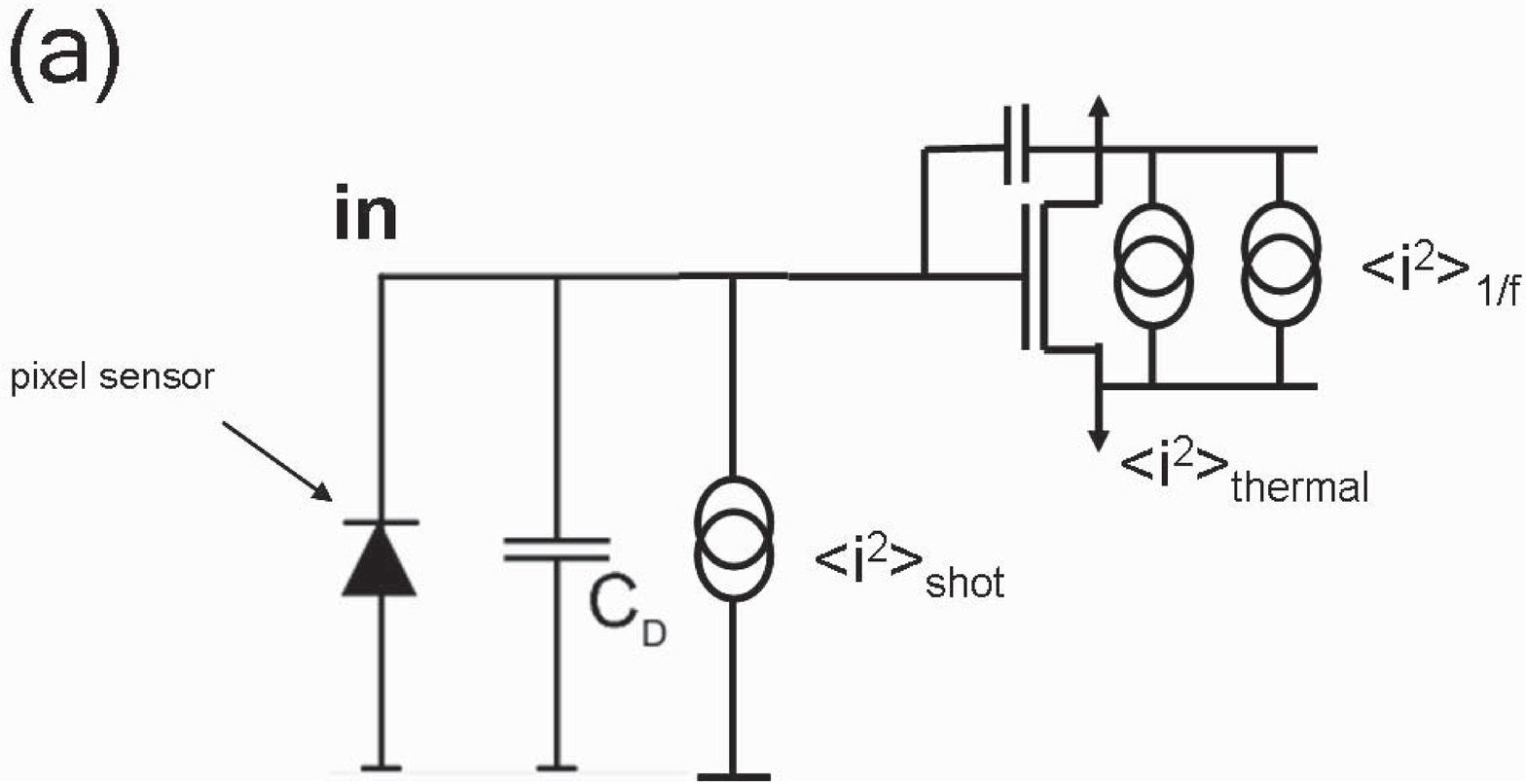}
\hskip 1cm
  \includegraphics[width=.45\textwidth]{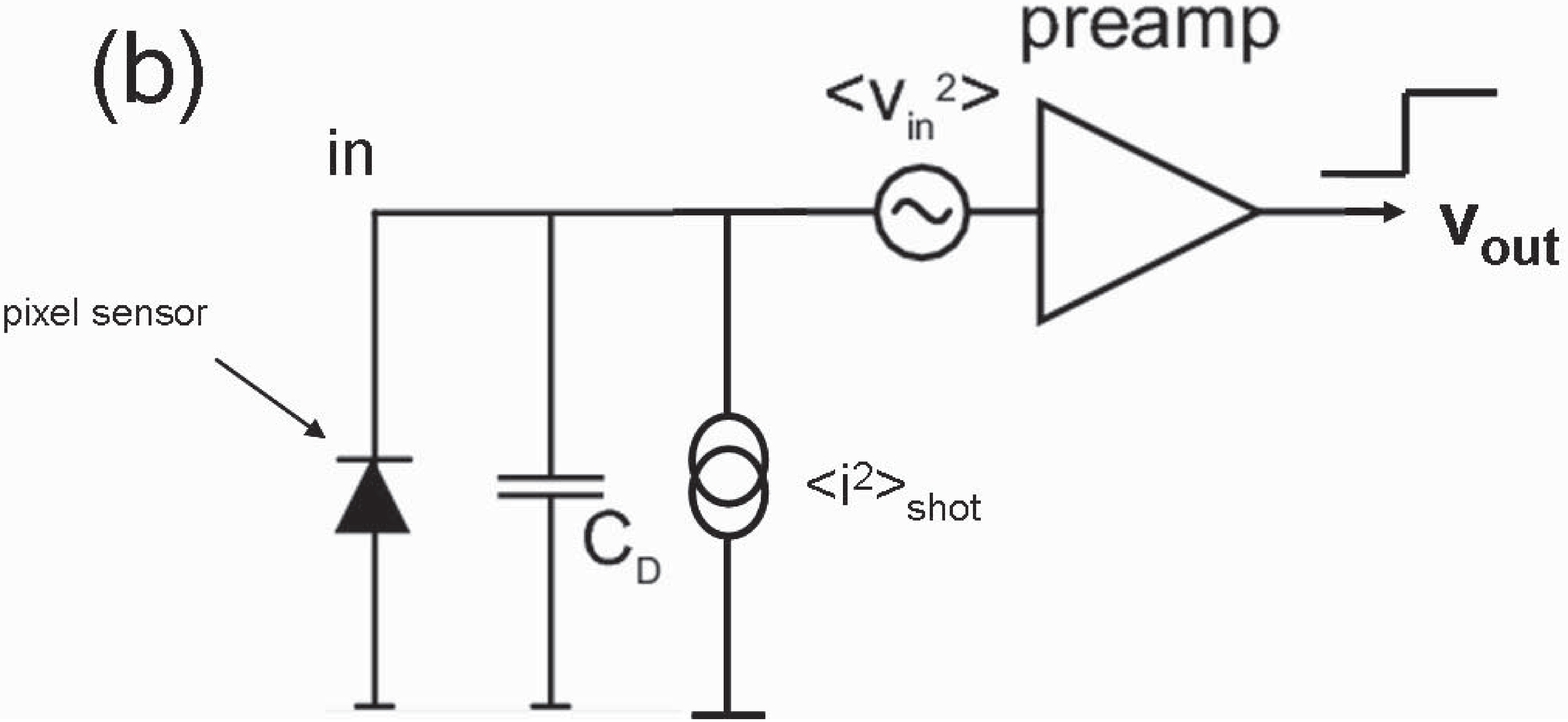}
\end{center}
\caption{\label{noise-1}} (a) A typical charge integrating pixel
detector readout, represented here by a single amplifying FET and
the three main noise sources: shot noise from detector leakage
current, and thermal as well as 1/f noise in the transistor
channel. (b) Transistor thermal and 1/f noise is treated as serial
voltage noise at the input of the preamplifier.
\end{figure}

Where do these noise sources appear in a typical pixel detector
readout chain (cf. Fig.~\ref{noise-1}) ?

Using $i^2 = (g_m v_{in})^2$ the current noise source inside the
transistor channel can be regarded as a voltage noise source in
series at the input of the amplifier (c.f. Fig.~\ref{noise-1}(b)).
The noise sources then can be written as
\begin{itemize}
\item series thermal voltage noise
\begin{eqnarray}
\langle v^2_{\mathrm therm} \rangle = \frac{8kT}{3g_m} \, {\mathrm
df}
\end{eqnarray}
\item series voltage 1/f noise
\begin{eqnarray}
\langle v^2_{\mathrm 1/f} \rangle = \frac{K_f}{C_{ox}WL}
\frac{1}{f}\, {\mathrm df}
\end{eqnarray}
\item and parallel current shot noise
\begin{eqnarray}{\label{shot noise}}
\langle i^2_{\mathrm shot} \rangle = 2 \, q \, I_{\mathrm leak} \,
{\mathrm df}
\end{eqnarray}
\end{itemize}
with \\
f = frequency \\ k = Boltzmann constant \\ T = temperature \\
g$_m$ = transconductance of the amplifying transistor \\ K$_f$ =
1/f
noise coefficient = 3-4 $\cdot$ 10$^{-32}$ C$^2$/cm$^2$ for n-channel MOSFETs\\
W, L = width and length of the transistor gate \\ C$_{ox}$ = gate
oxide capacitance
\\ q = elementary charge \\ i$_\mathrm{leak}$ = detector leakage
current \hfill\break

It is customary for charge integrating devices to express the
noise voltage output as an equivalent noise charge ENC, i.e. the
charge one would need to see at the input that produces the
observed output noise voltage $\sqrt{<v^{2\,noise}_{out}>}$
\begin{eqnarray}
\mathrm{ENC} = \frac{\mathrm{noise \ output \ voltage \
(rms)}}{\mathrm{signal \ output \ voltage \ for \ the \ input \
charge \ of \ 1 e^-}}
\end{eqnarray}
with the relation
\begin{eqnarray}
\mathrm{ENC}^2_\mathrm{tot} = \mathrm{ENC}^2_\mathrm{shot} +
\mathrm{ENC}^2_\mathrm{therm} + \mathrm{ENC}^2_\mathrm{1/f} \quad
.
\end{eqnarray}
The charge integrating amplifier transforms the noise input into
an output voltage, which depends on the feedback- $C_f$, input-
$C_{D}$ and load- $C_\mathrm{load}$ capacitances, as well as on
the characteristic feedback time constant
$\tau_f$~\cite{pixelbook}.
\begin{eqnarray}
ENC_\mathrm{shot} & = \sqrt{\frac{I_{\mathrm leak}}{2q} \tau_f}
\hspace{3.5cm}
  & =  56 e^- \times \sqrt{\frac{I_{\mathrm leak}}{\mathrm nA} \frac{\tau_f}{\mu
  s}} \nonumber \\
ENC_\mathrm{therm}
  &= \frac{C_f}{q} \sqrt{\langle v^2_{\mathrm{therm}} \rangle }
  = \sqrt{ \frac{kT}{q} \frac{2 C_D}{3q} \frac{C_f}{C_{load}}}
  & = 104 e^- \times \sqrt{ \frac{C_D}{100 {\mathrm{\, fF}}}
  \frac{C_f}{C_{load}}}\\
ENC_\mathrm{1/f}
 &\approx \frac{C_D}{q}
           \sqrt{ \frac{K_f}{C_{ox}W L} }
           \sqrt{ {\mathrm {ln}}{\left(\tau_f \frac{g_m}{C_{load}} \frac{C_f}{C_D}\right)} }
           & = 9 e^- \times \frac{C_D}{100 {\mathrm{\, fF}}}
           {\mathrm{(for\, NMOS \, trans.)}} \nonumber
\end{eqnarray}

\begin{figure}[h]
\begin{center}
  \includegraphics[width=.8\textwidth]{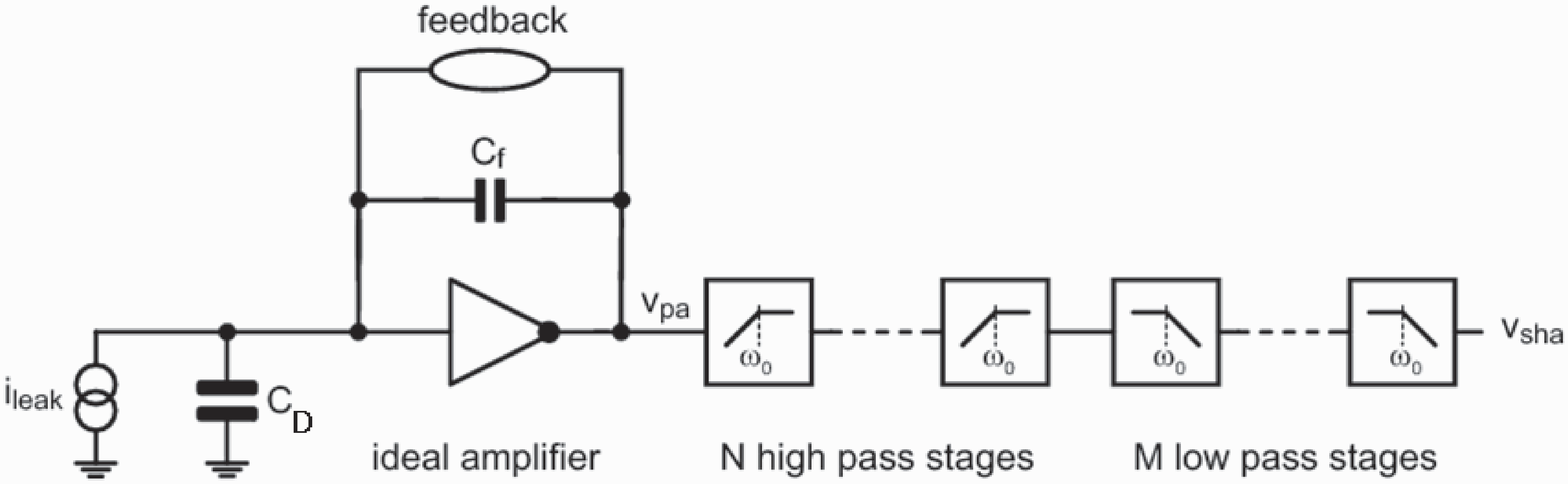}
\end{center}
\caption{\label{shaper}} Pixel detector readout with a charge
integrating preamplifier and a filter amplifier (shaper).
\end{figure}

If - as is often the case (see Fig.~\ref{shaper}) - an additional
filter amplifier (shaper) is the bandwidth limiting element and
not the preamplifier, the latter can be neglected in the noise
calculation. The shaper consists of a sequence of bandwidth
limiting elements, in the most general case N high pass stages and
M low pass stages (see Fig.~\ref{shaper}). The equivalent noise
charge behind the shaper is then given by~\cite{pixelbook}
\begin{eqnarray}
  \left(\frac{{\mathrm ENC}}{e^-}\right)^2
         &=& 115 \cdot \frac{\tau}{10 {\mathrm ns}} \cdot \frac{I_\mathrm{leak}}{1 {\mathrm nA}} \nonumber \\   
         &+& 388 \cdot \frac{10\mathrm{ns}}{\tau} \cdot \frac{\mathrm{mS}}{\mathrm{g_m}} \cdot \left(\frac{\mathrm{C_D}}{100 \mathrm{fF}}\right)^2\\
         &+& 74  \cdot \left(\frac{\mathrm{C_D}}{100\mathrm{fF}}\right)^2.
         \nonumber
  \label{ENC2}
\end{eqnarray}

\begin{figure}[h]
\begin{center}
  \includegraphics[width=.40\textwidth]{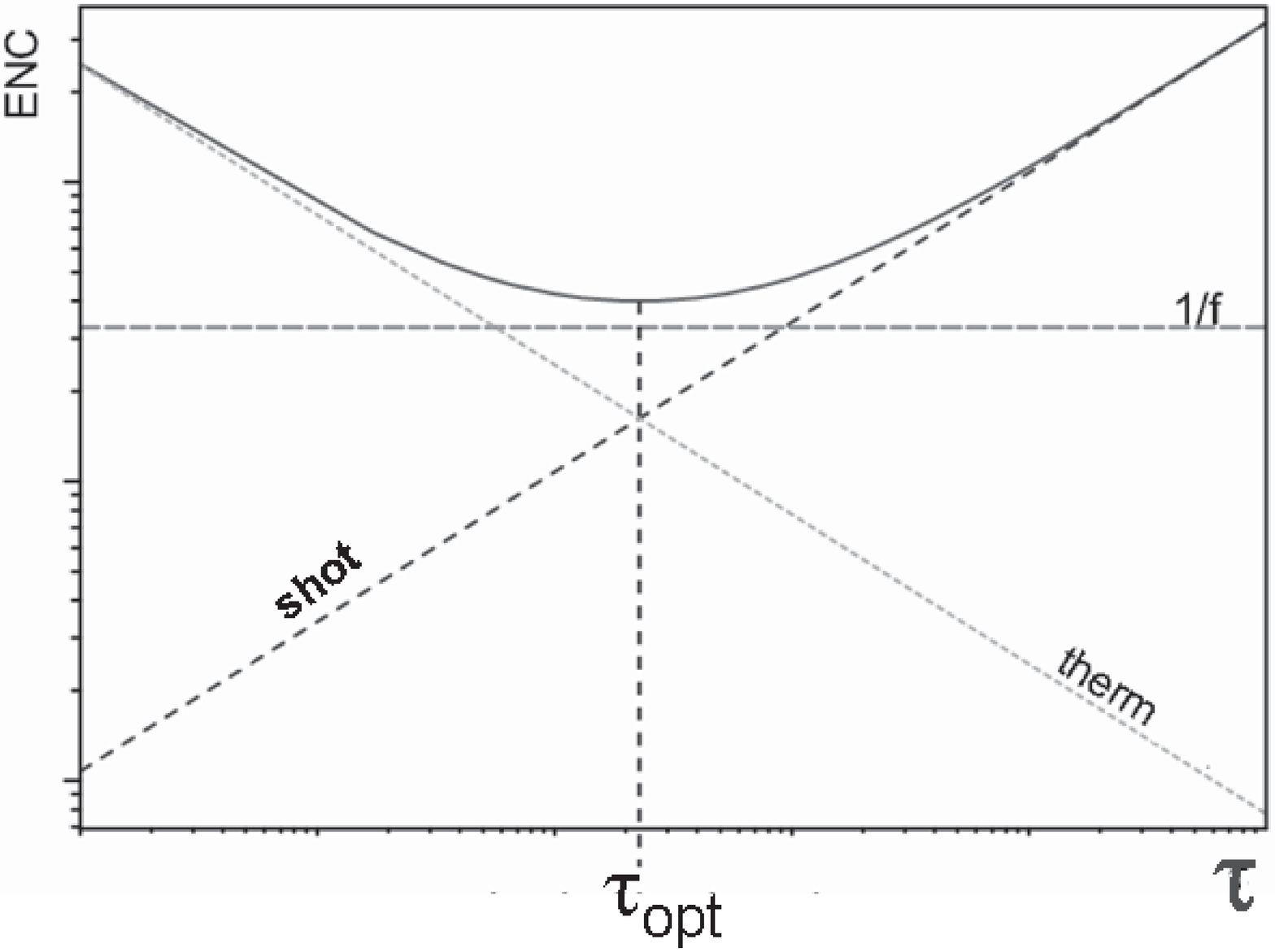}
\end{center}
\caption{\label{optshaping}} The three noise contributions
(thermal, 1/f and shot noise) as a function of the shaping time in
a logarithmic representation. The dependence results in a minimal
total noise at an optimal shaping time.
\end{figure}

Note that the noise depends on three important quantities to which
attention has to be paid to: the sensor leakage current
$I_\mathrm{leak}$ and the shaping time $\tau$, both of which
increase the parallel shot noise contribution, and the total input
capacitance, dominated by the detector capacitance C$_D$, which
increases the two parallel noise components (thermal and 1/f). The
shaping time decreases the thermal noise which results in the fact
that the three noise contributions to the detector/readout
configuration have a minimum at an optimum shaping time
$\tau_{opt}$, as is sketched in Fig.~\ref{optshaping}.

Pixel detectors have the advantage of small input capacitances in
the order of several hundred fF and thus have the potential of
being low noise devices. Typical figures for LHC pixel detectors
are

\begin{tabbing}
\hskip 3cm \= noise \quad \= = 150 e$^-$ \qquad \qquad \= initially \\
\hskip 3cm  \> \> = 200 e$^-$ \> after 10 years at LHC \\
\hfill \\
\hskip 3cm \> signal \> $\sim$ 19500 e$^-$ \> total
charge in 250
$\um$ silicon \\
\hskip 3cm \> \> = $\sim$ 13000 e$^-$ \> including charge sharing between pixels \\
\hskip 3cm \> \> = 6000 - 8000 e$-$ \> after 10 years at LHC
\end{tabbing}

hence guaranteeing a S/N ratio of more than 30 throughout the LHC
lifetime.

\section{Making a Pixel Detector \\
From the pixel sensor to the module-ladder} The particle fluence
that pixel detectors at 5cm distance from the IP at LHC have to
expect during 10 years of operation is about 10$^{15}$
n$_{eq}$/cm$^2$. This fact constitutes enormous challenges to
pixel sensors, which develop large leakage currents and need high
voltages for full depletion, as well as to the front end
electronics, which suffers from threshold shifts and generated
parasitic transistors; but also to the mechanic elements whose
material performance degrades, glue becomes brittle, etc. This has
required intensive irradiation tests over many years.

\subsection{Hybrid Pixel Assembly}
\begin{figure}[h]
\begin{center}
  \includegraphics[width=.50\textwidth]{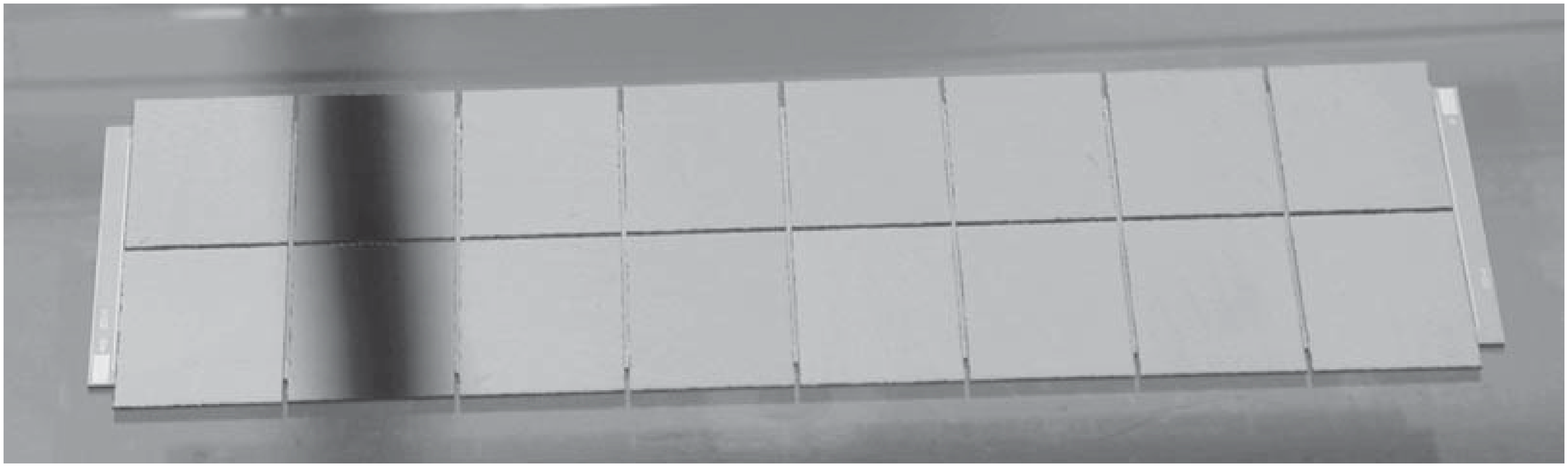}
\hskip 1cm
 \includegraphics[width=.35\textwidth]{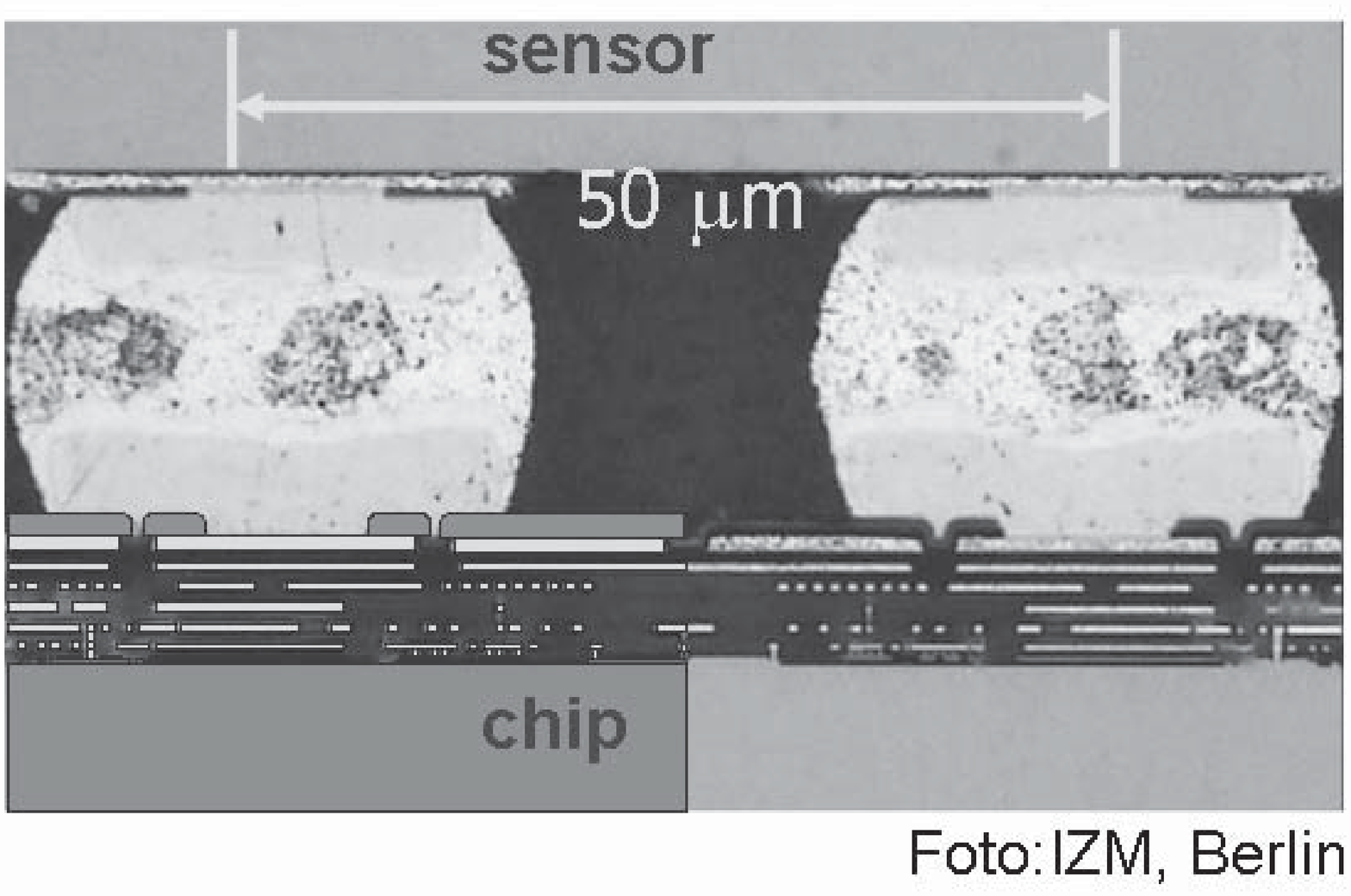}
\end{center}
\caption{\label{baremodule}} (Left) components of the ´'bare`'
module are the sensor, several FE-chips and the interconnection
between them. (Right) a cut through the bump row showing the bumps
and the metal layers of the chip electronics.
\end{figure}

A so called bare pixel module consists of the pixel sensors and a
number of electronics chips for amplification and readout (16 in
the case of ATLAS and CMS, 5 for ALICE) which are mated by means
of the bumping and flip-chipping technology~(\cite{IZM1,Fiorello}
and~\cite{pixelbook}). In the case of ATLAS the 10~cm sensor wafer
has three 2.1 x 6.4 cm$^2$ sized sensor tiles. The 20~cm chip
wafer instead has 288 7.1 x 11.4 mm$^2$ sized front end chips
which have been tested to be functional with an average yield of
82$\%$. The mating is done in part with indium
bumps~\cite{Fiorello} or solder bumps~\cite{IZM1,IZM2}. Each chip
has about 3000 such bumps. The chip wafers are thinned by backside
grinding after bumping from initially about 800$\um$ thickness
down to 180$\um$. Figure~\ref{baremodule} shows a bare module
after assembly and a cut through a bump row, which nicely shows
the module after the mating process (here solder). The bumps in
case of Indium are about 7$\um$ high and are applied on both,
sensor and chip wafer. After thermo-compression the remaining
thickness is about 10$\um$. Solder bumps are applied only to the
chip wafers and are 25$\um$ in diameter, and about 20$\um$ after
flip-chip and reflow (see Fig.~\ref{baremodule}).

\begin{figure}[!h]
\begin{center}
  \includegraphics[width=.45\textwidth]{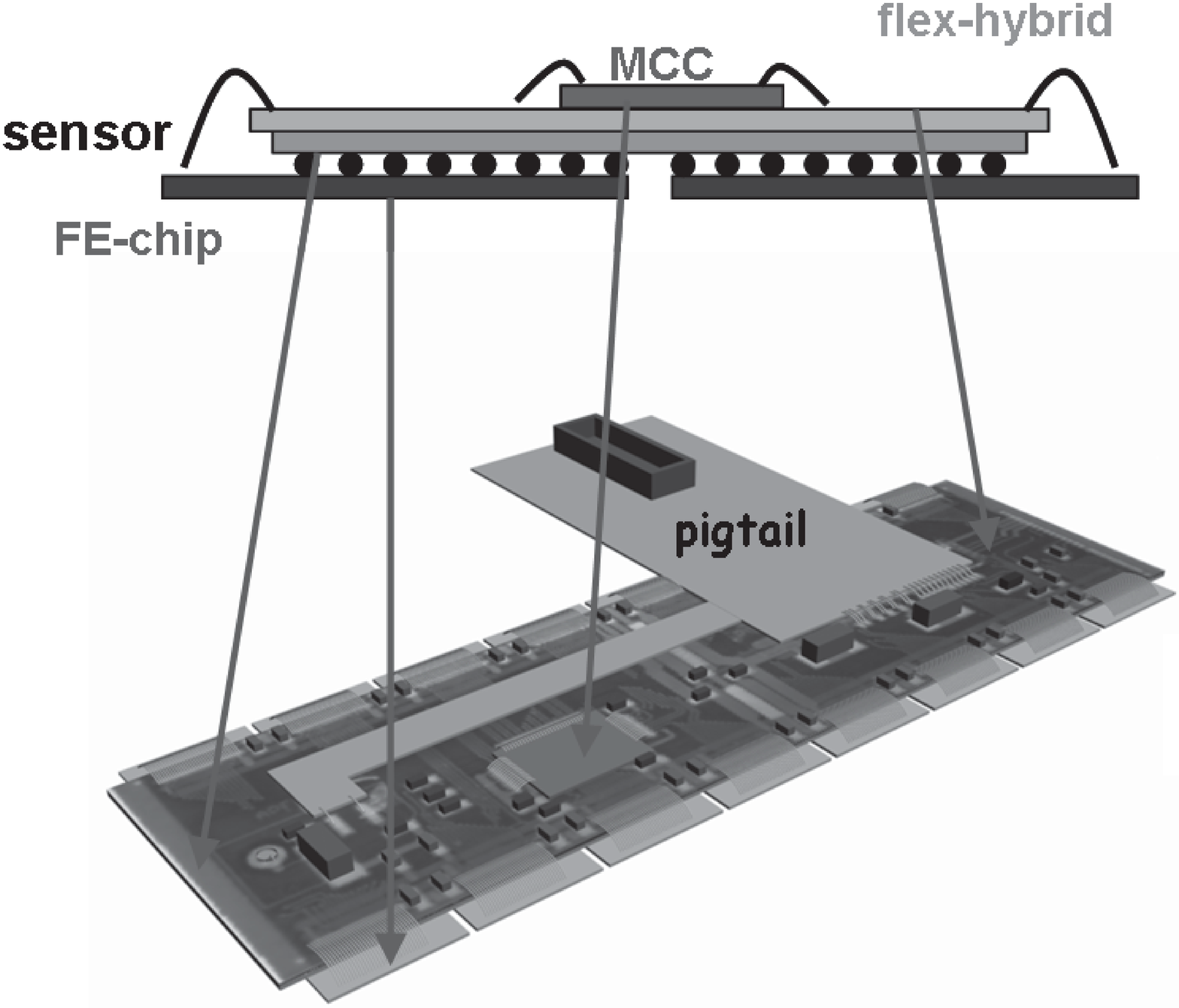}
  \hskip 1cm
  \includegraphics[width=.45\textwidth]{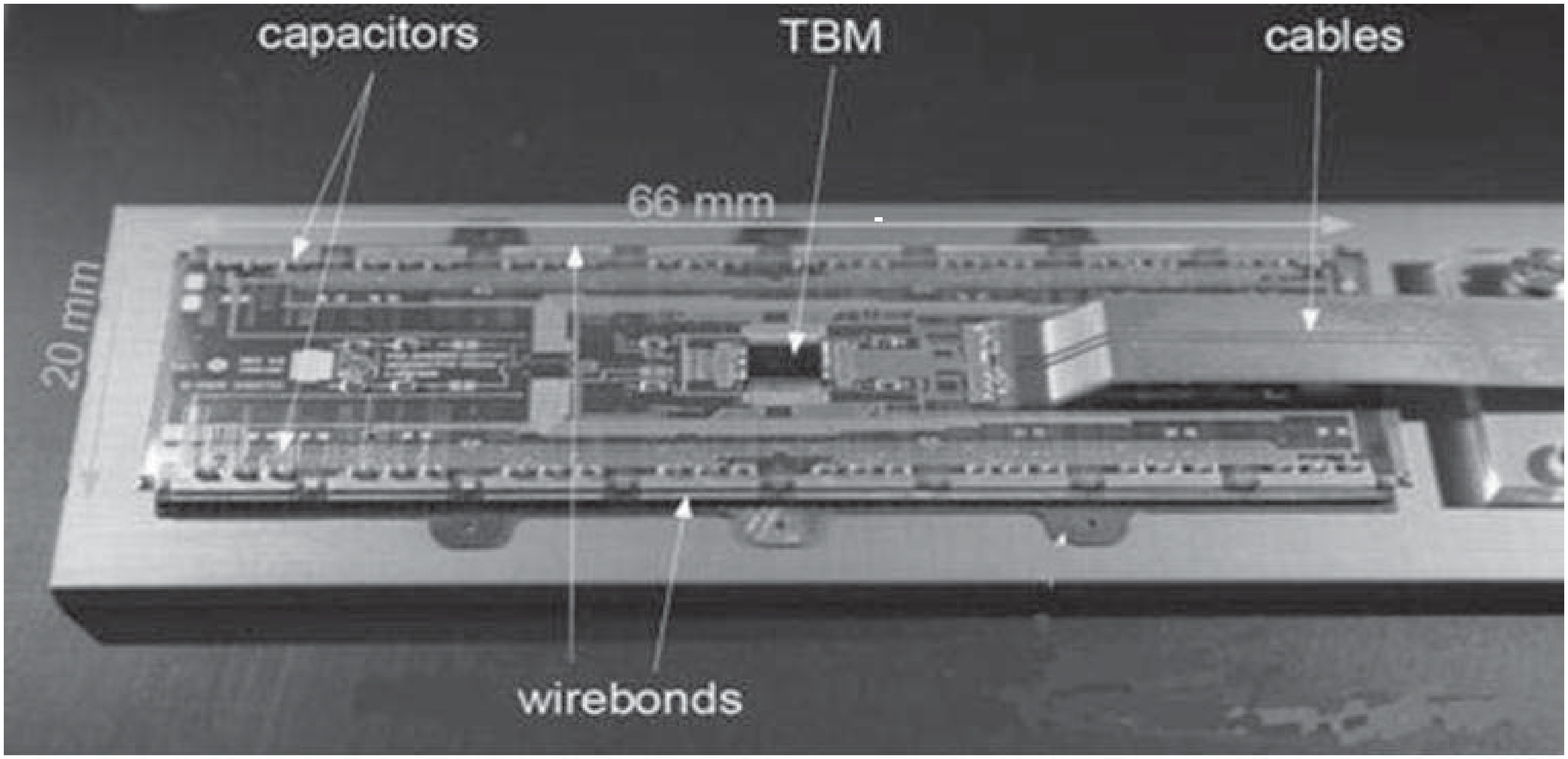}
\end{center}
\caption{\label{CMS-module}} (Left) Sketch of the composition of a
complete ATLAS module, including sensor and FE-chip connected by
bump bonds, a FLEX hybrid kapton foil and the module control chip
MCC. The ´`pigtail´' provides the connection to the outside;
(Right) Photograph of the CMS Pixel Module.
\end{figure}

To produce a fully functional module, additional assembly steps
are necessary as shown in Fig.~\ref{CMS-module}. The output lines
are connected via wire bonds to a flex-kapton hybrid foil with
fine-print lines. The signals are routed to a further chip, the
module control chip MCC and from there through another flex
component (pigtail) to the micro cables which run to the
electro-optical interface which is in ATLAS between about 20 cm
and 120 cm away from the module. Figure~\ref{CMS-module} also
shows a CMS-pixel module, fully assembled.

\subsection{Pixel Sensors in the LHC radiation
environment}\label{sensors}
\begin{figure}[h]
\begin{center}
  \includegraphics[width=.38\textwidth]{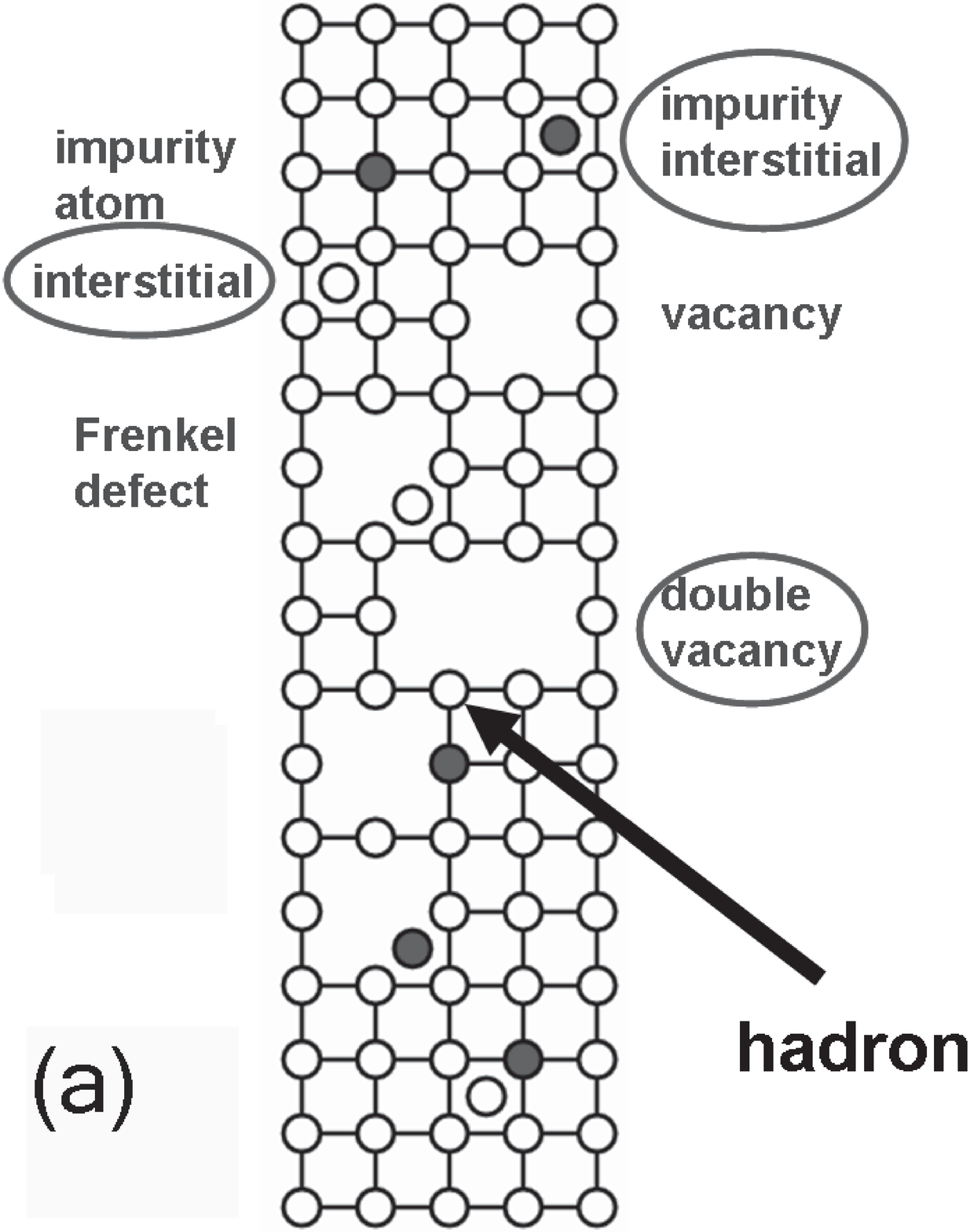}
  \hskip 1cm
  \includegraphics[width=.48\textwidth]{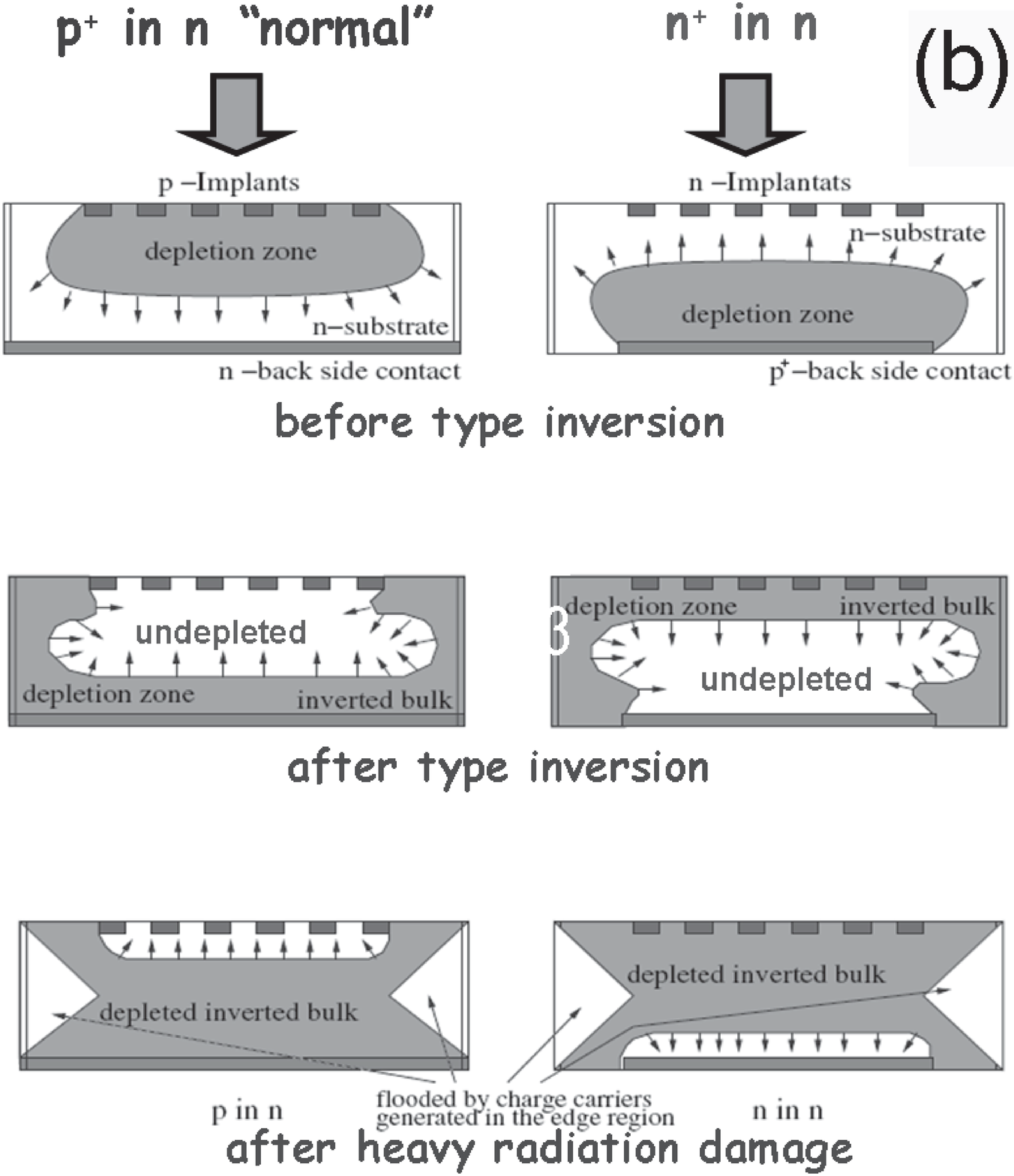}
\end{center}
\caption{\label{latticedamage}} (a) Sketch of some damage
processes that hadrons can cause to a lattice; (b) the effects of
radiation damage to ´`p in n´' (left) and to ´`n in n´' sensors
(right) in comparison.
\end{figure}

 The intense radiation at LHC requires
dedicated designs for both sensor and FE-chip. For the sensor,
mostly the non-ionizing energy loss which produces non-reversible
lattice damage is important in this respect.
Figure~\ref{latticedamage}(a) sketches the main defect
appearances. We distinguish so called point defects and defect
clusters, the latter originate from damage that a recoiling
lattice atom can cause. The size of the clusters typically is 10
nm $\times$ 200 nm. For pixels sensors the largest concern comes
from double vacancies and interstitials, both producing
generation/recombination levels deep in the band gap which lead to
an increase of the leakage current, and therefore to higher noise
(see eq.~\ref{shot noise}). Also the effective space charge in the
depletion region is changed leading to an increase of the
effective doping concentration and as a consequence an increase in
the necessary voltage for full depletion.
\begin{figure}[h]
\begin{center}
  \includegraphics[width=.43\textwidth]{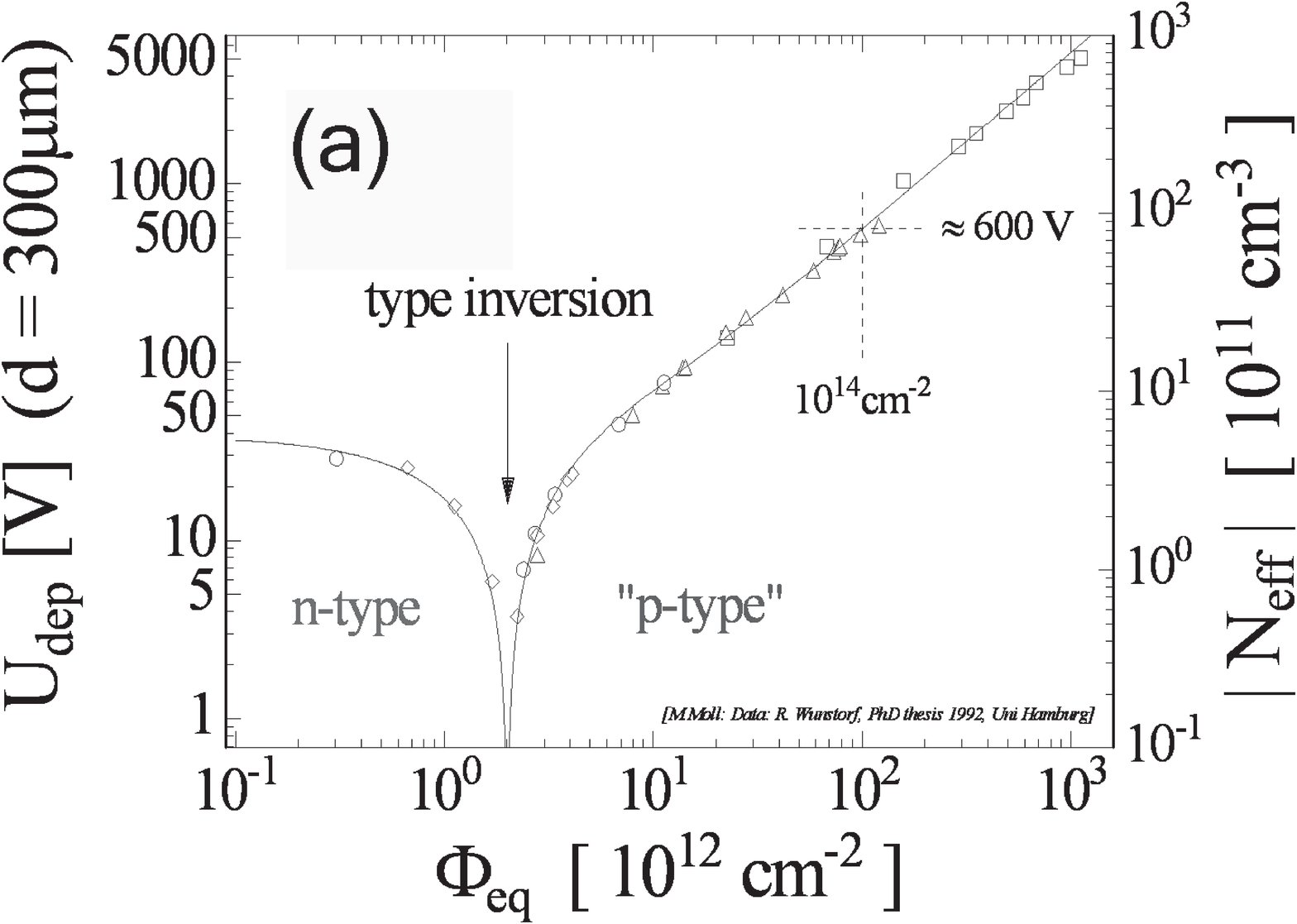}
  \hskip 1cm
  \includegraphics[width=.47\textwidth]{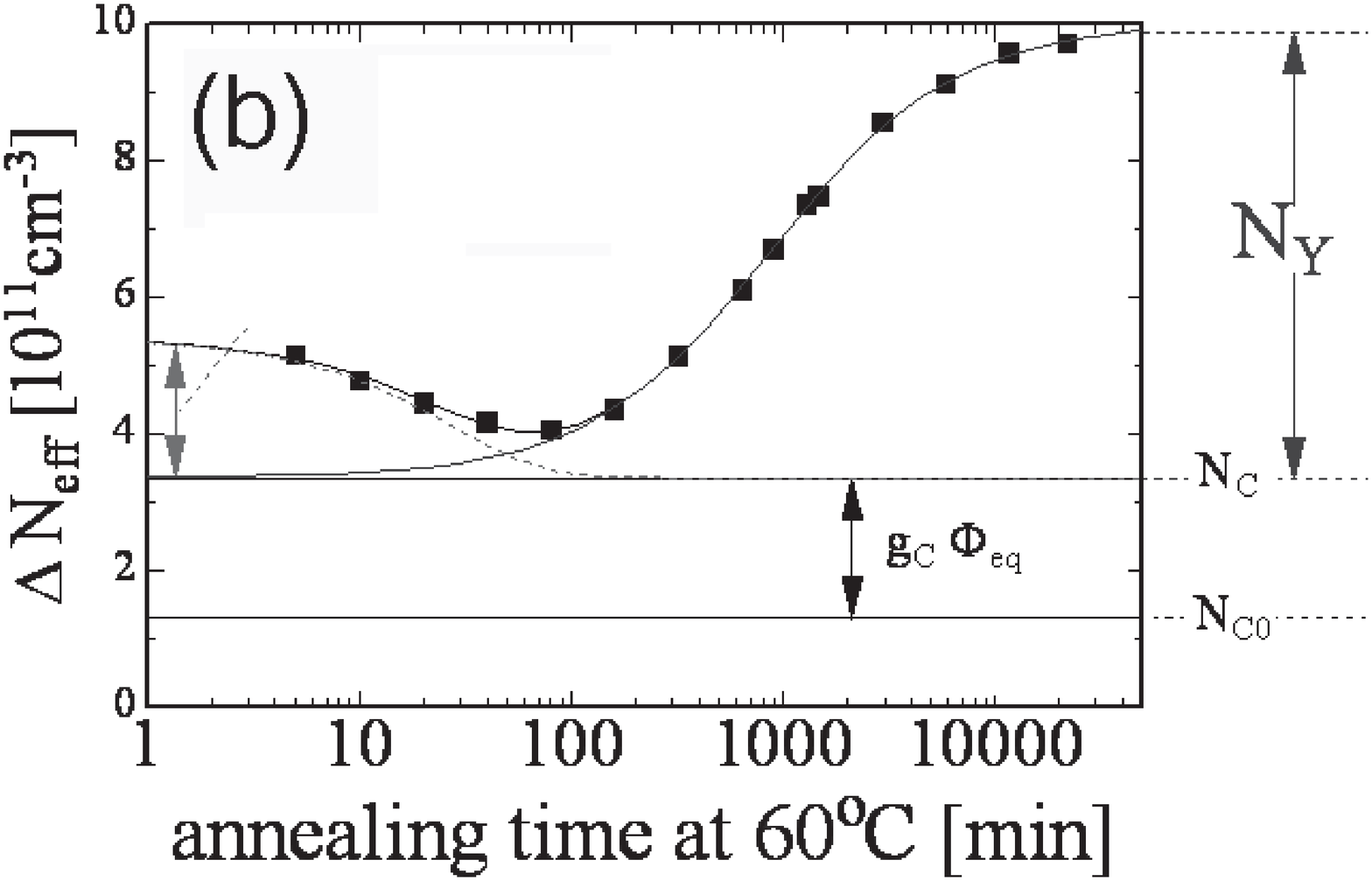}
\end{center}
\caption{\label{niel}} (a) Change of the effective doping
concentration with fluence showing type inversion; (b) Annealing
of irradiation damaged sensors with time showing the beneficial
annealing at the start and the reverse annealing at longer time
scales.
\end{figure}
This is shown in Fig.~\ref{niel}(a), where the effective doping
concentration as well as the full-depletion voltage are plotted as
a function of the fluence. N$_{eff}$ first decreases to zero and
then becomes almost intrinsic at a fluence of 2-5
$\times$10$^{12}$ cm$^{-2}$. It then rises again (type
inversion)~\cite{rwu}. The effective doping concentration after
irradiation damage also changes with time as shown in
Fig.~\ref{niel}(b): beneficial annealing occurs at short time
scales, so-called reverse annealing at longer time
scales~\cite{lindstroem01}. The latter is suppressed when keeping
the damaged sensors at low temperatures below the freezing point.
At -10$^\circ$ C the time constant for reverse annealing is about
500 years, while at +20$^\circ$ C it is only about 500
days~\cite{moll2006}.

\begin{figure}[h]
\begin{center}
  \includegraphics[width=.8\textwidth]{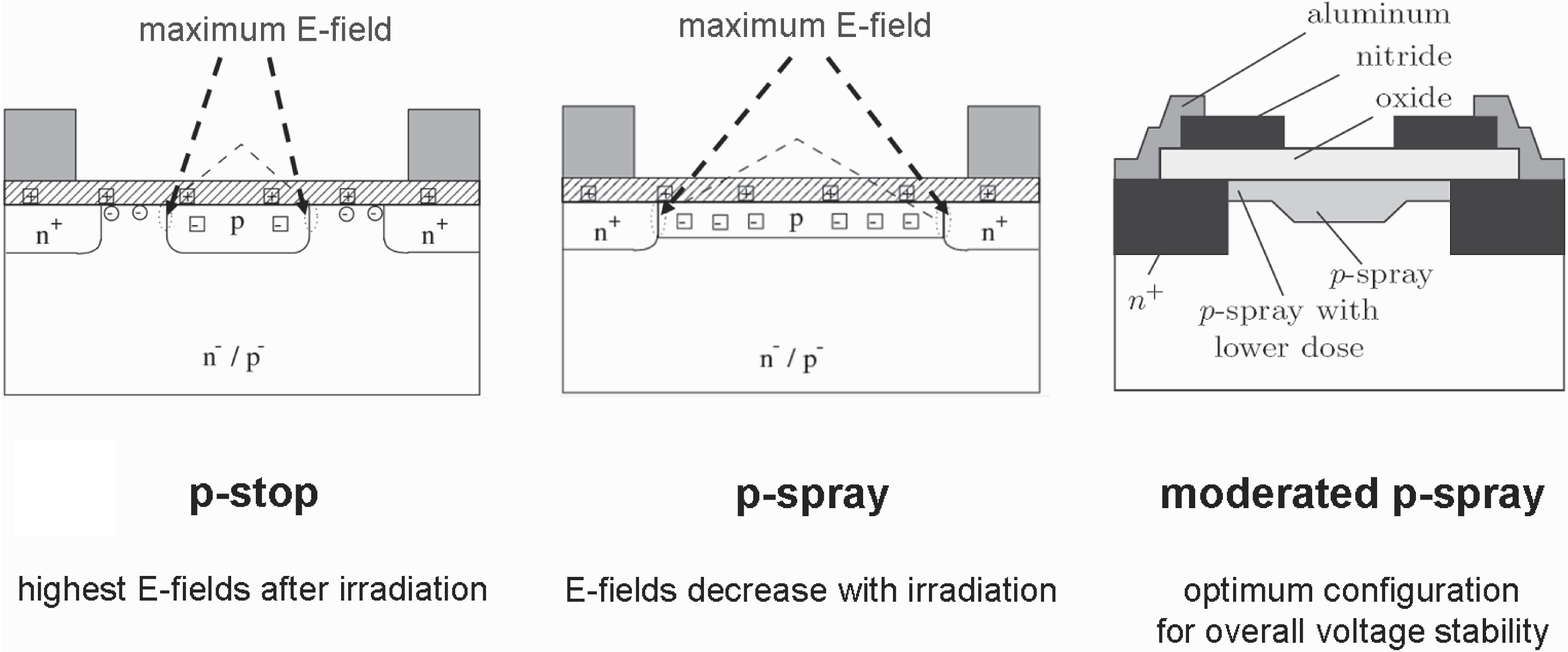}
\end{center}
\caption{\label{pspray}} Isolation measures at the n$^+$n
interface: left p-stop, center p-spray, right moderated p-spray.
\end{figure}

Finally, trapping centers are being created which can trap the
signal charge and thus lead to a loss in signal. Solutions to
irradiation damage in silicon have been found through intensive
R$\&$D, most notably by the CERN RD48 and RD50
collaboration~\cite{lindstroem01,moll-nimb,eremin,moll-current}.
For the fluences expected at LHC in the pixel detector oxygenated
float zone silicon has proven to stand the expected irradiation
from protons and pions within the operation specs of the detectors
(e.q. 600 V maximum bias voltage), while for neutrons which
produce less point defects but more cluster defects this is not
the case. Close to the interaction region, however, particle
fluence by pions is dominant. The pixel sensor design has also
been adapted to the development of the effective doping
concentration with increasing fluence by chosing n$^+$ pixel
implants on lower doped n-bulk material.
Figure~\ref{latticedamage}(b) shows the development of the
depletion region for p$^+$ in n and for n$^+$ in n pixel sensors
in comparison. The depletion zone grows from the higher doped
p-region into the lower doped n-bulk, in the presence of
type-inversion under irradiation. While. after type inversion, for
the standard p in n sensors the depletion region grows from the
non-pixellated bottom, which requires to always operate under full
depletion, for n in n sensors after type inversion the depletion
region grows from the electrode (i.e. pixel) side. This allows
operation also without full depletion which might be unavoidable
after several years of LHC-running. At the n-pixel n-bulk
interface at the top, measures are taken to prevent the electron
accumulation layer, which develops a the Si-SiO$_2$ interface
where there is no intrinsic depletion zone as is the case for a pn
interface. Figure~\ref{pspray} shows the different advantages and
disadvantages of isolation measures using either p-stop implants
between the n-implants, or a lighted doped boron p-spray in the
entire region between implants. The highest fields occur always at
the lateral pn junctions. Irradiation increases the oxide charges
and also the substrate doping, such that the backside bias must be
increased after irradiation. This generally increases also the
maximum field with irradiation. For the p-spray situation the
sensor develops high fields initially. The concentration of the
shallow p-layer can be chosen such that when the oxide charges go
into saturation the p-spray layer goes into depletion which
results in a decrease of the fields with irradiation. Finally, the
moderated p-spray, i.e. p-spray with a doping profile, keeps the
best of both worlds~\cite{rar96}.

\begin{figure}[h]
\begin{center}
  \includegraphics[width=.45\textwidth]{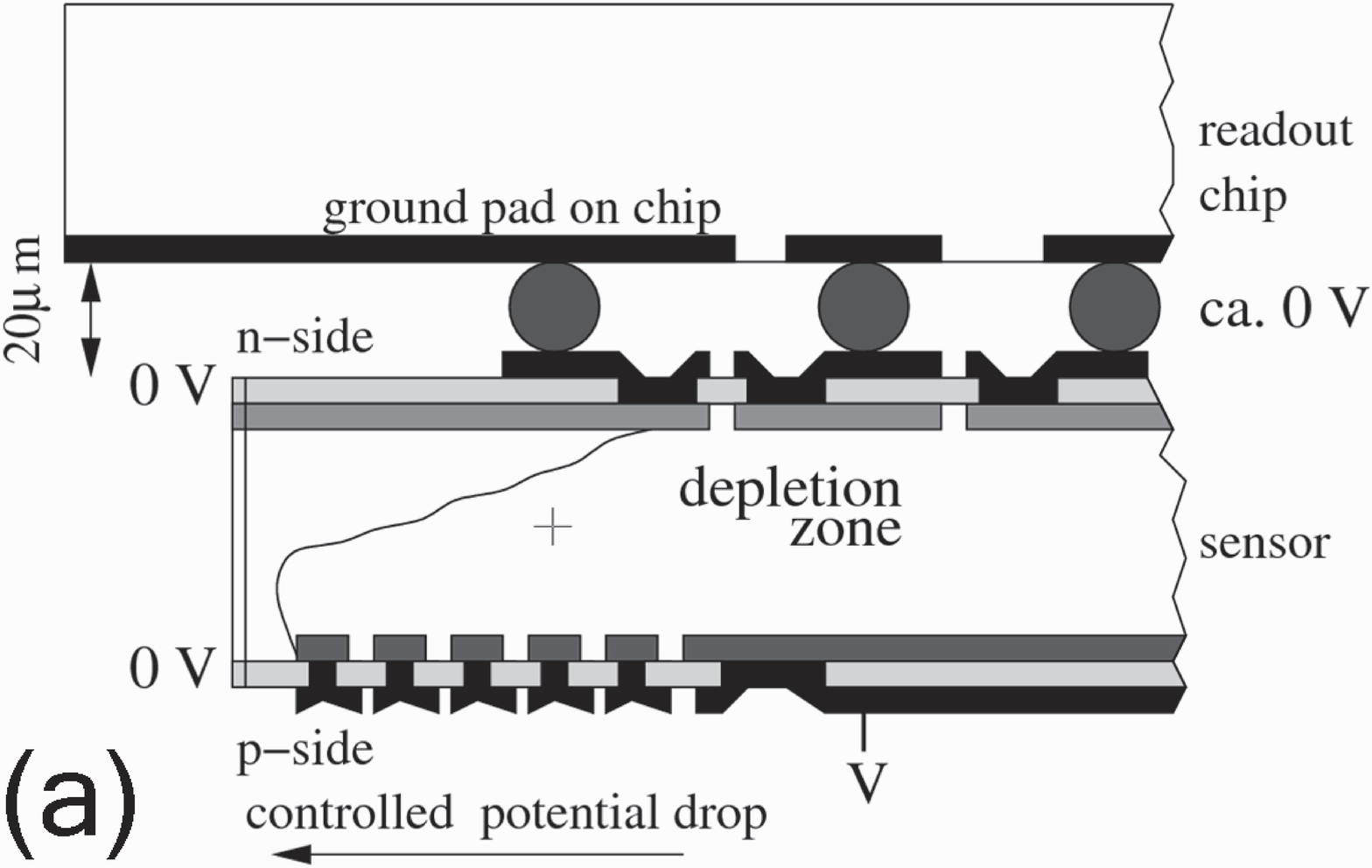}
  \hskip 1cm
  \includegraphics[width=.30\textwidth]{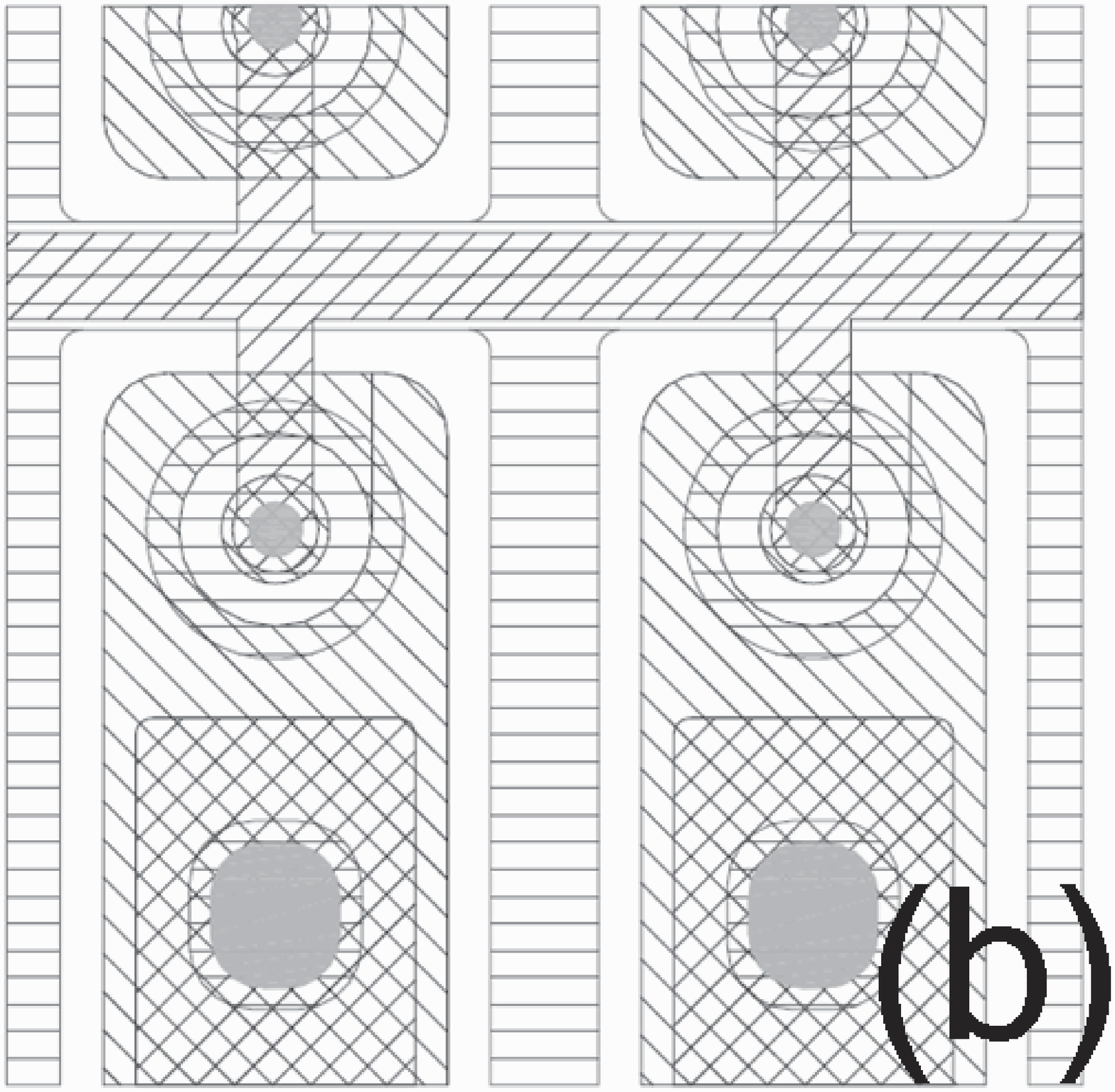}
\end{center}
\caption{\label{biasing}} (a) Edge of a pixel module showing the
controlled potential drop from the bias voltage to 0V via a series
of guard rings; (b) punch-through biasing of pixel implants by
means of a bias grid supplying every pixel.
\end{figure}

Figure~\ref{biasing}(a) shows the edge of a pixel module. As the
edge of the sensor is conducting the backside bias voltage must be
brought to 0V by a voltage dividing guard ring structure.
Otherwise the risk of discharge is present over the small gap
between sensor and chip. In normal operation the sensor pixels
have a defined potential by contact to the FE-chip. For testing
purposes of the sensor, however, before bonding has been done,
this is not the case. This problem has been overcome by
introducing a bias grid which runs around all pixels (see
Fig.~\ref{biasing}(b)). Applying the punch through voltage to the
grid via an external contact, all pixels are brought to a defined
potential by the punch-through effect (more details can be found
in~\cite{pixelbook}).
The bias grid of course produces additional implants in the pixel
cell which lead to a reduced charge collection efficiency in this
region by some 10-20$\%$. This is however tolerable in expense for
being able to thoroughly test the sensors before module assembly.

\begin{figure}[h]
\begin{center}
  \includegraphics[width=.55\textwidth]{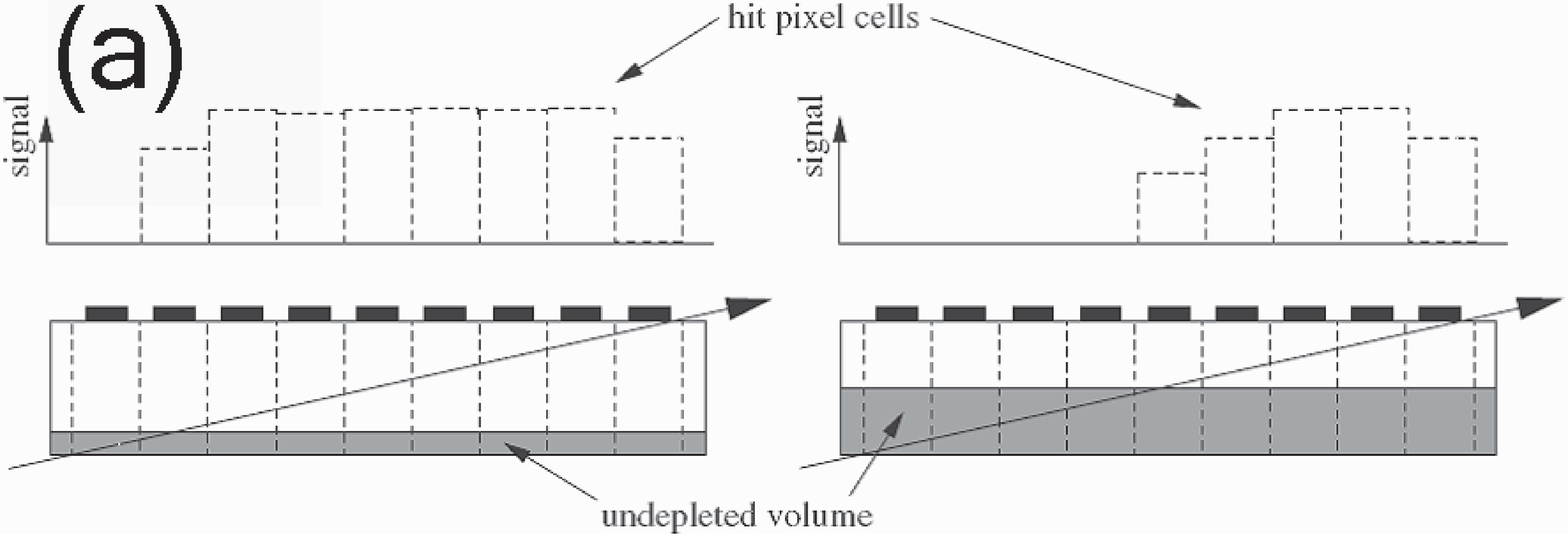}
  \includegraphics[width=.40\textwidth]{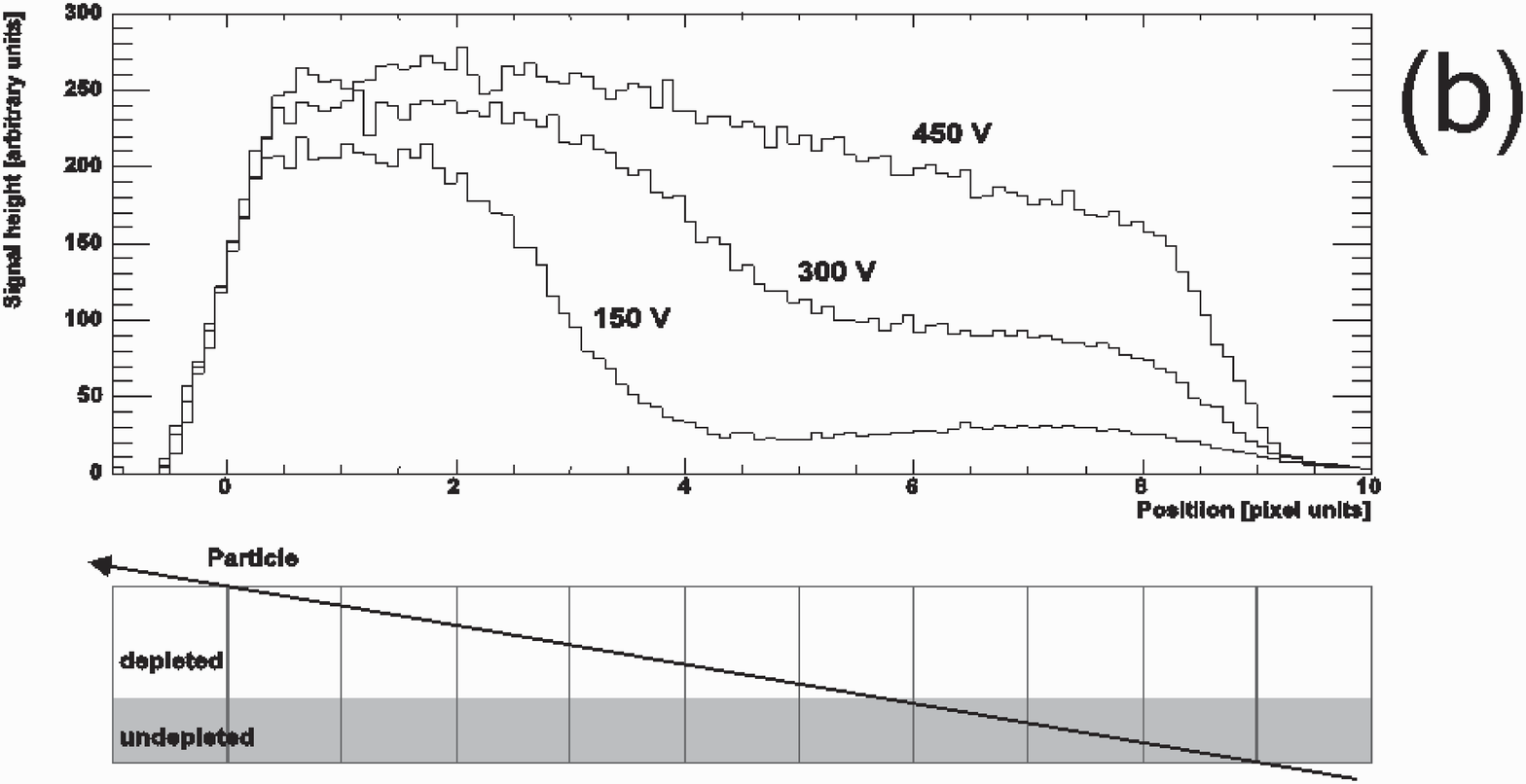}
\end{center}
\caption{\label{depdepth}} (a) Shallow angle method to measure the
depletion depth of a pixel sensor after irradiation. (b) observed
cluster size under shallow incidence angle, showing the dependence
of the cluster size on the depletion state of the sensor.
\end{figure}

Pixel detectors offer a nice way of measuring the depletion depth
after irradiation by exposing them to particle beams under a steep
angle of inclination, the shallow angle method~\cite{rohe-CMS}.
This is illustrated in Fig.~\ref{depdepth}(a). The observed
cluster size is a direct measure for the depletion depth.
Figure~\ref{depdepth}(b) shows the measured cluster size for
different bias voltages for CMS pixels, which correspond to fully
and less depleted pixel sensors. Full depletion after irradiation
is only obtained for bias voltages of more than 450 V.
Trapping effects due to irradiation can also be measured using the
shallow angle method as the track passes the pixel cells in a
varying depth~\cite{andreazza2006}. If trapping occurs, the
collected charge depends on the distance of the track passage to
the pixel electrode.

\subsection{The pixel front end chip}
\begin{figure}[h]
\begin{center}
  \includegraphics[width=0.9\textwidth]{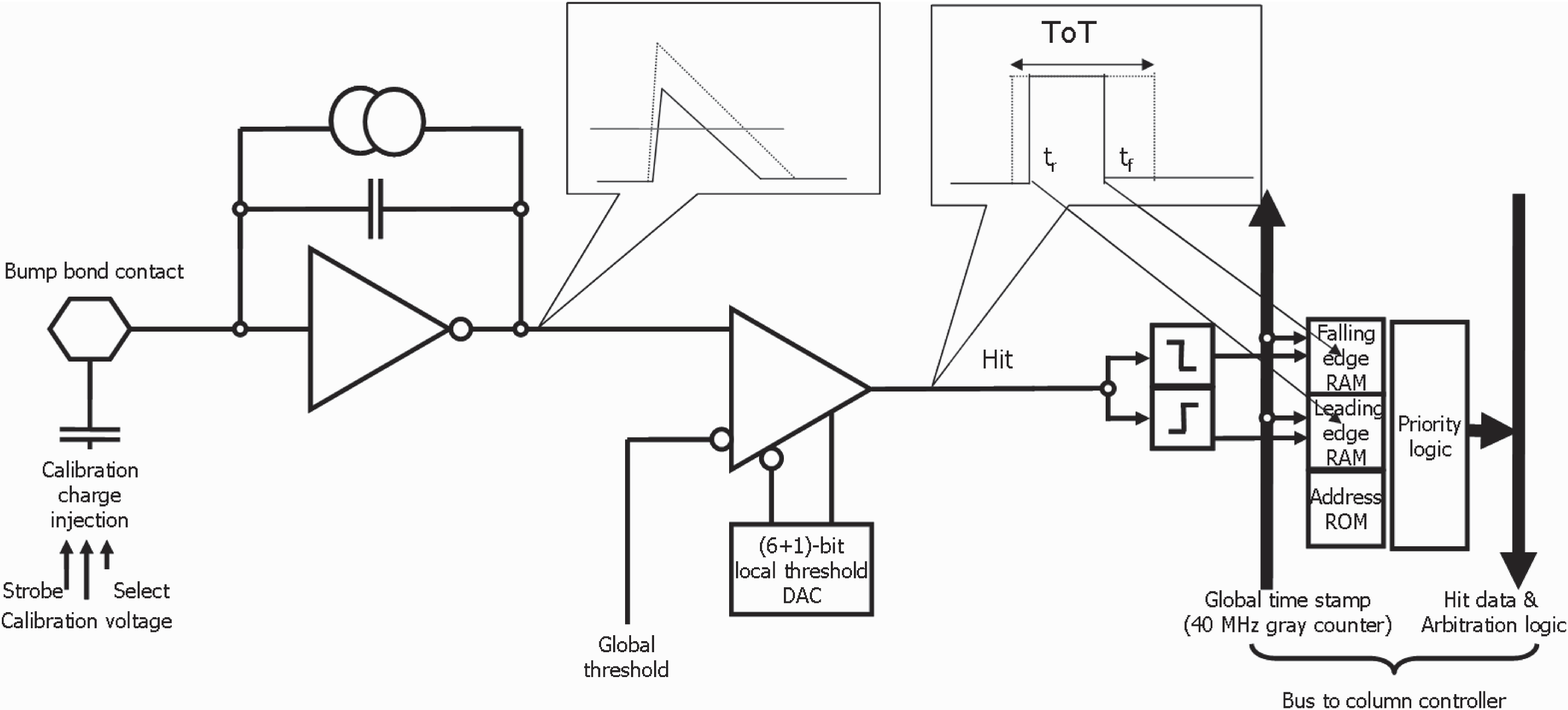}
\end{center}
\caption{\label{FEI3}} Functional block diagram of the ATLAS pixel
front end chip FEI3 (see text).
\end{figure}

\begin{figure}[h]
\begin{center}
  \hskip 2cm \includegraphics[width=0.55\textwidth]{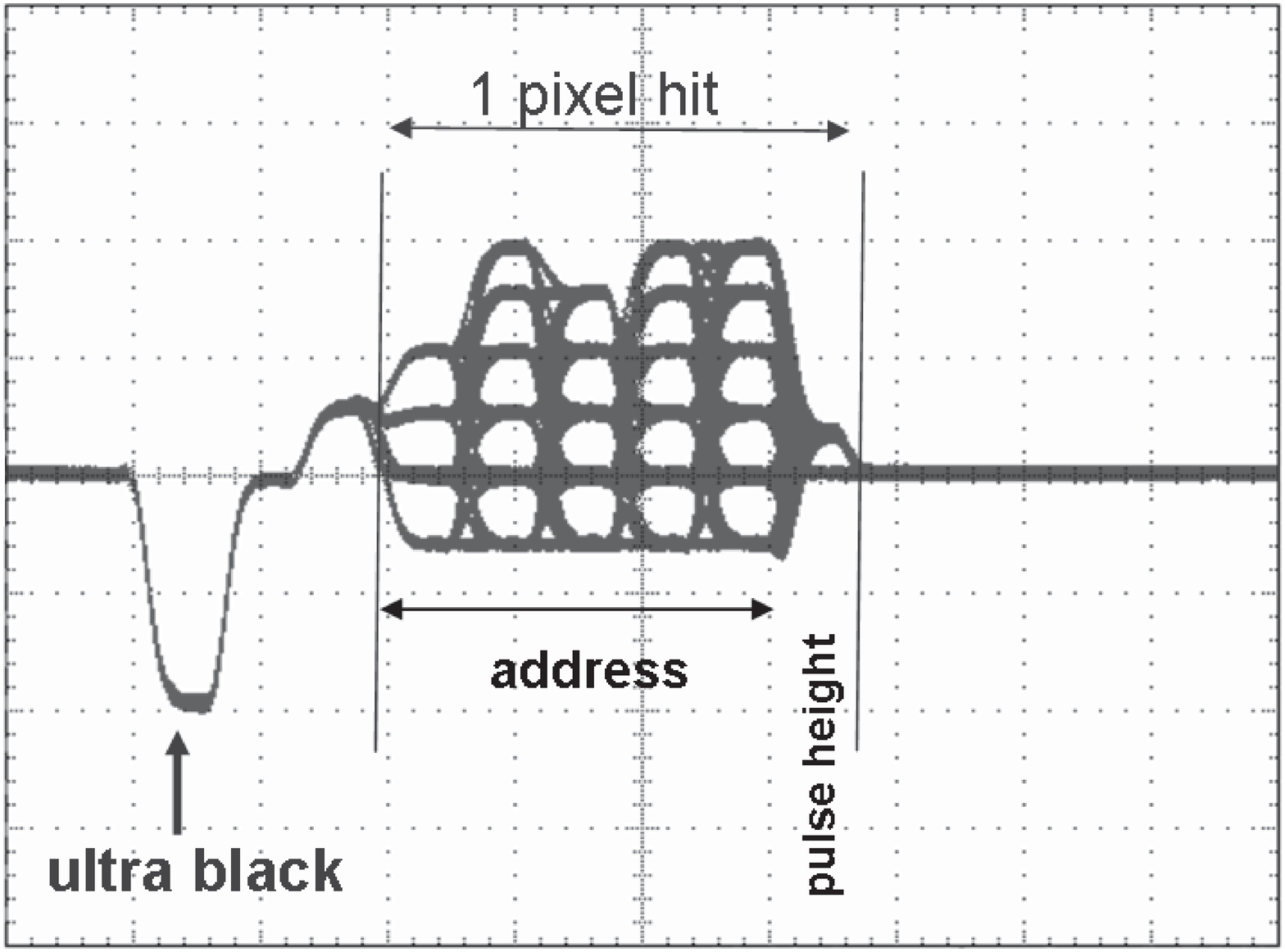}
\end{center}
\caption{\label{CMS-RO}} Analog encoding of a hit in the CMS Pixel
Detector displayed here by overlying 4160 readouts
(after~\cite{Kaestli2005}). After an ´'ultra black' five clock
cycles encode in analog levels 13 bits for the pixel address. The
analog pulse is transmitted with the following clock cycle.
\end{figure}

The pixel front end chips of ATLAS, CMS and ALICE are all
fabricated in the 0.25~$\um$ CMOS technology. The cell sizes of
ATLAS and ALICE are 50$\times$400~$\um^2$ while CMS has
100$\times$150~$\um^2$. In the case of ATLAS the
chip~\cite{blanquart03,blanquart01} is organized in 18 columns
time 160 rows, in total 2880 pixel cells, which are processed in
parallel. The readout is zero-suppressed. A block diagram of the
functionality is shown in Fig.~\ref{FEI3}. The current signal of a
hit is integrated by a charge sensitive preamplifier, whose
feedback capacitor is discharged by a constant current. This
results into a triangular pulse shape. The pulse then runs through
a discriminator which issues a norm pulse whose length is
determined by the times at which the rising and the falling edges
of the input cross the discriminator threshold. The length of the
output pulse is thus a measure of the time over threshold (ToT)
which in turn is a measure of the total charge. The address of the
pixel cell and the time stamps are stored into RAM storage at the
periphery. A fast scanner through a double column picks up the
valid hit addresses and time stamps and stores them into buffers
at the bottom of the pixel columns where they remain waiting for
the ATLAS first level trigger. The hits are removed from the
buffers if no trigger occurs in coincidence with the time stamps.
If the trigger time agrees with the time stamp a hit is read out.
Requirements to the front end chip are that the noise hit rate be
kept small compared to the real hit rate. It is demanded that the
quadratic sum of the noise value and the spread of the thresholds
over the chip be more than a factor of five below the chosen
threshold. For a nominal threshold setting of 3000 e$^-$ this
means less than 600 e$^-$. The obtained noise figures are in the
order of 50-60e$^-$. The untuned threshold dispersion is $\sim$
600 e$^-$. It is possible to tune (correct) the threshold settings
by a 7-bit digital trim DAC. After tuning the dispersion is
drastically reduced however to values much lower than 100e$^-$,
thus easily meeting the above requirement. The functional blocks
of the CMS Pixel chip~\cite{erdmann03,baur-CMS} are similar.
However, CMS additionally stores the analog pulse height in a
sample and hold stage, i.e. have a direct analog readout, while
the ATLAS readout is analog through the ToT trick. In CMS the
amplitude as well as the row and column addresses are then coded
in analog levels.
Figure~\ref{CMS-RO} shows the overlay of 4160 pixel readouts in
the CMS system~\cite{Kaestli2005}. Every event starts with an
´'ultra black' followed by seven clock cycles, five of which
encode 13 bits for the pixel address. The analog pixel pulse
height then follows in the seventh clock cycle.

\subsection{Radiation effects in the front end electronics}
\begin{figure}[h]
\begin{center}
  \includegraphics[width=0.35\textwidth]{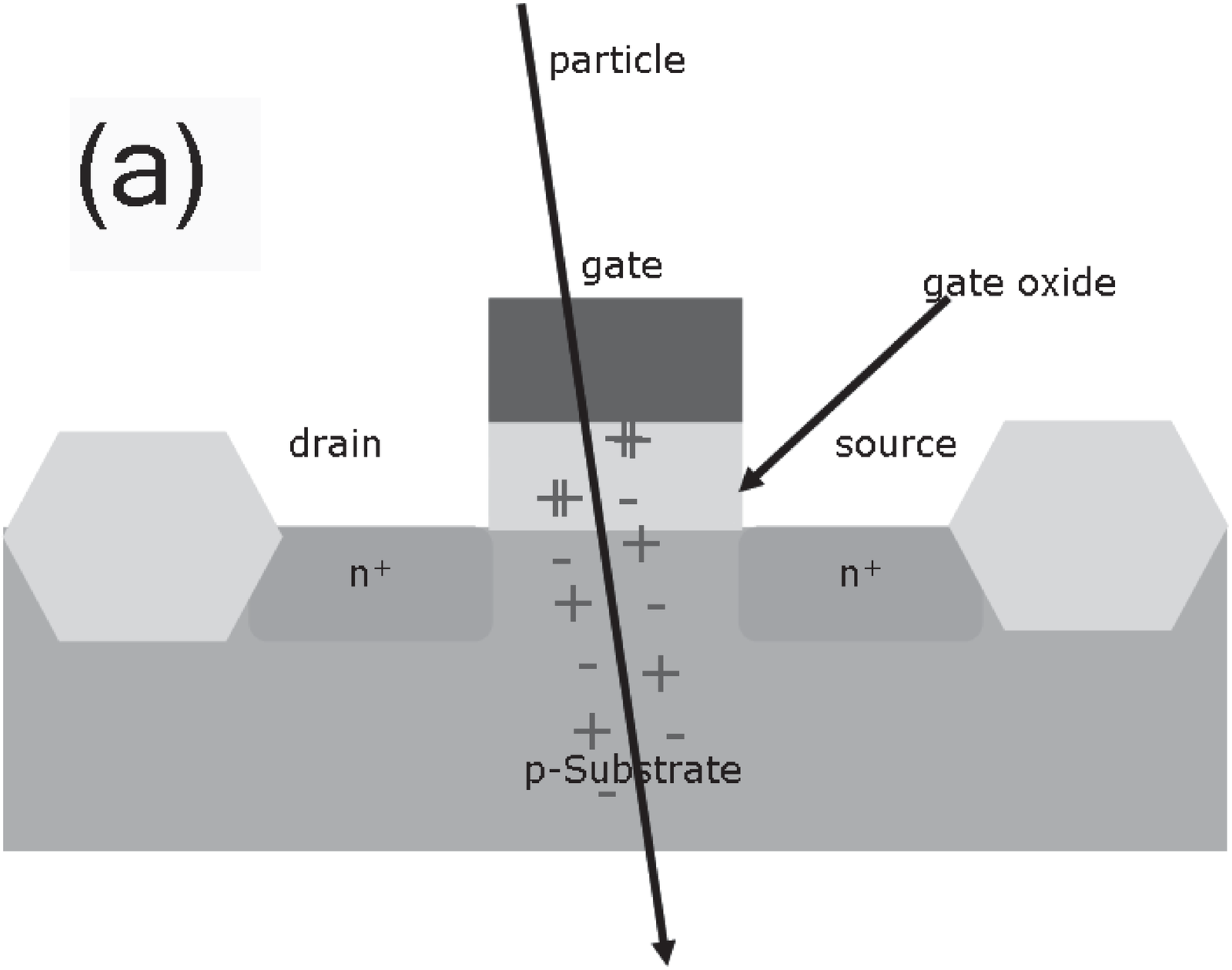}
  \hskip 3cm
  \includegraphics[width=0.25\textwidth]{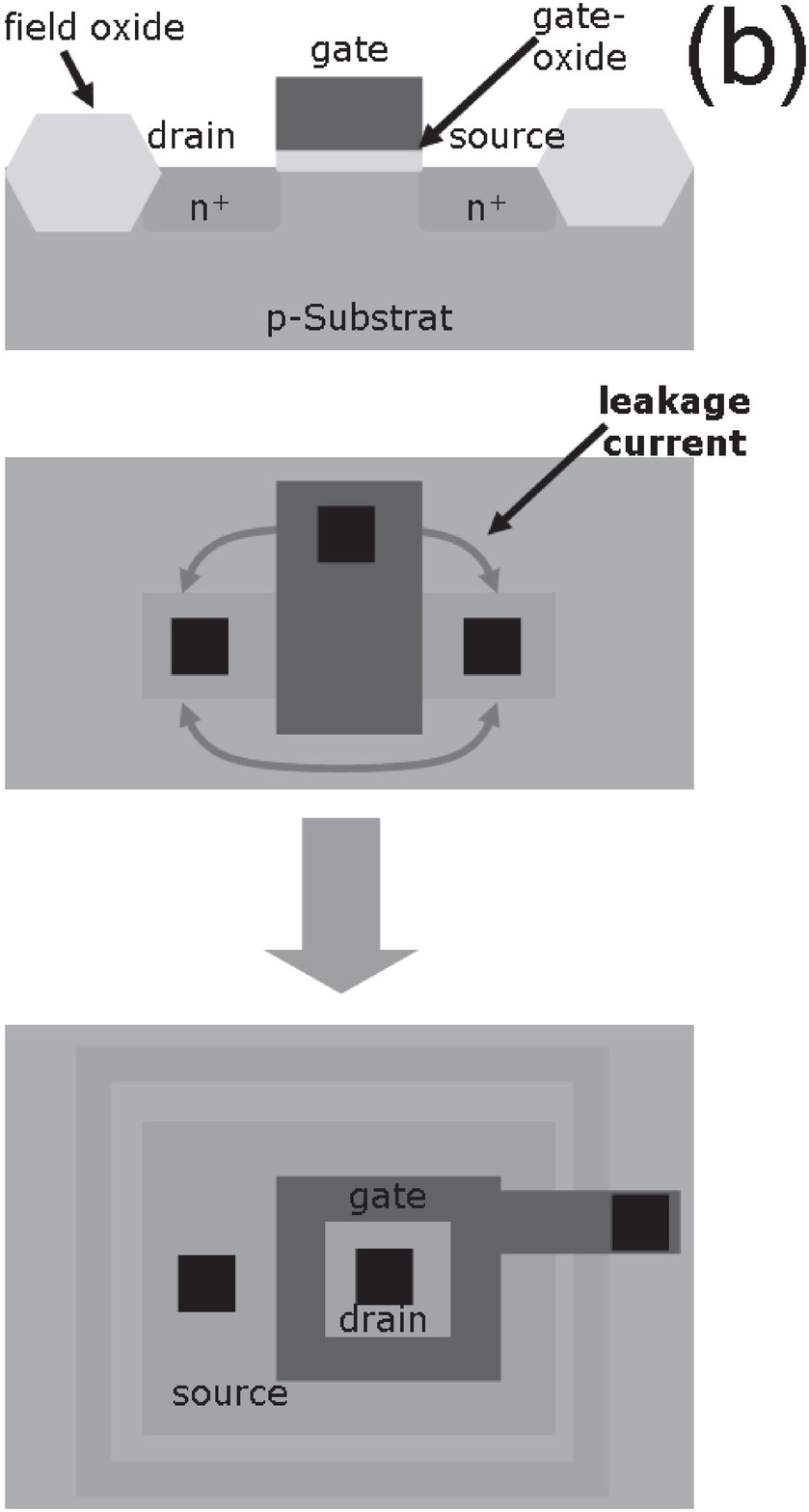}
\end{center}
\caption{\label{ChipRadiation}} (a) Effect of ionizing radiation
on the performance of CMOS ICs. Positive oxide charges lead to
transistor threshold shifts; (b) thin gate oxide in deep
sub-micron technologies as well as annular transistors cure the
main irradiation problems of CMOS ICs.
\end{figure}

The radiation effects that pixel chips suffer from differ from
those already described for the pixel sensors. For CMOS chips it
is mostly the ionizing radiation that generates positive charges
in the SiO$_2$ and also creates defects in the Si-SiO$_2$
interface as shown in Fig.~\ref{ChipRadiation}(a). The mobility of
electrons (high) and holes (low) is vastly different in SiO$_2$
such that charge pairs created by ionization produce low mobility
positive charges in the gate oxide while the electrons have
escape. This creates a net positive oxide charge and as a
consequence a shift in the transistor threshold. Another effect is
the generation of leakage currents under the field oxide, which in
effect creates parasitic unwanted transistors. Both effects have
been much reduced since so-called deep sub-micron CMOS
technologies have become available for HEP customers. The thin
gate oxide ($\sim$5nm in 0.25$\um$ CMOS) arranges for the low
mobility holes to tunnel out, thus curing or at least reducing the
effect of threshold shifts (see Fig.\ref{ChipRadiation}(b,top).
Transistor leakage currents can be eliminated by designing
transistors with annular gate electrodes and additional guard
rings as is also shown in Fig.~\ref{ChipRadiation}(b).

\begin{figure}[h]
\begin{center}
  \includegraphics[width=0.30\textwidth]{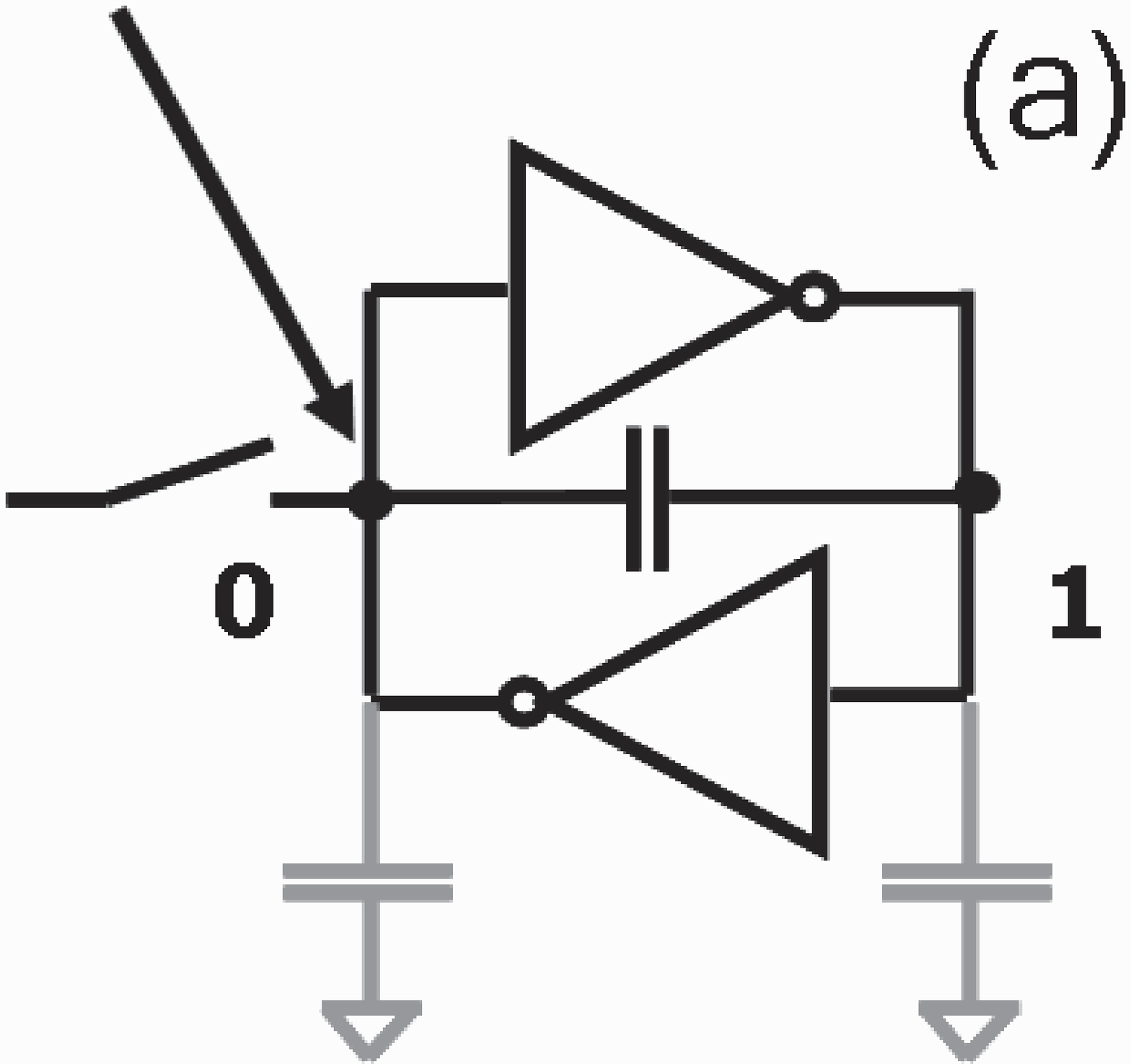}
  \hskip 2cm
  \includegraphics[width=0.30\textwidth]{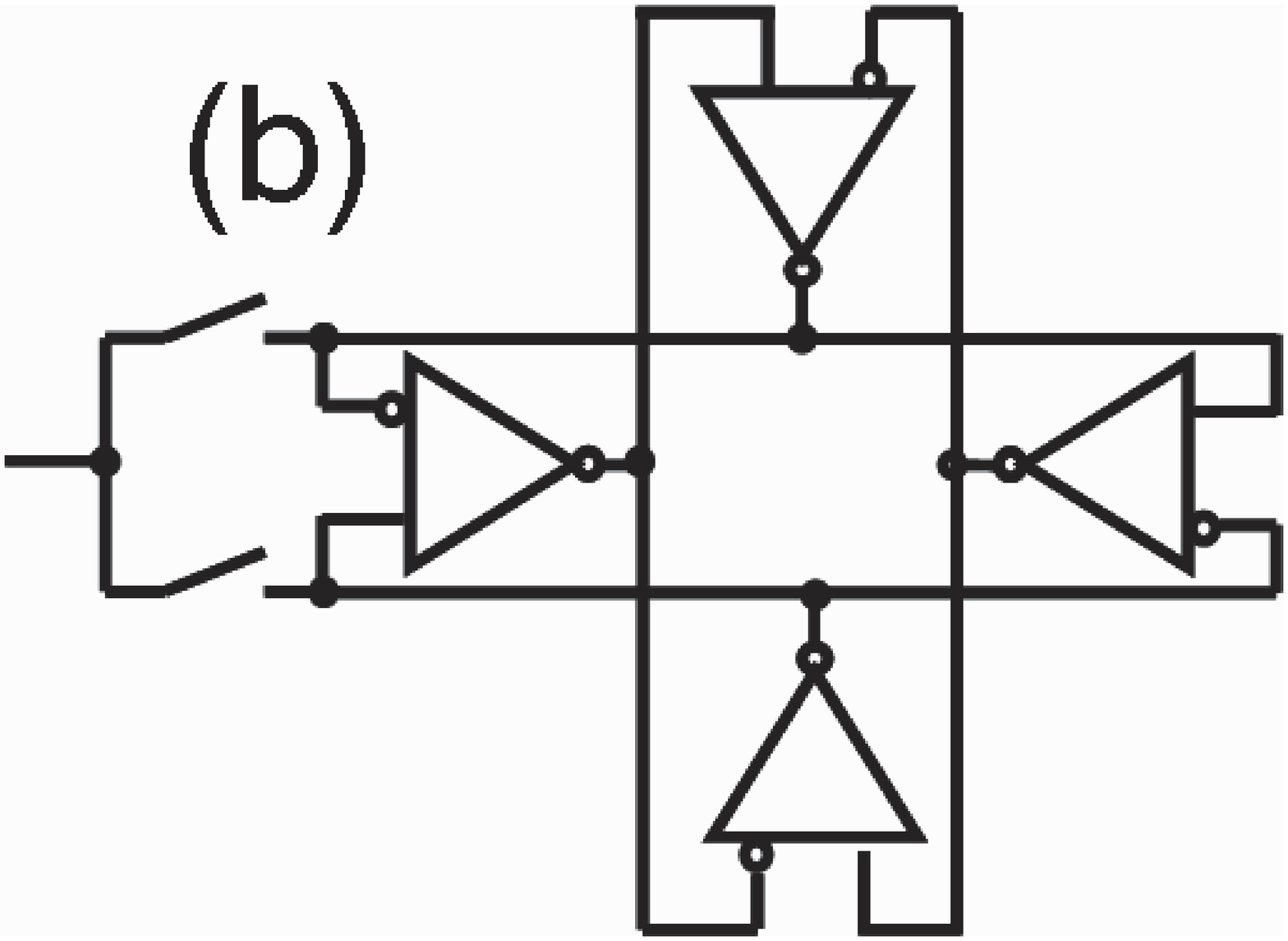}
\end{center}
\caption{\label{SRAM}} (a) Standard SRAM cell made more SEU
tolerant by introducing a capacitance between input and output
node. (b) The SEU tolerant DICE SRAM cell.
\end{figure}

\begin{figure}[h]
\begin{center}
  \includegraphics[width=0.35\textwidth]{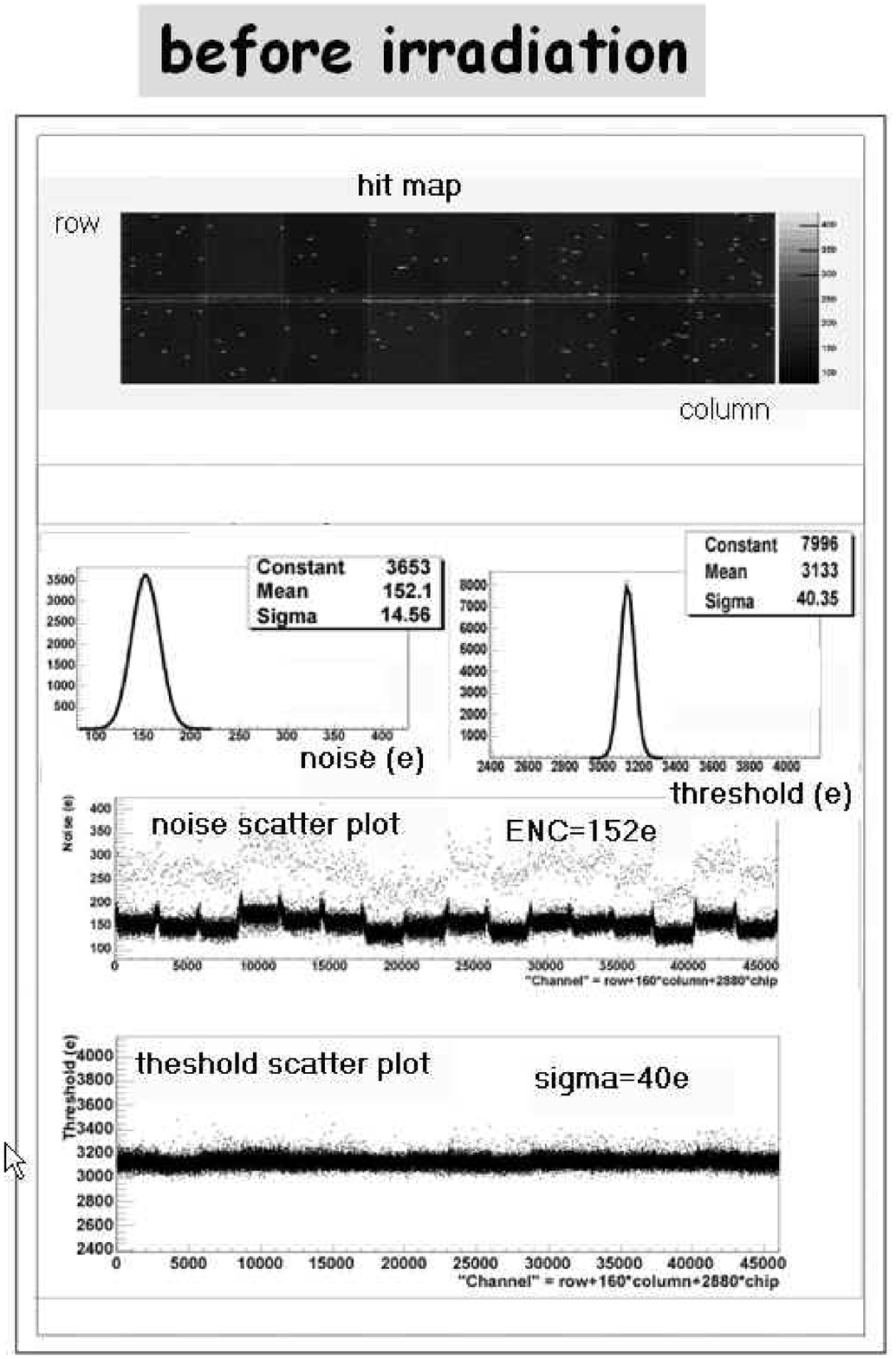}
  \hskip 2cm
  \includegraphics[width=0.35\textwidth]{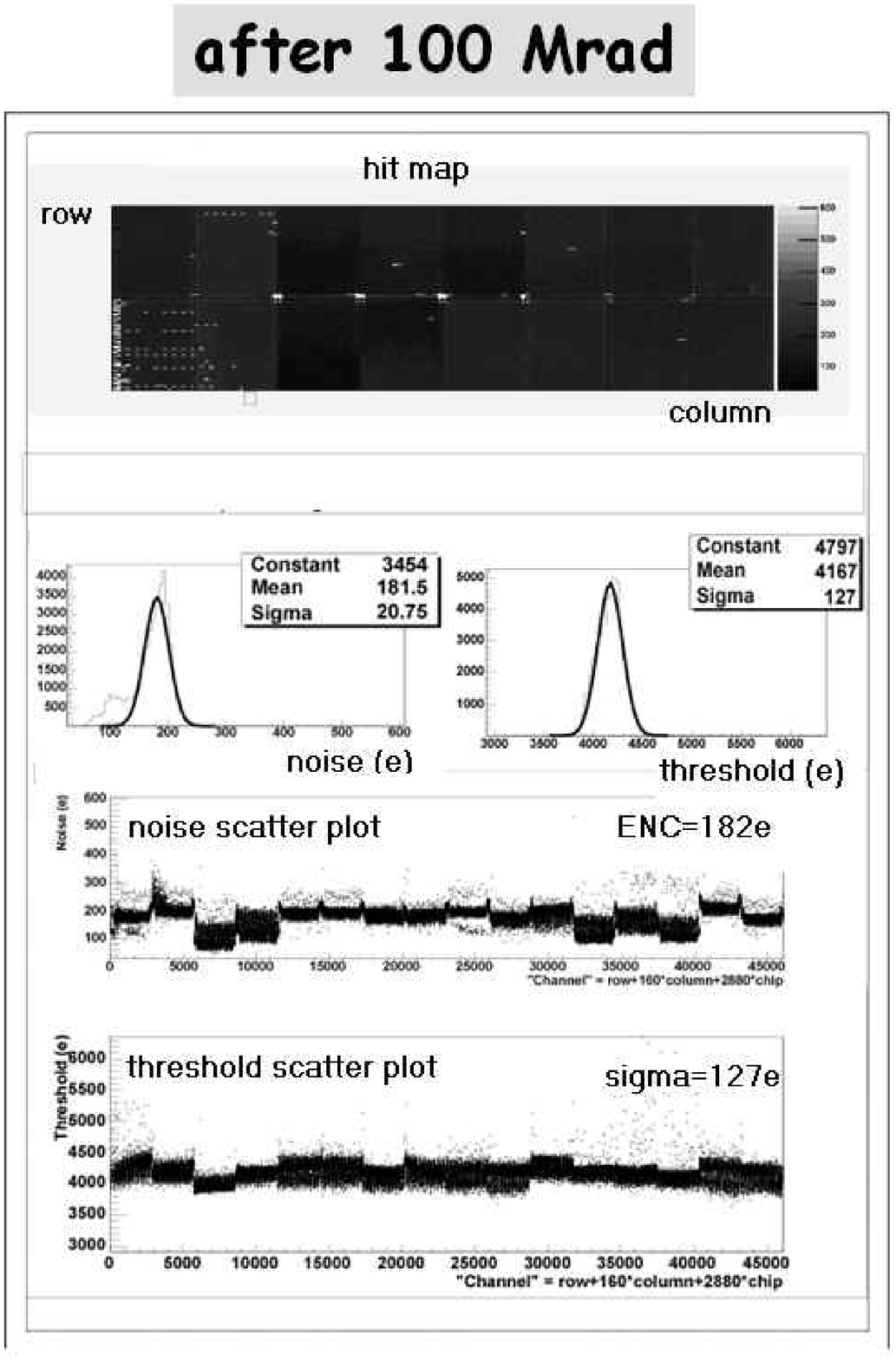}
\end{center}
  \caption{Comparisons of ATLAS pixel modules before and after
  irradiation to fluences of about 2.5$\times$10$^{15}$ p/cm$^2$, i.e. ionisation loss doses of
  about 100 Mrad/1 MGy in 250$\um$ silicon.
}
  \label{10yearsLHC}
\end{figure}

Also the digital parts of the front end electronics are affected
by the radiation at LHC. Most prominently radiation induced bit
errors, so called ´'single event upsets' (SEU) are reasons of
concern. Large amounts of charge on certain circuit nodes, created
for instance by nuclear reactions or high track densities can
cause a bit-flip. Two examples of more error resistant logic cells
are given in Fig.~\ref{SRAM}. A capacitance between input and
output node (Fig.~\ref{SRAM}(a)) slows down the response time and
makes the standard SRAM cell more tolerant against short charge
glitches on the input as long as the critical charge value
Q$_{crit}$ = V$_{thresh}$ $\cdot$ C is not crossed. The CMS ROC
chip uses such protections. The ingenious DICE SRAM cell
(Fig.~\ref{SRAM}(b)) stores the information and its inverse on 2+2
independent and cross-coupled nodes, such that a temporary flip of
one node cannot permanently flip the cell. Such a cell is used in
the ATLAS FE-chip on sensitive nodes.
Pixel modules have been irradiated to 10-year LHC doses and
beyond. Figure~\ref{10yearsLHC} shows the comparison of an ATLAS
pixel module with 16 chips before and after irradiation to 1 MGy
or 2.5 $\times$ 10$^15$ p/cm$^2$ which, in fact, corresponds to a
total dose of about 20 years at LHC. The figure shows on the top a
hit map and below distributions of the measured noise and the
thresholds determined by respective threshold scan measurements.
It can be noted that while the noise and threshold distributions
vary more from chip to chip, which can be seen in the bottom
scatter plots, the quadratic sum of their averages still remains
well under 300$e^-$, i.e. comfortably away from the nominal
threshold setting of 3000$e^-$. The spatial resolution of
irradiated modules decreases by about 40$\%$ and the in-time hit
efficiency - the efficiency of hit detection within the 25 ns time
window of the LHC bunch crossing - decreases from 99.9$\%$ before
irradiation to 97.8$\%$ after 600 kGy.

\subsection{Support structures and material issues}
\begin{figure}[htb]
\begin{center}
\includegraphics[width=1.\textwidth]{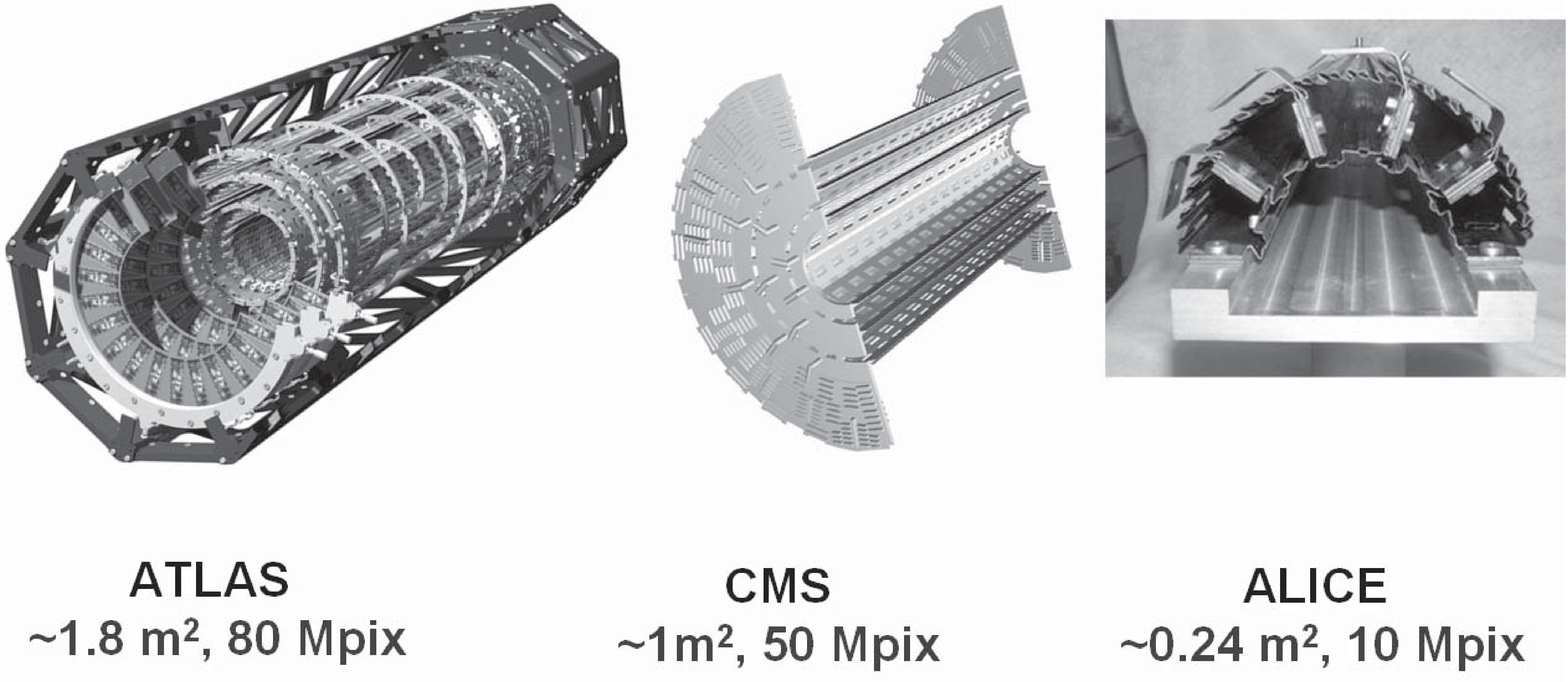}
\end{center}
  \caption{Support structures of the ATLAS (carbon-carbon), CMS (carbon-fibre),
  and ALICE (carbon-fibre) pixel detector global support structures.}
  \label{supports}
\end{figure}

The pixel modules are loaded on ladders and disks to complete the
detector. The main goal for detector builders in this respect is
to minimize the amount of material in terms of radiation length
imposed by the support structure and the cooling mechanics, which
is particularly difficult to achieve when the operation
temperature is required to be about -5$^\circ$ to -10$^\circ$~C in
order to block the reverse annealing effect (see
chapter~\ref{sensors}). Figure~\ref{supports} shows the mechanical
support structures of the pixel detectors for ATLAS, CMS and
ALICE.
In central heavy ion collisions, the environment of the ALICE
detector, up to 8000 charged particles per unit of rapidity are
produced, generating hit densities of about 80 hits / cm$^2$. As
the luminosity for heavy ion collisions is much smaller than for
pp collisions (L $\approx$ 10$^{27}$ cm$^{-2}$ s$^{-1}$ for Pb-Pb
collisions, L $\approx$ 10$^{29}$ cm$^{-2}$ s$^{-1}$ for Ar-Ar or
O-O collisions) the radiation levels are only $\sim$ 5 kGy or
6$\times$10$^{12}$ n$_{eq}$ / cm$^2$. Hence, operation at
temperatures below zero to suppress the reverse annealing effect,
are not mandatory. Instead, operation at room temperature is
possible. ALICE has exploited this~\cite{pepato2006} for the
benefit of reducing the cooling needs and the material. Using a
light weight carbon fibre structure in a chamber type fashion
rather than ladders and shell structures, and using C$_4$F$_{10}$
as coolant with very thin cooling pipes (PHYNOX) of 40$\um$ wall
thickness only, they arrive at a total material budget at $\eta$ =
0 of only 0.9$\%$ of a radiation length. For comparison, ATLAS and
CMS have about 2.5$\%$.

\begin{figure}[h]
\begin{center}
  \includegraphics[width=0.55\textwidth]{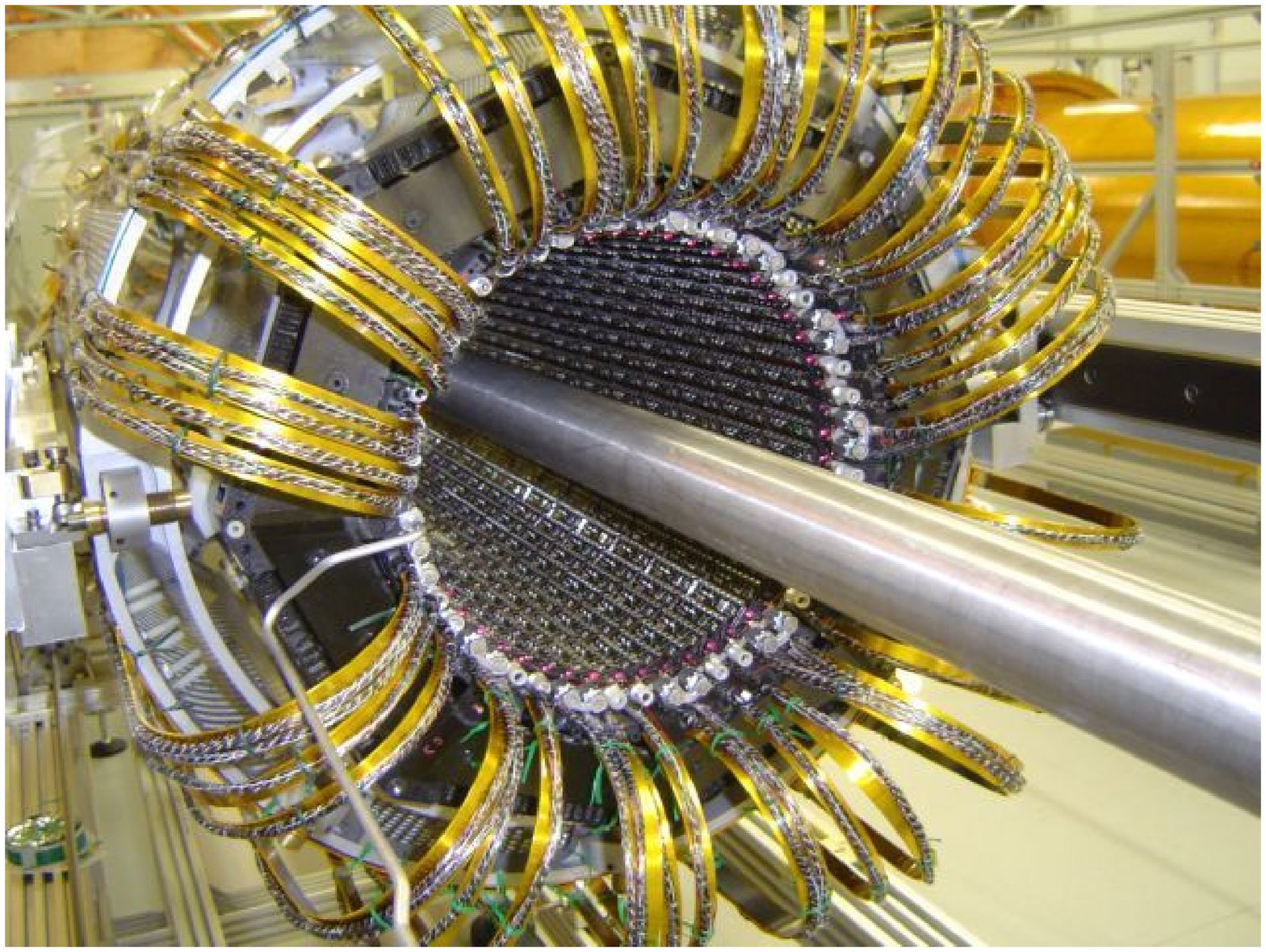}
  \hskip 1.5cm
  \includegraphics[width=0.30\textwidth]{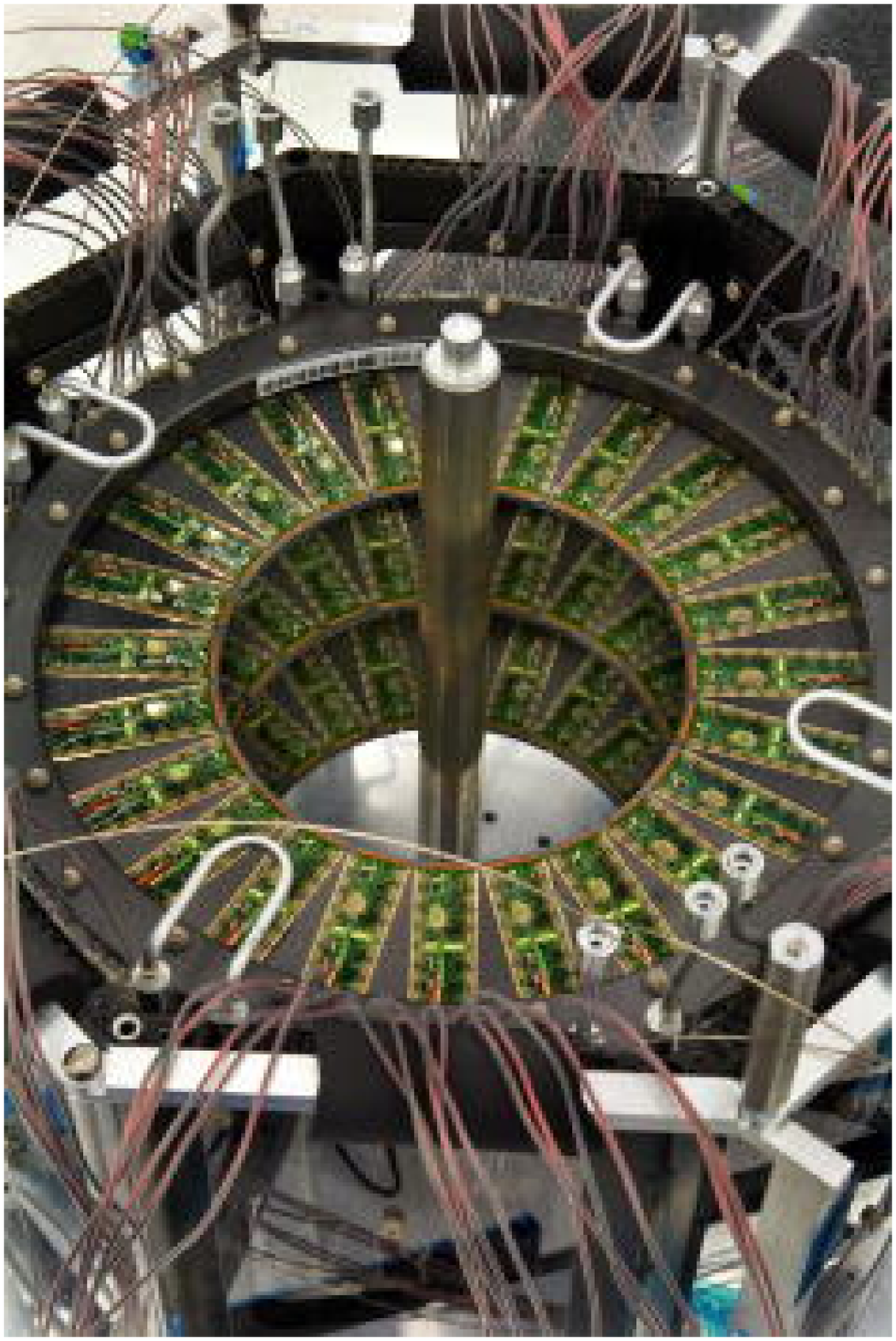}
\end{center}
\caption{\label{ATLASPix}} Photographs of the ATLAS pixel detector
fully assembled. (a) Pixel barrel layer-2, (b) one stack of disks.
\end{figure}

At the time of writing the ATLAS detector nears completion for
assembly. Figure~\ref{ATLASPix} shows photographs of the barrel
layer-2 and the disk stack, completely assembled in its support
structure.

\subsection{Summary on Hybrid Pixel Detectors}
Hybrid pixel detectors constitute the state of the art in pixel
detectors at present. The have proven to be suited for LHC
conditions in terms of readout speed as well as radiation
hardness. Every cell already provides complex signal processing
including zero suppression. Hits are temporarily stored during the
trigger latency of several~$\um$. The 3D capability renders pixel
detectors at LHC superior to any other detector regarding pattern
recognition capability. The spatial hit resolution is about
10-15~$\um$ in the transverse direction.

As disadvantages one would probably have to consider the
comparatively large material budget of $\sim$2.5$\%$ in ATLAS and
CMS and $\sim$1$\%$ in ALICE, mostly due to cooling, support
structure and services. The module production is quite laborious
including many production steps such as bump bonding and
flip-chipping, which are also a cost issue.

\section{Pixel R\&D for Future Colliders}

\subsection{Challenges imposed by a Super-LHC}
The data rate and also the radiation levels expected at an LHC
upgrade, called Super-LHC or sLHC, are a factor of up to ten
higher than at the LHC, i.e. about
2$\times$10$^{16}$n$_{eq}$/cm$^2$ at a radius of 4 cm. Concerning
irradiatio issues there are mainly three effects as a consequence
(see review~\cite{moll2006}).
\begin{enumerate}
\item[1.] A change of the effective doping concentration (higher
depletion voltage necessary, under-depletion) \item[2.] An
increase of leakage current (increase of shot noise, thermal
runaway) \item[3.] An increase of charge carrier trapping (loss of
charge)
\end{enumerate}
Several routes to cope with this are being pursued, among them the
development of even more radiation hard silicon based on
oxygenated float-zone (DOFZ), Czochralski (Cz), n$^+$ in p silicon
sensors, as well as epitaxial silicon~\cite{moll2006}. For this
lecture I would like to address in this context two new approaches
which are more linked to pixel detectors: diamond pixel detectors
and 3D-silicon devices.

\subsection{Diamond Pixels}
\begin{figure}[h]
\begin{center}
\includegraphics[width=0.55\textwidth]{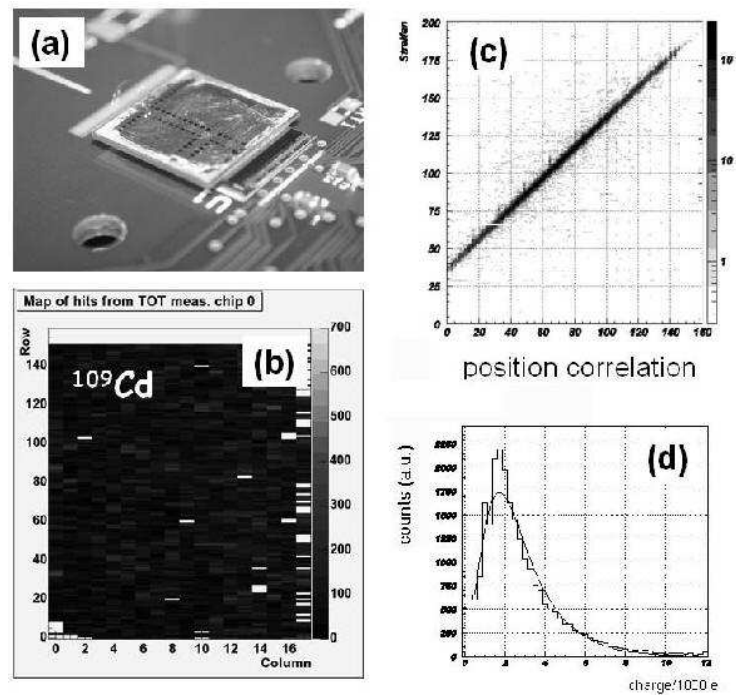}
\end{center}
\caption[]{(a) Single chip diamond pixel module using ATLAS
front-end electronics, (b) hit map obtained by exposure to a
$^{109}$Cd radioactive source (22 keV $\gamma$), (c) scatter plot
of position correlation between the diamond pixel detector and a
reference beam telescope, and (d) measured Landau distribution in
a CVD-diamond pixel detector.} \label{diamond-single}
\end{figure}
\begin{figure}[!h]
\begin{center}
\includegraphics[width=0.8\textwidth]{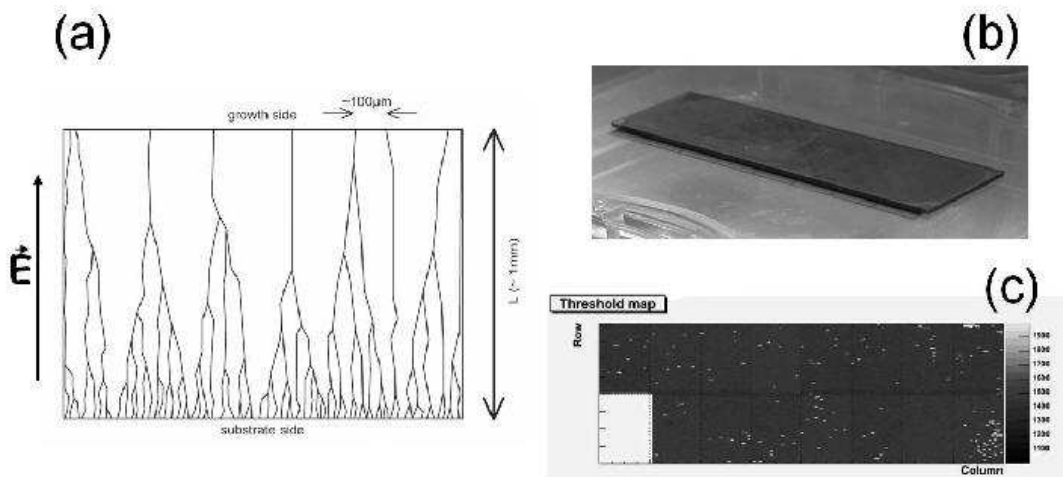}
\end{center}
\caption[]{(a) Grain structure of CVD-diamond sensors. (b) a full
size CVD-diamond module from a CVD diamond wafer bump bonded to 16
ATLAS FE-chips (on the bottom), (c) threshold map after tuning of
the module showing its full functionality.}\label{diamond-grains}
\end{figure}

CVD-Diamond as a sensor material has been developed by the CERN
R$\&$D group RD42 for many years~\cite{RD42}. Charge collection
distances approaching 300 $\mu$m has also triggered the
development of a hybrid pixel detector using diamond as sensors
\cite{diamond-pixels,kagan_PIX2005}. The non-uniform field
distribution inside CVD-diamond, which originates from the grain
structure in the charge collecting bulk
(cf.~Fig.~\ref{diamond-grains}(a)) introduces polarization fields
inside the sensor due to charge trapping at the grain boundaries
which superimpose on the biasing electric field. This results in
position dependent systematic shifts in the track reconstruction
with a typical average grain size of 100$\mu$m -
150$\mu$m~\cite{lari04}. Diamond sensors with charge collection
distances in excess of 300$\mu$m have been fabricated and
tested~\cite{Kagan_Hiroshima}. Single chip pixel modules as well
as a full size wafer scale 16-chip module assembled using ATLAS
front-end chips have been built and tested.
Figure~\ref{diamond-single}(a) and (b) show the diamond pixel
detector and a hit response pattern obtained by exposing the
detector to a $^{109}$Cd source of 22 keV $\gamma$ rays, which
deposits approximately 1/4 of the charge of a minimum ionizing
particle. The single chip module has been tested in a high energy
(180 GeV) pion beam at CERN, the module in a $\sim$4 GeV electron
beam at DESY. Figure~\ref{diamond-single}(c) and (d) show position
correlation and the charge distribution of the diamond pixel
detector in a high energy beam, respectively. A spatial resolution
of $\sigma$ = 12$\mu$m has been measured with the single chip
module at high energies with 50$\mu$m pixel pitch. A technical
challenge to produce a wafer scale module lies in the
hybridization process, i.e. bump deposition and flip-chipping.
Figure~\ref{diamond-grains} shows the 16-chip diamond module
(Fig.~\ref{diamond-grains}(b), chips are on the bottom) and its
tuned threshold map (Fig.~\ref{diamond-grains}(c)) with a very
small dispersion of only 25e$^-$ and good bump yield homogeneity.
One chip was damaged during test beam by electrostatic discharge.
The rms of the position residuals was measured to 24~$\mu$m in the
DESY 6 GeV beam~\cite{kagan_PIX2005} including a large multiple
scattering contribution from the beam telescope.

\subsection{3D silicon sensors}
So-called 3D silicon detectors have been
developed~\cite{3D-parker} to overcome several limitations of
conventional planar Si-pixel detectors, in particular in high
radiation environments, in applications with inhomogeneous
irradiation and in applications which require a large
active/inactive area ratio such as protein crystallography.
\begin{figure}
\begin{center}
\includegraphics[width=0.45\textwidth]{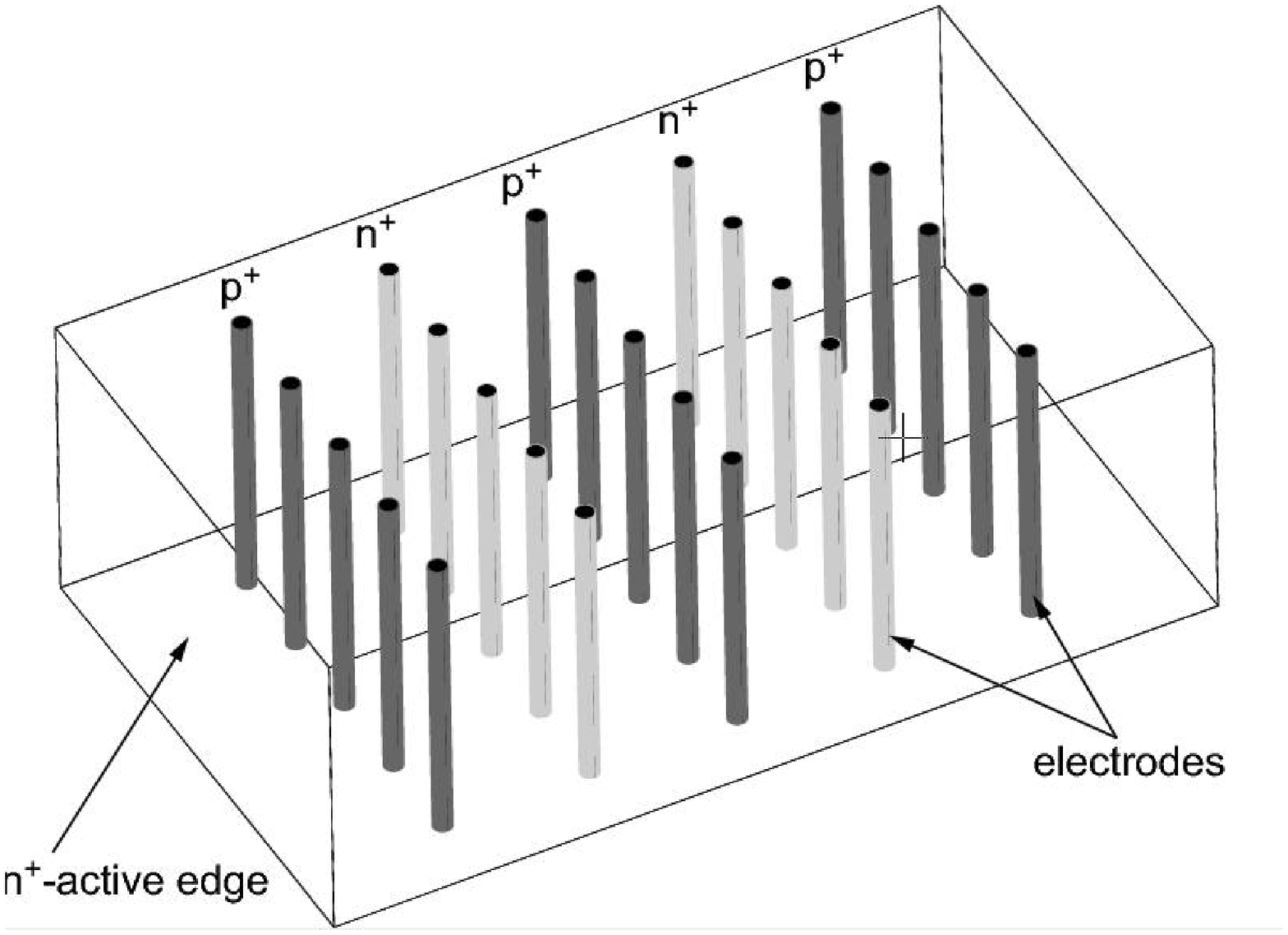}
\vspace{0.4cm}
\includegraphics[width=0.45\textwidth]{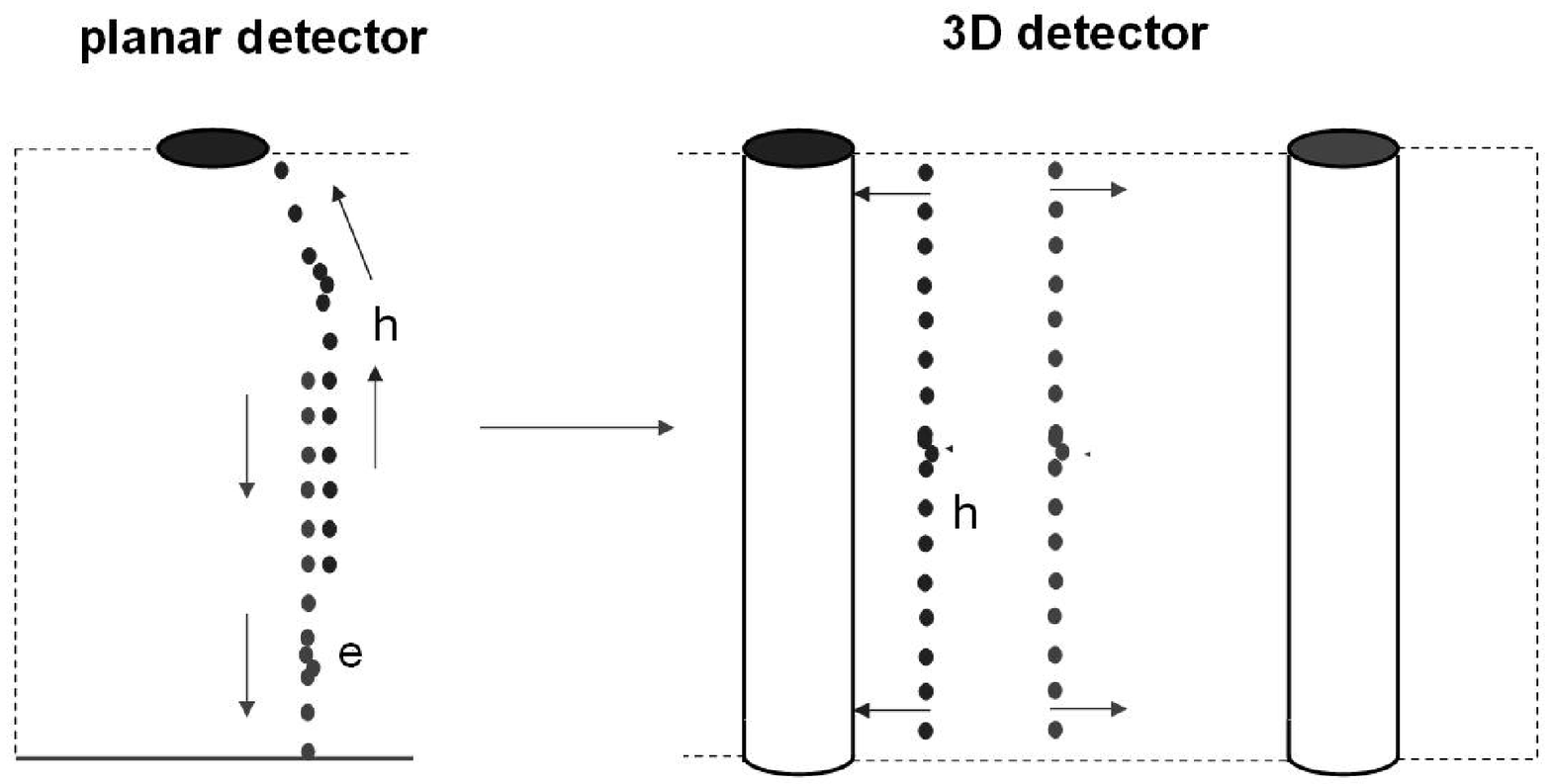}
\end{center}
\caption[]{(top) Schematic view of a 3D silicon detector, (bottom
left) comparison of the charge collection in a conventional planar
electrode silicon detector, (bottom right) a 3D-silicon detector.}
\label{3D}
\end{figure}
A 3D-Si-structure (Fig.~\ref{3D}(a)) is obtained by processing the
n$^+$ and p$^+$ poly silicon electrodes into the detector bulk
rather than by conventional implantation on the surface. This is
done by combining VLSI and MEMS (Micro Electro Mechanical Systems)
technologies. Charge carriers drift inside the bulk parallel to
the detector surface over a short drift distance of typically
50$\mu$m. Another feature is the fact that the edge of the sensor
can be a collection electrode itself thus extending the active
area of the sensor to within few $\mu$m to the edge. Edge
electrodes also avoid inhomogeneous fields and surface leakage
currents which usually occur due to chips and cracks at the sensor
edges. The main advantages of 3D-silicon detectors, however, come
from a different way of charge collection (cf. Fig.~\ref{3D}(b))
and the fact that the electrode distance is short (50$\mu$m) in
comparison to conventional planar devices at the same total
charge. This results in a fast (1-2 ns) collection time, low ($<$
10V) depletion voltage and, with edge electrodes in addition, a
large active/inactive area ratio of the device. The technical
fabrication is much more involved than for planar processes and
requires a bonded support wafer and reactive ion etching of the
electrodes into the bulk. Prototype detectors using strip or pixel
electronics have been fabricated~\cite{parker_pix2005} and show
encouraging results with respect to speed (3.5 ns rise time) and
radiation hardness ($\gg$ 10$^{15}$
protons/cm$^2$)~\cite{3DPortland}. 3D-pixel detectors with ATLAS
frontend electronics have also been successfully built.
Preliminary results have been presented in~\cite{cinzia2006}.

\subsection{Monolithic and Semi-Monoli\-thic Pixel Detectors}
Monolithic pixel detectors, in which amplifying and logic
circuitry as well as the radiation detecting sensor are one
entity, are in the focus of present developments for future
experiments. To reach this ambitious goal, optimally using a
commercially available and cost effective technology, would be
another breakthrough in the field. The present developments have
been much influenced by R$\&$D for vertex tracking detectors at
future colliders such as the International Linear $e^+e^-$
Collider (ILC) \cite{TESLA-TDR}. Very low ($\ll$1$\%$ X$_0$)
material per detector layer, small pixel sizes
($\sim$20$\mu$m$\times 20 \mu$m) and a high rate capability (80
hits/mm$^2$/ms) are required, due to the very intense
beamstrahlung of narrowly focussed electron beams close to the
interaction region, which produce electron positron pairs in vast
numbers. High readout speeds with typical line rates of several
$10$ MHz and a 40$\mu$s frame readout time are necessary. At
present, two developments have already reached some level of
maturity: so called CMOS active pixels and DEPFET pixels.
\subsubsection{CMOS active pixels}
\begin{figure}[h]
\begin{center}
\includegraphics[width=0.6\textwidth]{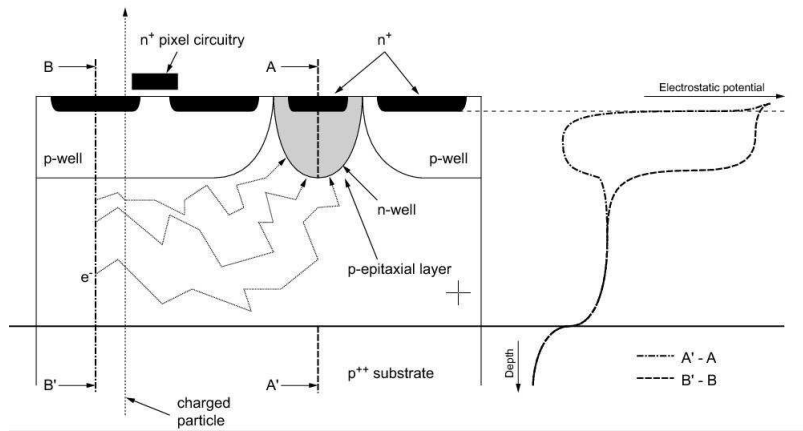}
\end{center}
\caption[]{Principle of a Monolithic Active Pixel Sensor
(MAPS)~\cite{MAPS1}. The charge is generated and collected by
diffusion in the very few $\mu$m thick epitaxial
Si-layer.}\label{MAPS}
\end{figure}
In some CMOS chip technologies a lightly doped epitaxial silicon
layer of a few to 15$\mu$m thickness between the low resistivity
silicon bulk and the planar processing layer can be used for
charge collection~\cite{meynants98,MAPS1,MAPS2}. The generated
charge is kept in a thin epi-layer atop the low resistivity
silicon bulk by potential wells that develop at the boundary and
reaches an n-well collection diode by thermal diffusion (cf. Fig.
\ref{MAPS}). With small pixel cells collection times in the order
of 100~ns are obtained. The charge collecting epi-layer is --
technology dependent -- at most 15$\mu$m thick and can also be
completely absent. The attractiveness of active CMOS pixels lies
in the fact that standard CMOS processing techniques are employed
and hence they are potentially very cheap.
\begin{figure}[!h]
\begin{center}
\includegraphics[width=1.0\textwidth]{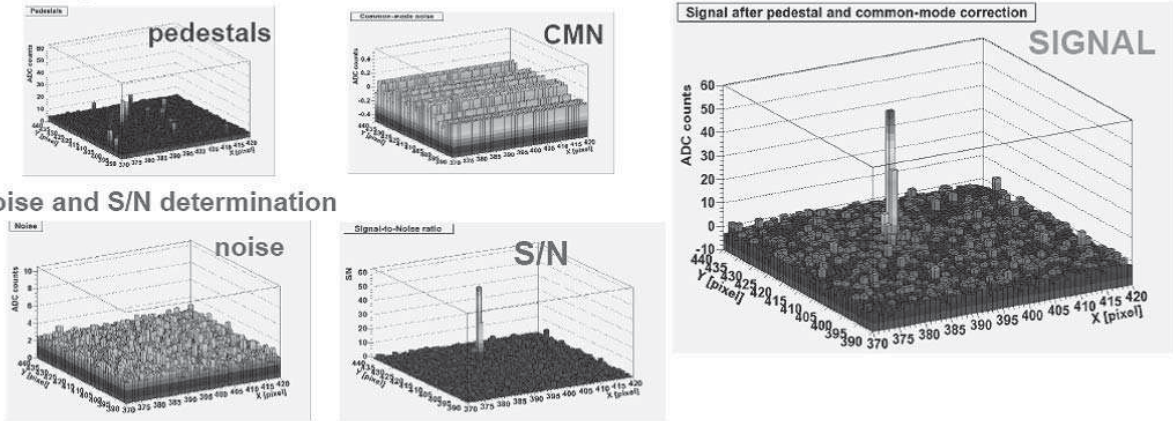}
\end{center}
\caption[]{Readout of a CMOS active pixel matrix (see
text)}\label{MAPS-CDS}
\end{figure}
CMOS active pixel sensor development are pursued by many groups
which partially collaborate in various projects (BELLE-upgrade,
STAR-upgrade, ILC, CBM at GSI) who use similar approaches to
develop large scale CMOS active pixels, also called MAPS
(Monolithic Active Pixel Sensors)~\cite{MAPS1}. The sensor is
depleted only directly under the n-well diode. The signal charge
is hence very small ($<$1000e) and full charge collection is
obtained only in the depleted region under the n-well electrode.
Low noise electronics is therefore the challenge in this
development.

Matrix readout of MAPS is performed using a standard 3-transistor
circuit (line select, source-follower stage, reset) commonly
employed by CMOS matrix devices, but can also include current
amplification and current memory~\cite{Dulinski03}. In the active
area only nMOS transistors are permitted because of the
n-well/p-epi collecting diode which does not permit other n-wells.
For an image two complete frames are subtracted from each other
(correlated double sampling, CDS) to eliminate base levels, 1/f
and fixed pattern noise (see Figure~\ref{MAPS-CDS}). In a second
step pedestals and common mode noise are subtracted to extract the
signal and to determine the remaining noise. Detector sizes up to
19.4$\times$17.4 mm$^2$ with 1M pixels have been tested. Frame
speeds of 10$\mu$s for 132x48 pixels have been reached for the
BELLE development, with a noise figure of
30-50e$^-$~\cite{varner_pix2005}. With other pixel matrices with
slower readout noise values of 15-20~e$^-$, S/N ratios larger than
20 and spatial resolutions of 1.5$\mu$m (5$\mu$m) for 20$\mu$m
(40$\mu$m) pitch have been measured~\cite{winter_PIX2005}.
The presently favored technology is the AMS 0.35$\mu$m OPTO
process, which possesses a 10$\mu$m thick epitaxial layer.
Regarding radiation hardness MAPS appear to sustain non-ionizing
radiation (NIEL) to $\sim$10$^{12}$n$_{eq}$. The effects of
ionizing radiation damage (IEL), the main damage source at the
ILC, are threshold shifts and leakage currents in and between nMOS
transistors. The damage effects are less severe when short readout
integration times ($\sim$10$\mu$s) are used. This way doses of
about 10 kGy can be tolerated~\cite{winter_PIX2005}.

The present focus of further development lies in improving the
radiation tolerant design, making 50$\mu$m thin detectors, making
larger area devices for instance by stitching over reticle
boundaries~\cite{MAPS2}, and increasing the charge collection
performance in the epi-layer.

\subsubsection{DEPFET pixels}
\begin{figure}[!h]
\begin{center}
\includegraphics[width=0.50\textwidth]{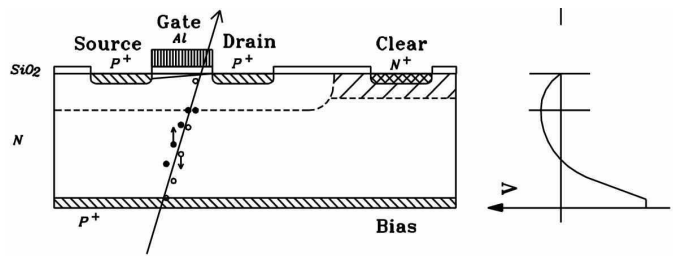}
\end{center}
\caption[]{Principle of operation of a DEPFET pixel structure
based on a sidewards depleted detector substrate material with an
imbedded planar field effect transistor. Cross section (left) of
half a pixel with symmetry axis at the left side, and potential
profile (right).} \label{DEPFET_principle}
\end{figure}
In so-called DEPFET pixel sensors~\cite{kemmer87} a FET transistor
is implanted in every pixel on a sidewards depleted~\cite{gatti84}
bulk. Electrons generated by radiation in the bulk are collected
in a potential minimum underneath ($\sim 1 \mu$m) the transistor
channel (internal gate) thus modulating its current
(Fig.~\ref{DEPFET_principle}). Electrons collected in the internal
gate are completely~\cite{sandow-kreuth05} removed by a clear
pulse applied to a dedicated contact outside the transistor.
Amplification values of $\sim$300 pA per collected electron in the
internal gate have been achieved. Further current amplification
and storage enters at the second level stage. The bulk is fully
depleted yielding large signals and the small capacitance of the
internal gate offers low noise operation, for a very large S/N
ratio. This in turn can be used to fabricate very thin devices.
Thinning of pn-diodes to a thickness of 50$\mu$m using a
technology based on wafer bonding and deep anisotropic etching has
been successfully demonstrated~\cite{laci04}.
\begin{figure}[htb]
\begin{center}
\includegraphics[width=0.7\textwidth]{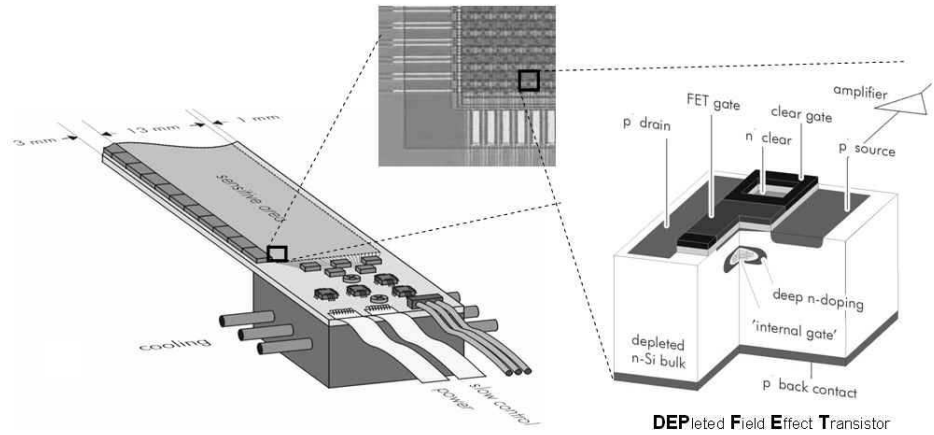}
\vskip 0.5cm
\includegraphics[width=0.6\textwidth]{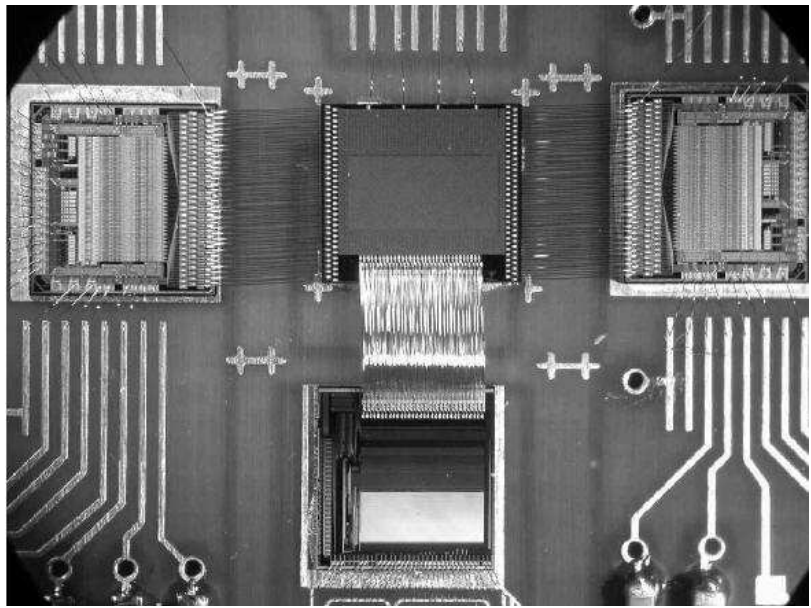}
\end{center}
\caption[]{(top) Sketch of a ILC first layer module with thinned
sensitive area supported by a silicon frame. The enlarged view
show a DEPFET matrix and a DEPFET double pixel structure,
respectively, (bottom) photographs of a DEPFET matrix readout
system (left). The sequencer chips (SWITCHER II) for select and
clear are placed on the sides of the matrix, the current readout
chip (CURO II) at the bottom; (right) stack of the hybrid together
with readout, ADC, and control boards operated in the testbeam.}
\label{DEPFET_TESLA}
\end{figure}

Readout of a DEPFET matrix is done by selecting a row by a gate
voltage from a sequencer chip (SWITCHER) to the external gate. The
drains are connected column-wise delivering their current to a
current-based readout chip (CURO) with amplification and current
storage at the bottom of the
column~\cite{Trimpl02,DEPFET_Portland}. A sketch of a module made
of DEPFET sensors is shown in Fig.~\ref{DEPFET_TESLA}(top). Figure
\ref{DEPFET_TESLA}(bottom) shows a DEPFET pixel matrix readout
system used in the test beam.

The radiation tolerance, in particular against ionizing radiation,
which is expected to doses of 2 kGy due to beamstrahlung at the
ILC, again is a crucial question. Irradiation with 30 keV X-rays
up to doses of $\sim$10 kGy, about five times the amount expected
at the ILC, have lead to transistor threshold shifts of only about
4 V. Threshold shifts of this order can be coped with by an
adjustment of the corresponding gate voltages supplied by the
SWITCHER chip. The estimated power consumption for a five layer
DEPFET pixel vertex detector at the ILC -- assuming a power duty
cycle of 1:200 -- is in the order of $\sim$5-10W. Such a
performance renders a very low mass detector without cooling pipes
feasible. A DEPFET pixel matrix with 128$\times$64 pixels has been
tested in a 6 GeV electron beam at DESY~\cite{wermes_LECC2005}.
The noise values obtained for the full system in the test beam
including sampling noise of the CURO chip is 200-300e$^-$. The
signal in a 450$\um$ thick DEPFET sensor was about 35000e$^-$.
At 6 GeV beam energy the spatial residuals are still multiple
scattering dominated. Residuals on the order of 10$\mu$m are
obtained, while with the large S/N value of 144 true space
resolutions in the order of 2$\mu$m should be possible and has
been recently demonstrated using high energy test beams at
CERN~\cite{jaap_private}.

\subsection*{Acknowledgments} I would like to thank the directors
and organizers of the 2006 SLAC Summer Institute for their effort
and efficiency in organizing this highly motivating and
interesting research school.

\bibliographystyle{unsrt}

\end{document}